# The ASD2 Chip for the Upgrade of the ATLAS MDT Chamber Readout at HL-LHC

S. Abovyan[+], A. Baschirotto[*], V. Danielyan[+], M. Fras[+], O. Kortner[+], S. Kortner[+], H. Kroha[+], M. de Matteis[*], F. Resta[*], R. Richter[+], Y. Zhao[+]

[+]) Max-Planck-Institut for Physics, Munich, [*]) University of Milano-Bicocca

Contact: richterr@mpg.mpp.de

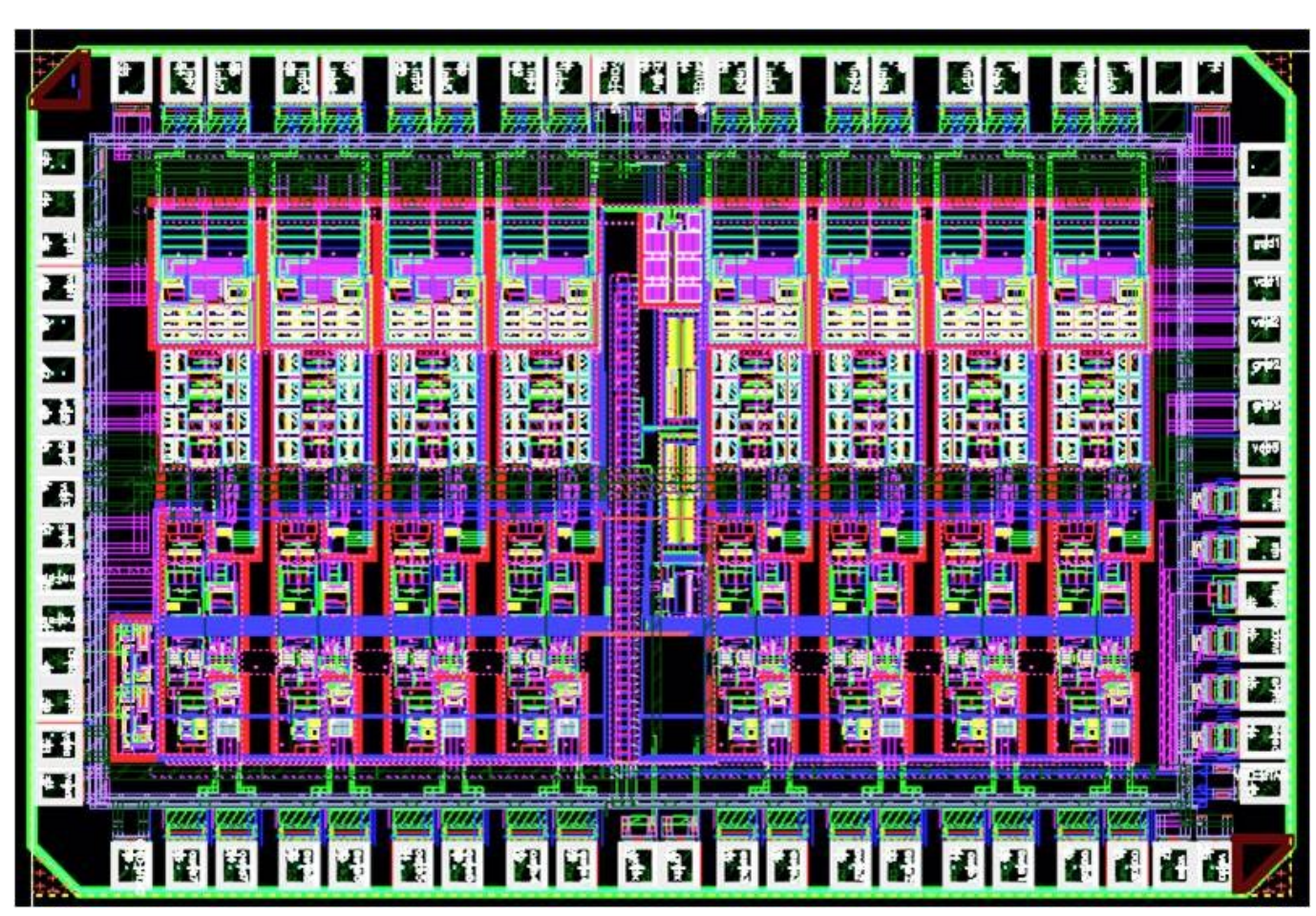

Abstract

Upgrading the ATLAS detector for operation at the High-Luminosity LHC requires a new, more selective trigger scheme to control the readout of MDT drift-tube chambers by incorporating RPC trigger information. This necessitates replacing the MDT readout electronics, including the front-end boards that house the ASD amplifiers and the custom-designed TDC. The latter buffers timing measurements for transmission to a chamber-mounted data concentrator (CSM), which, in turn, communicates with the MDT Data Processor, where MDT tracking data are combined with RPC timing information, to specifically identify high-energy muon tracks. In this article, we report on the new ASD2 preamplifier, discussing its architecture, design details, functionality and measured performance. We also present test results for chips from MPW runs, the engineering run and the volume production of 80,000 chips.

Unlike the ASD1 preamplifier, currently used for MDT readout, which relies on 500 nm chip technology, the ASD2 is manufactured using the more advanced 130 nm technology. This offers a range of technical improvements that enhance critical performance parameters such as signal rise time, noise levels, and the reproducibility of threshold settings across a chip's channels. Comparative measurements are presented to verify this improved performance. Finally, we discuss the robustness of ASD2 against environmental conditions like radiation exposure and potential high-voltage discharges within the MDT drift tubes.

# Table of Contents

# List of Figures

# List of Tables

# 1 The ATLAS Muon Spectrometer and the Upgrade to High Luminosity

The LHC collider is currently being modified to provide significantly higher luminosity to the experiment, which leads to higher event rates and, ultimately, greater sensitivity for processes with low cross-section.

Increasing the luminosity by approximately an order of magnitude necessitates a corresponding increase in readout bandwidth and storage capacity in the data processing chain, together with greater trigger system selectivity for specific event topologies. In the current system, high-energy muon tracks—with a transverse momentum exceeding 20 GeV—are identified solely by specialized trigger chambers, whereas in the upgraded trigger system, information from the MDT (Monitored Drift Tube) chambers will be incorporated into the Level-1 trigger decision. Thanks to their high spatial resolution, data from MDT chambers enhance the selectivity for these tracks [32]. In the new trigger system, the merging of track and trigger information takes place in the MDT Data Processor [29][31], which follows the data concentrator (CSM) mounted directly on the chamber. An overview of the architecture of the new Level-1 trigger system compared to the current system is provided in [30].

To handle the increased hit rates, a fundamental redesign of the MDT front-end electronics is required. This entails higher readout bandwidth while simultaneously adhering to the maximum permissible latency for data arrival at the MDT Data Processor. The newly developed TDC, housed on the frontend board and processing 24 MDT channels, meets these requirements by implementing, for instance, multiple parallel buffers instead of the single buffer used in the predecessor model [15].

Replacing the frontend boards necessitated new amplifiers and there-by offered the opportunity to profit from new chip technologies. IBM's 130 nm technology was selected for the design of the ASD2 chip, which, compared to the 500 nm technology from HP, on which the ASD1 design was based, offered significant performance improvements. In addition to smaller feature size, the 130 nm technology incorporates technical innovations such as copper rather than aluminum interconnects to reduce ohmic resistance, low-k dielectrics to minimize parasitic capacitance, and thinner gate oxides (2 nm) for fast switching times. Collectively, these innovations improved signal rise time, noise characteristics, and process control, the latter innovation resulting in better matching of threshold voltage parameters for identical structures on the same chip.

# 2 MDT system overview

## 2.1 The MDT precision tracking Chambers

The Monitored Drift Tube (MDT) spectrometer represents the outermost shell of the ATLAS experiment at the LHC collider at CERN. It is segmented in 1084 rectangular and trapezoidal chambers made out of 370000 drift tubes and covers an area of about 5500 m2. Shape and size of the chambers follow the projective geometry of the spectrometer and contain between 192 and 432 individual tubes. The early concept of the Muon Spectrometer is described in the TDR from 1997 [1], while an up-to-date description of the present ATLAS detector and of MDT readout electronics is given in [2] and [3], respectively. The LHC collider will go into a new mode of high-luminosity operation in 2025, providing about 7 times higher luminosity ("Phase-II"). The corresponding higher particle rates require the complete replacement of the MDT readout electronics. A detailed description of this upgrade is presented in section 2.4 of [4].

The MDT chamber design is optimized for precision tracking. The MDT tubes are pressurized at 3 bar to reduce diffusion effects and increase the rate of primary electrons, while the position of each chamber inside the spectrometer – crucial for the accuracy of momentum measurement - is monitored by an optical alignment system. The main operating parameters of the MDT are summarized in Table 2.

A schematic diagram of the MDT tube supplies and readout is given in Figure 1, where HV supplies are located at the right and components for signal readout on the left. On the right side, the wire is terminated with the tube impedance of 380 Ω in order to avoid signal reflection. On the readout side, signals from 24 tubes are collected by a passive distribution board ("Hedgehog board") and channeled to a readout board ("Mezzanine card"), containing three Amplifier/Shaper/ Discriminator (ASD) chips, a single 24-channel TDC and control circuitry.

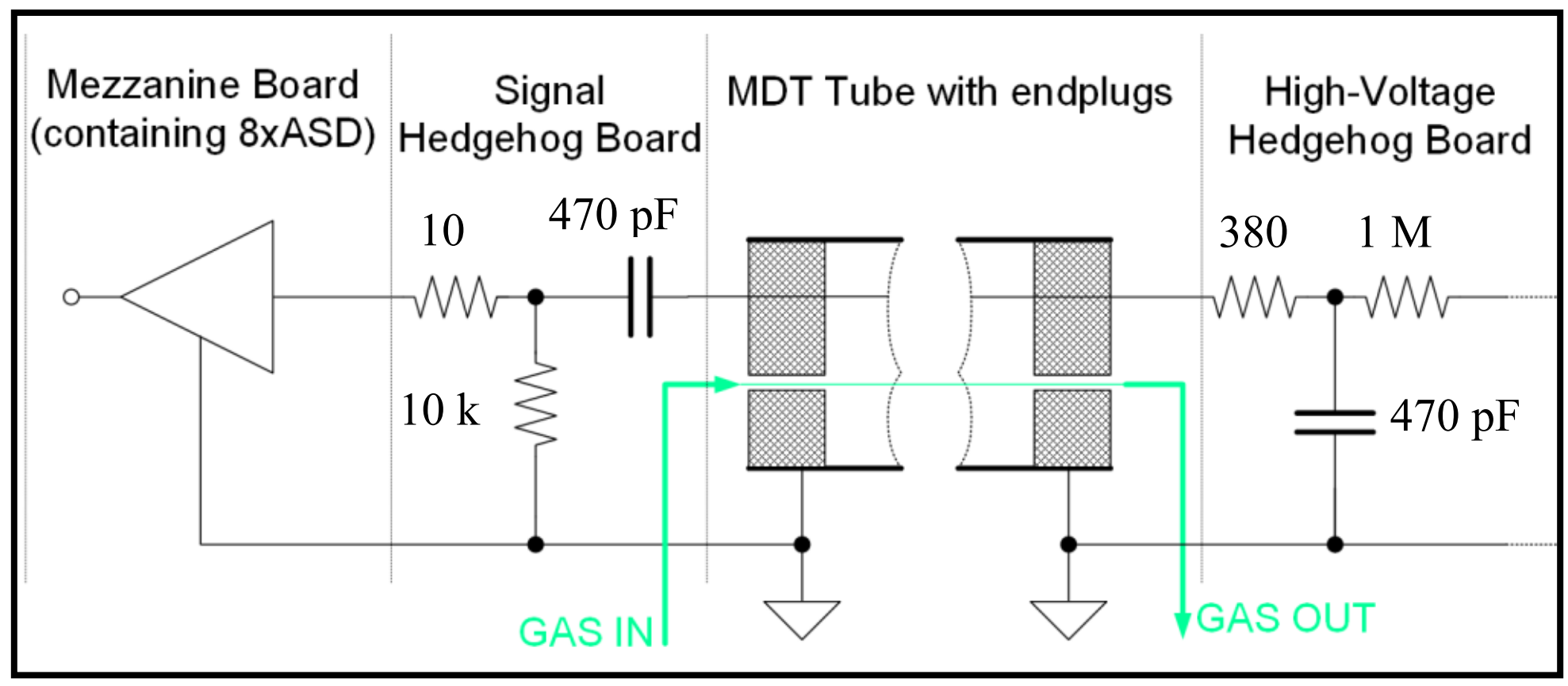

*Figure 1 - Interface of the MDT-ASD with the tube*

A single MDT chamber may have up to 432 drift tubes, serviced by 18 hedgehog/mezzanine board sets. Data are read out from each TDC via a 80 Mbit/s serial link[1] to a Chamber Service Module (CSM), multiplexing the (up to) 18 serial links into an optical fiber for transmission to the ATLAS DAQ.

A daisy-chain JTAG bus permits downloading of parameters to ASDs and TDCs. Each multilayer (3 or 4 tube layers) is shielded at both ends by a Faraday Cage. All AC signals exchanged with the CSM are low-level differential signals (LVDS), while DC supplies are distributed from the CSM to each mezzanine board. MDT chambers are electrically isolated from support structures and (metallic) gas services, being grounded to a single common ground point to avoid ground loops. DC power supplies are "floating", i.e. their grounds are not connected among each other or to any other ground (e.g. safety ground).

*Table 1 - Component numbers in the MDT Readout for 30 mm and 15 mm tube chamber in Phase-II*

| | Large tubes | Small tubes | Total |
|---|---|---|---|
| MDT Chambers | 958 | 126 | 1.084 |
| Tubes | 305.952 | 63.514 | 369.466 |
| ASDs | 38.244 | 8.106 | 46.350 |
| Mezz. cards | 12.748 | 2.702 | 15.450 |
| CSMs | 958 | 240 | 1.198 |

While MDT tubes with a diameter of 30 mm are used in the largest part of the Muon detector, a special MDT type with half-diameter tubes is used in regions of high particle rates ("sMDT"). The smaller diameter leads to shorter drift time and lower acceptance for background hits (caused by converted ?'s and neutrons), improving hit efficiency and accuracy at high rates. Details about the sMDT are given in [4] and [5]. Table 1 contains numbers for the main components of the MDT readout system.

[1] In the upgraded version of the readout electronics, the bandwidth of this link will be increased to 320 Mbit/s [4].

## 2.2 Characteristics of the MDT tube signals

An ionizing track crossing a MDT tube generates a string of primary electrons which subsequently drift towards the central anode wire, see Figure 2. Depending on the distance of the track from the wire, the drift time may extend up to 750 ns, leading to a sequence of signals at the input of the amplifier of corresponding duration. Depending on the spatial distribution of primary electron clusters along the path of the track, described by Poisson and Landau statistics, more than one crossing of the discriminator threshold may occur due to one single track, while only the first crossing (corresponding to the electrons closest to the wire) is relevant for the determination of the track-to-wire distance. Subsequent crossings do not contain useful information and lead to an unwanted increase of data flow and the corresponding load to the DAQ. To discard all but the first threshold crossing, a veto for the discriminator has been installed, the length of which can be programmed in the range of about 150 to 750 ns.

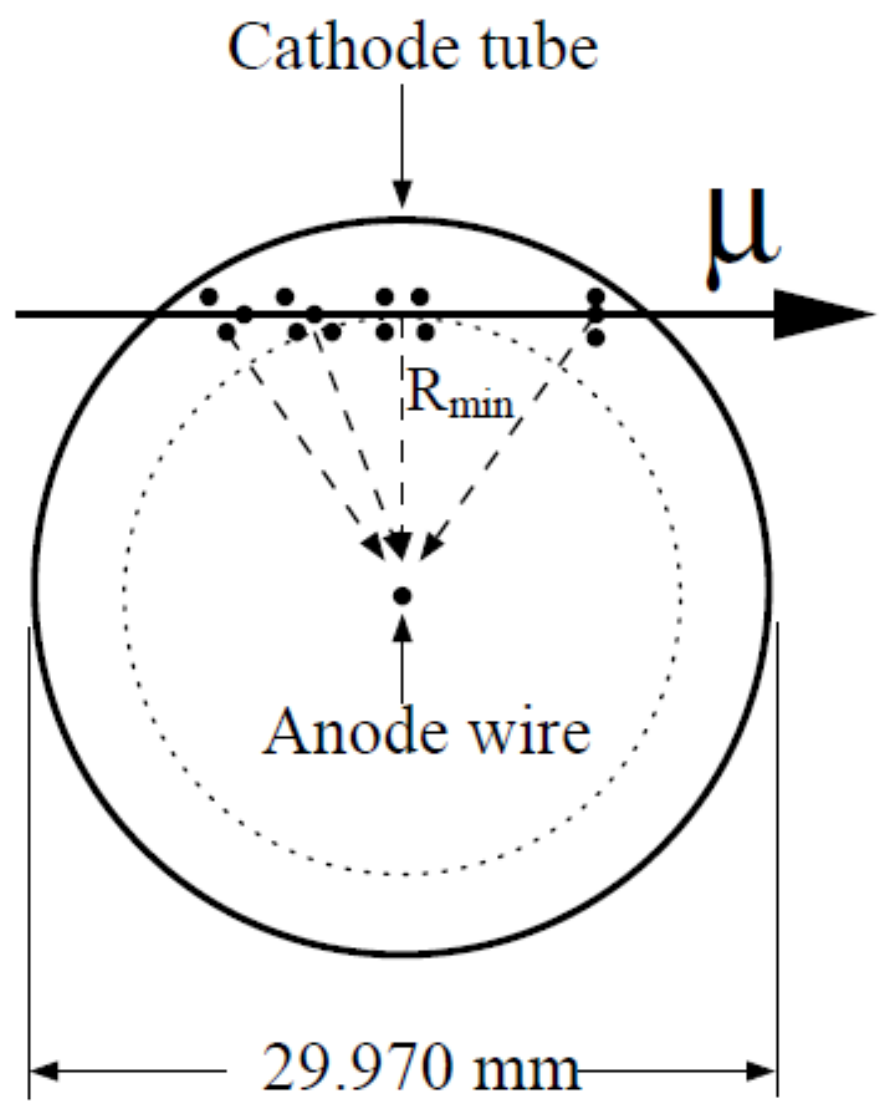

*Figure 2 - Only the primary electrons of an ionizing track at $R_{min}$ are relevant for the drift time measurement*

*Table 2 - MDT properties*

| Parameter | Large tubes | Small tubes |
|---|---|---|
| Length | From 1.5 to 6 m | 1.5 to 2.50 m |
| Diameter | 30 mm | 15 mm |
| Wire diameter | 50 µm | idem |
| Wire resistance | 44 Ω / m | idem |
| Impedance ($Z_0$) | 380 Ω | 340 Ω |
| Termination | 380 Ω in series with 470 pF | 340 Ω w. 470 pF |
| AC coupling capacitor | 470 pF | idem |
| Drift gas | $Ar/CO_2$ (93%/7%) | idem |
| Maximum hit rate | 500 Hz/cm$^2$ | 30 kHz/cm$^2$ |
| Nominal operating voltage | 3090 V | 2730 V |
| Electron avg. drift velocity | 20 µm/ns | 42 µm/ns |
| Maximum drift time | 750 ns | 180 ns |

## 2.3 Performance requirements of the Front-end electronics

For an accurate measurement of the drift time, the design of the ASD must address a number of critical requirements:

- With a gas gain of about $2{\times}10^4$, the expected average signal is 1500 electrons (0.25 fC) per primary electron. The standard trigger threshold is 5 primary electrons, equal to 1.25 fC at the input of the ASD. The noise contribution from the frontend should not exceed 1500 electrons. The frontend noise depends on the performance of active elements in the ASD, but also on passive components in the MDT, like the terminating resistor and the total capacitive load of the tube.
- Because of the high signal rates in chambers close to the beam pipe (up to 400 kHz/tube), bipolar shaping is chosen to avoid deterioration of efficiency due to baseline fluctuations.
- The channel to channel crosstalk is specified to be less than 1%.
- A pre-amp peak time of <15 ns is specified to minimize time slewing effects due to pulse height variations.
- To further reduce the error due to pulse height variations, an ADC is implemented in each of the 8 channels to measure the signal charge following the initial threshold crossing. The signal charge is encoded into the pulse width using the dual-slope "Wilkinson" technique (see section 3.4.6). This charge information allows to improve the accuracy of the time measurement by a charge dependent correction [3][4], thus improving the spatial resolution of the detector. The charge information is also used to monitor gas amplification and for other diagnostic purposes.

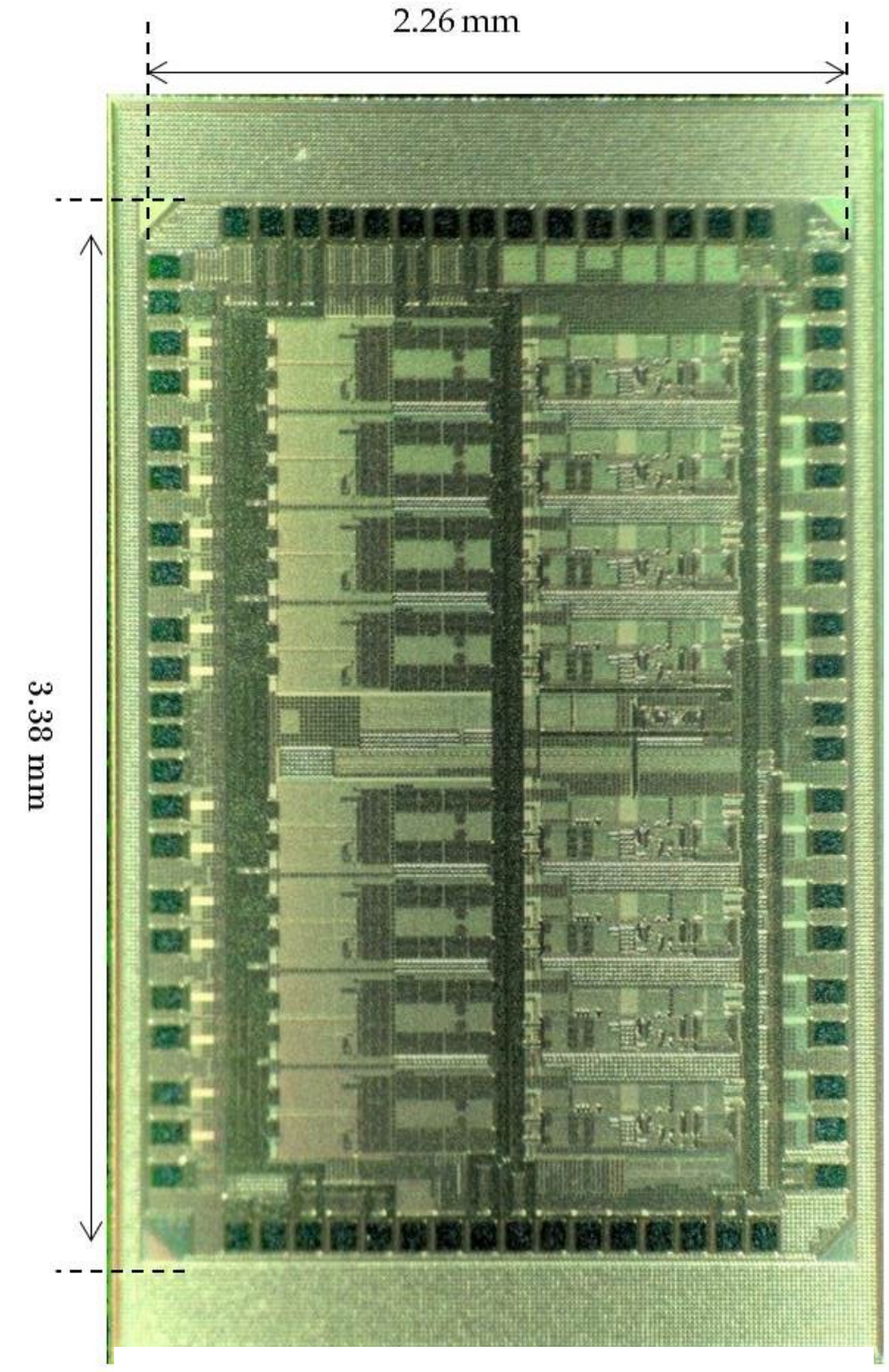

*Figure 3 - Photo of the ASD die.*

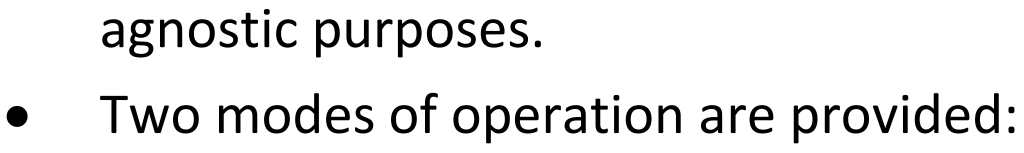

- Two modes of operation are provided:
  (a) Time over Threshold (ToT) mode: the length of the ASD output corresponds to the time of the input signal above threshold.
  (b) ADC mode: the length of the ASD output corresponds to the charge, contained in a time window of 8-16 ns after threshold crossing, thus representing an approximate measure of the peak amplitude of the incoming signal.
- To suppress multiple threshold crossings for a single track passing a tube, a “dead time” (DT) is introduced, disabling the discriminator for a programmable time after a threshold crossing. The DT can be selected in the range 180-900 ns.

Accuracy of the TDC:

The ATLAS Muon Spectrometer aims for a resolution of the transverse momentum ($p_T$) of 10% for 1 TeV muons, which translates into an RMS-resolution requirement of < 80 µm for a single tube. The binning error of the TDC should be negligible besides errors due to physics effects, like diffusion, distribution of primary electrons along the track and statistics of gas amplification. Given the electron drift velocity of 20 and 42 µm/ns in large and small MDT tubes, the 0.78 ns binning error of the time measurement results in an error of 5 and 10 µm (RMS) for large and small tubes, respectively. The least significant bit (LSB) of the TDC measurement, used in the MDT readout, is defined by the LHC clock of 25 ns divided by 32.

# 3 The Design of ASD2

The ASD2 is an octal CMOS Amplifier/Shaper/Discriminator, which has been optimized for the ATLAS MDT chambers. The MDT use the RC net shown in Figure 1 to suppress signal reflection from the “far end” of the tubes. The noise contribution from the terminating resistor is the dominant noise source in the ASD implementation [10][11]. For reasons of design flexibility and cost, implementation as an ASIC in a high-quality analog CMOS process has been selected for this device.

## 3.1 Overview and specifications

The structure of one analog channel of the ASD is shown in Figure 4. It is a fully differential structure with a Charge Sensitive Preamplifier (CSP) and three shaping stages (DA1-DA3), followed by a discriminator and a leading-edge charge integrator to be described later. The negative input of the CSP (Figure 5) is connected to the signal source of the MDT tube (i.e. the central wire), while the positive one is fixed with an internal resistor divider to a proper bias voltage. The capacitance formed by wire and tube corresponds to a parasitic capacitance of about 30 pF. At the same node, a 66 pF capacitor guarantees matching between the two differential inputs.

The analog specifications for the ASD1 [15] are summarized in Table 3 and were also used as a starting point for the design of ASD2. For comparison, the performance parameters measured in tests of the finished chip, are shown in the right column. Details of the corresponding measurements are presented in section 4.

The ASD2 was designed in the 130 nm technology, which provided an important number of technical innovations, which could be expected to improve its performance w.r.t. several critical parameters.

*Table 3- Specifications or AD1 and ASD2 and measured parameters for ASD2*

| | Specification for ASD1 and ASD2 | Measured performance of ASD2 |
|---|---|---|
| Input impedance | $Z_{IN} = 120\ \Omega$ | |
| Shaping function | Bipolar | |
| $\sigma_{noise}$ at the ASD input | 6000 electrons or ~ 5 prim. $e^-$ | |
| Shaper peaking time | 15 ns | 12 ns |
| Sensitivity at discriminator DA4 for a δ-pulse to the ASD input | 1.65 mV/$e^-$ or 8.9 mV/fC | 4.8 mV/$e^-$ or 20 mV/fC |
| Linear range within 10% | 1.5 V | 0.65 V |
| Threshold setting for ~ 5 $\sigma_{noise}$ | 40 mV | 24 mV |

It should be noted that the specification of a linear range of 1.5 V, (Table 3 line 6) was unrealistic for both types of ASDs. Simulation in section 3.4.8 and measurements in 4.1.3 (Figure 30) show linearity at the level of 10% up to about 50 fC, which corresponds to 650 mV for the ASD2 and to 420 mV for ASD1. For the accuracy of the time-of-arrival measurement, which is the figure of merit of the MDT readout, only the linearity in the range of small signals around the threshold is relevant, as the threshold for triggering the discriminator is in the range of a few fC.

## 3.2 Fabrication Process

The ASD chip has been realized with the 130 nm process (CMRF8RF) by Global Foundry (previous IBM), using the option for eight metal layers. This implies a sheet resistance of 71 mΩ/sq for the first metal (M1) that reduces down to 7 mΩ/sq for the last one (MA). The selected transistors for the design are nfet33 and pfet33, which operate at 3.3 V supply voltage[2] and a nominal threshold of 380 mV and -320 mV, respectively. The key electrical parameters of these devices are listed in Table 3

*Table 4 - Global Foundry 130nm Process Parameters of Field Effect Transistors (FETs)*

| Parameter | 3.3V I/O nfet33/pfet33 | Unit |
|---|---|---|
| Nominal supply voltage $V_{DD}$ | 3.3 | V |
| Gate oxide thickness $T_{OX}$ | 5.2 | nm |
| $L_{DES,MIN}$ | 0.40 | µm |
| $L_{EFF}$ | 0.335 | µm |
| $Vt_{SAT}$ | 380 / -320 | mV |
| ON current $I_{ON}$ | 740 / -380 | µA/µm |
| OFF current $I_{OFF}$ | 30 / -30 | pA/µm |
| Max supply voltage | 3.6 | V |

## 3.3 Topology and Architecture

The ASD2 was conceived to match or exceed the performance of the ASD1 [15]. It consists of a fully differential chain (Figure 4), optimized for the IBM 130 nm CMOS 8 RF-DM technology. A fully differential Charge Sensitive Preamplifier (CSP) is the best design for optimizing the immunity against external noise sources, in particular - rejection against Common Mode on the signal inputs as well as rejection of other noise on the power supply lines. The CSP and the three subsequent differential amplifier stages (DA1, DA2 and DA3) compose the analog part of the channel. The negative input ($IN_{CSP}^{-}$) of the CSP is connected to the MDT and the positive input ($IN_{CSP}^{+}$) is biased on-chip to about 783 mV for the purpose of input matching. The CSP amplifies the input charge, generating a proportional voltage signal which is amplified and shaped in the subsequent stages (DA1 to DA3).

The fully differential structure was chosen for noise immunity, while bipolar shaping was selected to reduce the effects of baseline shift at high signal rates [3][5][8]. For each of the eight channels, the output of the DA3 shaping stage is fed into the discriminator DA4 as well as into the Wilkinson Analog-to-Digital Converter (WADC). The DA4 compares the output of DA3 with the programmable DISC1 threshold. At the moment of threshold crossing, the 2 outputs of the DA4 change polarity, thus defining the "leading edge" of the ASD output. At this moment the WADC stage starts a charge-to-time conversion, based on the amount of input charge supplied to a capacitor during a short, programmable time span, and the Dead Time generator starts to inhibit the DA4 for a fixed, programmable time.

The operation of the ADC proceeds in the following steps:

- Integration of the DA3 output on a holding capacitor for a programmable time span, defined by the "Integration Gate Width" parameter (see 3.5.1.2).
- Discharge of the holding capacitor at a constant, programmable current, defined by the "Rundown Current" parameter (see section 3.5). Once the capacitor discharged, the output of DA4 changes polarity again, going back from HIGH to LOW, thus defining the “trailing edge” of the ASD output.

[2] The supply voltage to be used in the experiment was later reduced to 3 V, see discussion in sect. 4.3.2.

The integration and discharge times determine the width of the Wilkinson ADC output (the delay between leading and trailing edge), encoding the charge in the first 10 to 30 ns of the incoming signal, which is an approximate measure of the signal amplitude. This information can be used to compensate for the effect that small pulses take longer to reach the discriminator threshold than large pulses, allowing for a pulse height dependent correction in data analysis (“time slewing correction”).

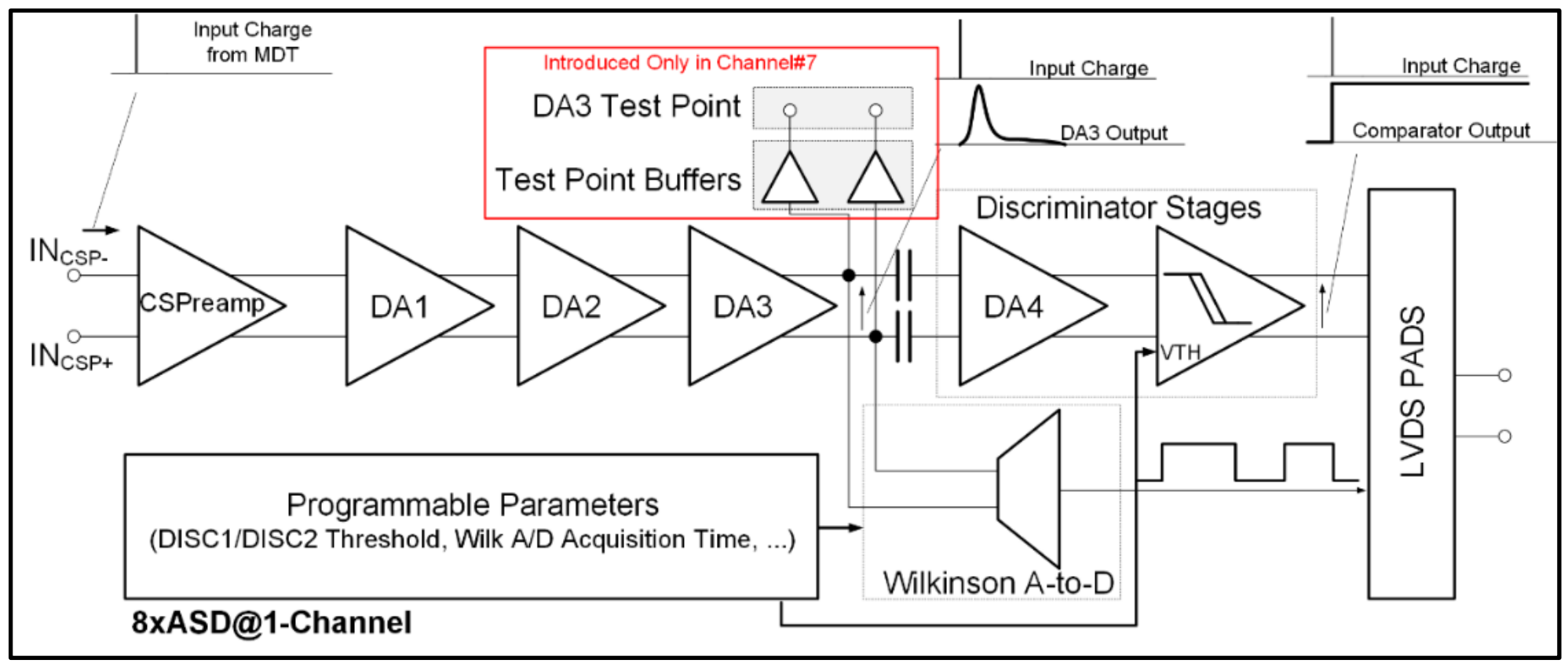

*Figure 4 - Block Diagram of one ASD channel*

Differential logic controls the operation of the Wilkinson ADC, leading to good immunity against substrate coupling. Coupling between the digital and analog domains is strongly reduced by the implantation of a highly resistive BFMOAT layer in the substrate. These „Moats” of insulating Boron-Fluoride implants are geometrically closed trenches in the substrate, preventing ohmic coupling across the trench. Further decoupling is obtained by strictly separating voltage and ground supplies among analog and digital domains.

The output of the discriminator stage is sent to the LVDS PADS cell and converted to external low-level signals (about ±200mV signal swing into 100Ω). Each ASD channel draws approximately 15 mA from a 3.3 V supply, thus dissipating about 50 mW per channel (Table 14 and Table 15). The operation of the ASD sub-cells is described in the following sections. The corresponding measurements of the analog behavior of the finished chip are presented in section 1.

## 3.4 The Analog Signal Chain

### 3.4.1 Charge Sensitive Preamplifier (CSP)

The design of the ASD was based on specifications given in Table 3, summarized below.

- Power dissipation of the CSP: 12.9 mW (~ 3.9 mA @ 3.3 V)
- $Z_{IN}$: < 120 ohms (DC & AC/dynamic)
- Charge gain: ~ 8.9 mV/fC
- DC voltage gain: > 100 (40dB)
- Input noise density: 1.3 nV/√Hz
- ENC (with 380 Ω termination): 6000 e- rms
- Peak time < 15 ns for a delta pulse, generated by a voltage step, capacitively injected into the ASD input.

The Charge Sensitive Preamplifier has been implemented with a cascade input differential stage and an optimized M4-RS-M5 source follower as shown in Figure 5. This is the key block of the chain, and its performance strongly depends on the parasitic capacitance $C_D$ of 60 pF connected to the negative input. The large value of $C_D$ is due to the capacitance between sense wire and metallic wall of the tubes and to parasitic capacitance along the signal path (Figure 1). This affects the closed-loop-gain and the bandwidth and leads to a reduction of signal rise-time and sensitivity of the CSP.

The transistors of the input differential pair (M1A and M1B in Figure 5) are crucial for an optimum trade-off between noise, speed and power consumption. For this reason, they are very large (W = 2 mm and L = 400 nm), sink about 1.8 mA of current and have a transconductance $g_m$ of about 34 mA/V. The values of currents, voltages, dimensions of passive/active components are reported in Figure 5 and in Table 5.

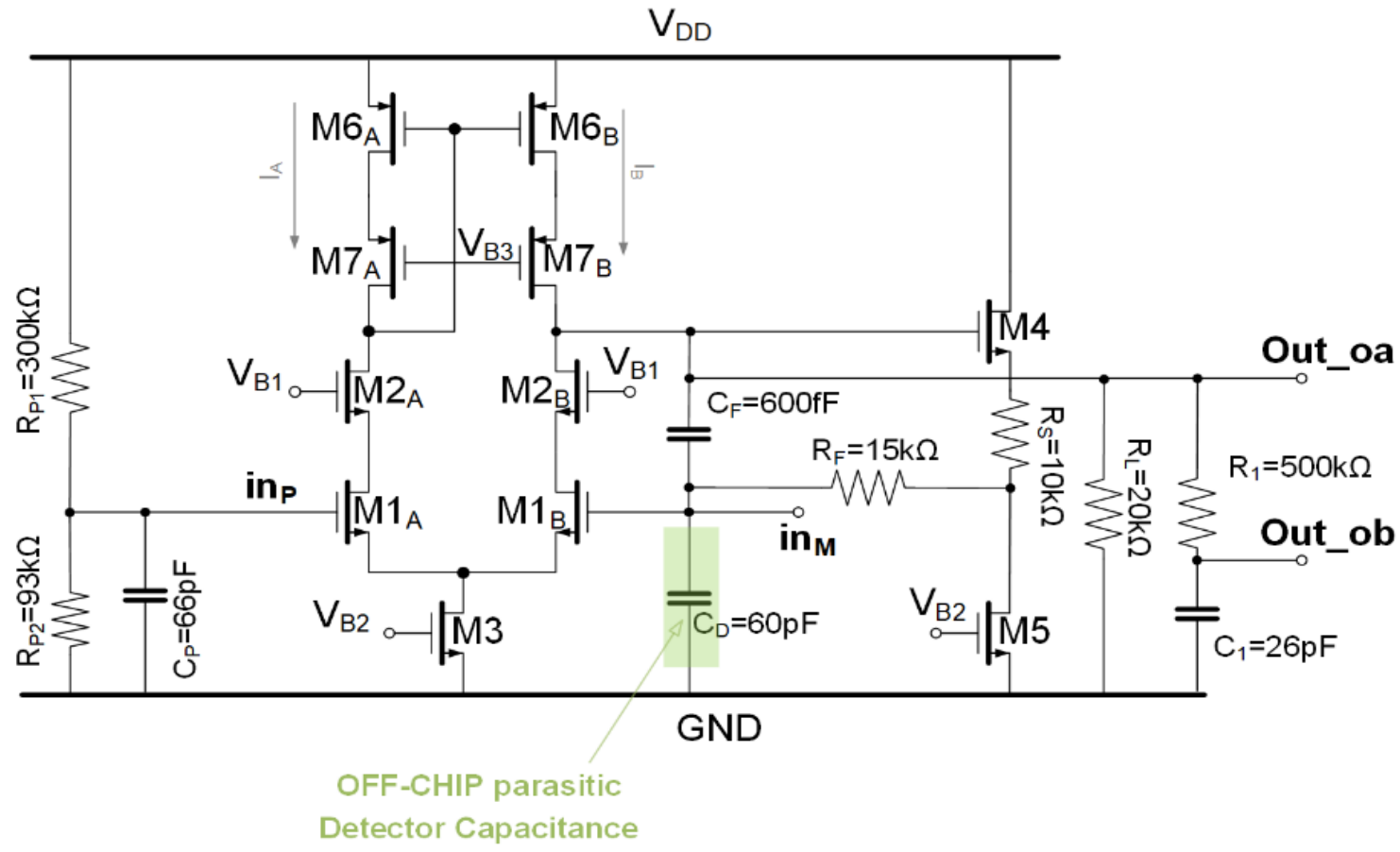

*Figure 5 - Simplified Scheme of the Charge Sensitive Preamplifier (CSP)*

The positive input has been fixed by a resistor divider (RP1 and RP2) to 0.783 V and connected to an integrated capacitance $C_P$ of 66 pF, reducing the mismatch introduced by $C_D$ when the tube is connected. The currents $I_A$ and $I_B$ are implemented through low voltage cascade current mirrors ($M6_{A-B}$ and $M7_{A-B}$ in Figure 5). They mirror the current from a reference of 31 µA and contribute to determine an equivalent impedance at the node Out_oa ($R_{OUT}$) of about 9.8 kΩ. It is given by the parallel between

- the load resistance ($R_L$) of 20 kΩ
- the $M_{1B}$-$M_{2B}$ output impedance ($R_{M2B-M1B}$) of about 32kΩ and
- the mirror impedance ($R_{IB}$) of about 46 kΩ.

The resulting closed-loop gain is 50 dB. The large input device sizes result in approximately 2 pF of gate-source parasitic capacitance, which, however, is negligible w.r.t. the detector parasitic capacitance $C_D$. The feedback network has been implemented with an equivalent resistor $R_F = R_{F1} + R_{F2}$ of 25 kΩ and a capacitance $C_F$ of 600 fF. This way, the $M_3$ - $R_{F2}$ - $M_4$ source-follower optimizes the common-mode voltage for the output node, leading to a better operation of the following DA1 stage, which has NMOS input transistors (and for thermal noise minimization).

*Table 5 - Parameters of the most important CSP transistors*

| MOS | W/L | $g_m$ – | $r_{DS}$ – [kΩ] |
|---|---|---|---|
| $M1_A$ | 2mm/400nm | 34.36 | 1.03 |
| $M1_B$ | 2mm/400nm | 33.36 | 0.978 |
| $M2_A$ | 150µm/400nm | 17.38 | 3.14 |
| $M2_B$ | 150µm/400nm | 16.74 | 2 |
| M3 | 928µm/3µm | 18.41 | 505.7 |
| M4 | 80µm/500nm | 1.187 | 118 |
| M5 | 16µm/3µm | 0.348 | 549 |

The CSP transfer function[3] depends on the feedback network ($R_F$ and $C_F$), the transconductance of the input transistor ($M_1$), the output impedance ($R_{OUT}$) and the detector capacitance ($C_D$). It can be approximated by Eq. 1 [20]. This equation can be further approximated by Eq. 2, replacing the ratio $R_{OUT}/R_F$ and neglecting the higher frequency zero (since $g_{m_{M1}} \gg 1/R_F$ and the zero is around 9 GHz).

[3] A concise introduction to the concept of transfer functions and to the application of the Laplace Transform for the analysis of electronics networks is given in [14].

$$T(s) \cong -R_F \frac{1 + s\frac{C_F}{g_{m_{M1}}}}{\left(1 + sC_F R_F \left(1 + \frac{C_D}{C_F}\left(1 + \frac{R_{OUT}}{R_F}\right)\frac{1}{1 + g_{m_{M1}} R_{OUT}}\right)\right)\left(1 + s\frac{C_D}{g_{m_{M1}}}\right)} \quad \text{Eq. 1}$$

$$T(s) \cong -R_F \frac{1}{\left(1 + sC_F R_F \left(1 + 1.39\,\frac{C_D}{C_F}\frac{1}{g_{m_{M1}} R_{OUT}}\right)\right)\left(1 + s\frac{C_D}{g_{m_{M1}}}\right)} \quad \text{Eq. 2}$$

$$\tau_{P1} \cong \tau_{P1_{IDEAL}} \left(1 + 1.39\,\frac{C_D}{C_F}\frac{1}{g_{m_{M1}} R_{OUT}}\right) \cong 1.4 \cdot C_F R_F \cong 21\,ns \quad \text{Eq. 3}$$

$$S_{CSP} \cong S_{IDEAL} \left(\frac{1}{1 + 1.39\,\frac{C_D}{C_F}\frac{1}{g_{m_M 1} R_{OUT}}}\right) \cong 0.7 \cdot \frac{1}{C_F} \cong 1.1\frac{mV}{fC} \quad \text{Eq. 4}$$

$$\tau_{P2} \cong \frac{C_D}{g_{m_{M1}}} \cong 1.7\,ns \quad \text{Eq. 5}$$

The values of $C_D$, $C_F$, $g_{mM1}$ and $R_{OUT}$ determine the constant time of the dominant pole ($\tau_{P1}$) and the sensitivity ($S_{CSP}$) according to Eq. 3 and Eq. 4 respectively. The variations from the ideal case depend mainly on a non-negligible parasitic detector capacitance ($C_D$). Moreover, the time constant of the non-dominant pole is fixed by the input transistor transconductance (see Eq. 5) and guarantees a good time response.

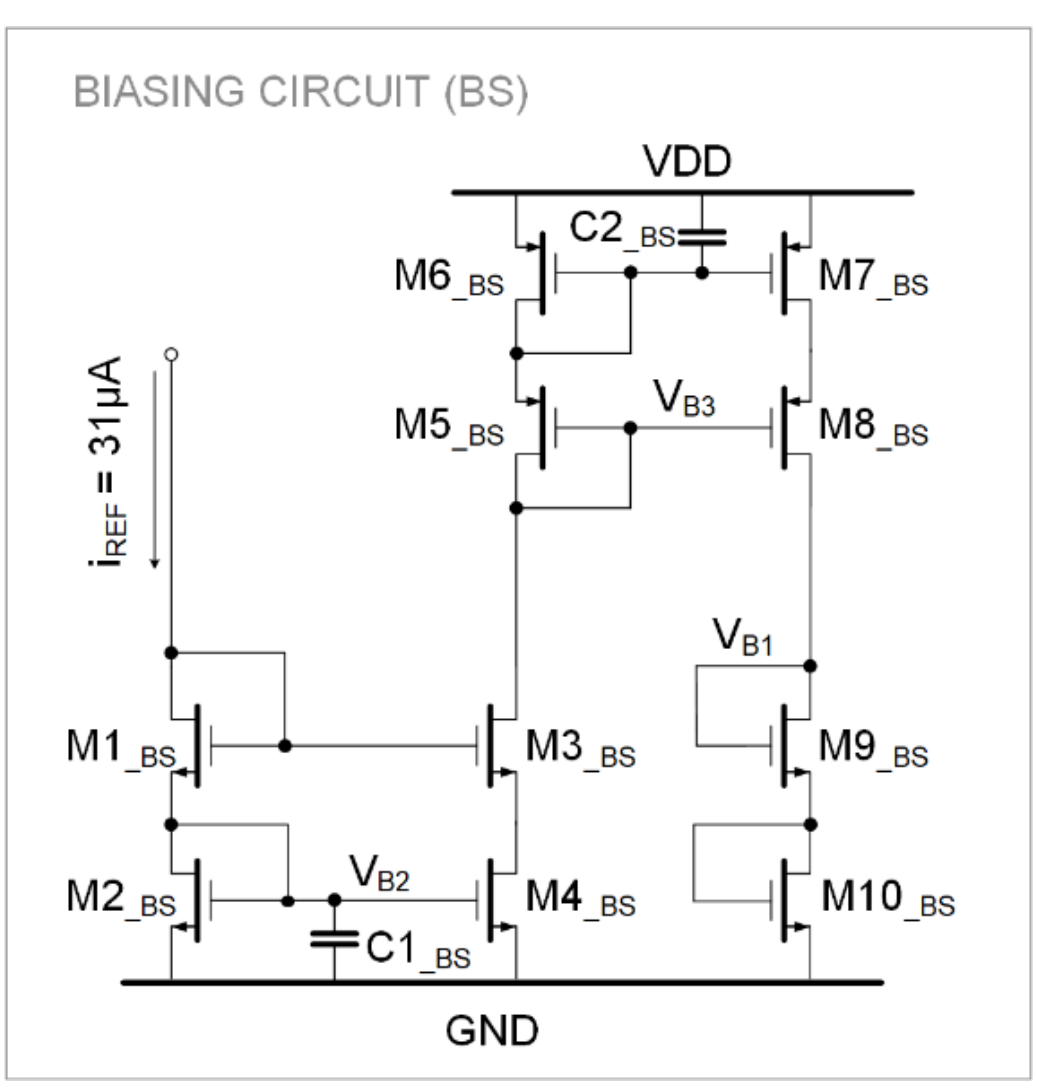

*Figure 6 - Scheme of the CSP Biasing Circuit*

In a similar way, the input impedance transfer function is approximately given by Eq. 6. It is almost constant ($Z_{IN}(0) \cong 57\ \Omega$) for all in-band frequencies since it is fixed by ($R_F$, $R_{OUT}$, $g_{mM1}$) at low frequencies and by ($C_F$, $C_D$) at higher frequencies.

$$Z_{IN}(s) \cong \frac{T(s)}{g_{m_{M1}} R_{OUT}} \quad \text{Eq. 6}$$

Out_oa and Out_ob nodes in Figure 5 represent the CSP outputs and are biased at the same voltage through a proper R1 - C1 net. Instead, VB1 bias voltage is provided by a bias circuit (Figure 6) exploiting a cascade current mirror.

### 3.4.2 The Differential Amplifiers

The CSP preamp is followed by a chain of three differential amplifiers, DA1 through DA3, implemented with the identical circuit topology (Figure 7). The DA1 to DA3 chain amplifies and filters the CSP output [20][21]. The bipolar shaping at DA3 determines the fall-down time of the signal. This is achieved by replacing the ohmic impedances Z1 and Z2 in DA1 by a combination of resistors and capacitances. The $M4_A$ – $M4_B$ and $M5_A$ – $M5_B$ transistor pairs implement the common mode feedback of the stage and work in the linear region with about 40 mV of $V_{DS}$.

Shaping is implemented by the stages DA2 and DA3 using Z1-RC networks. All values are selected to cancel the very long time constant component of the positive ion MDT pulse [7].

Each stage has a gain given by the ratio between Z2 and Z1, while the bandwidth is fixed by the resistive and capacitive loads connected to the output nodes. The design values of Z1, Z2, I1 (equal to I2) in the stages DA1 to DA3 are listed in Table 6 together with current and power consumption.

Figure 7 also shows the biasing circuit. A mirror current biases the gates of M3 and M2 transistors, mirroring the proper current to the input differential pair ($M1_A$ – $M1_B$).

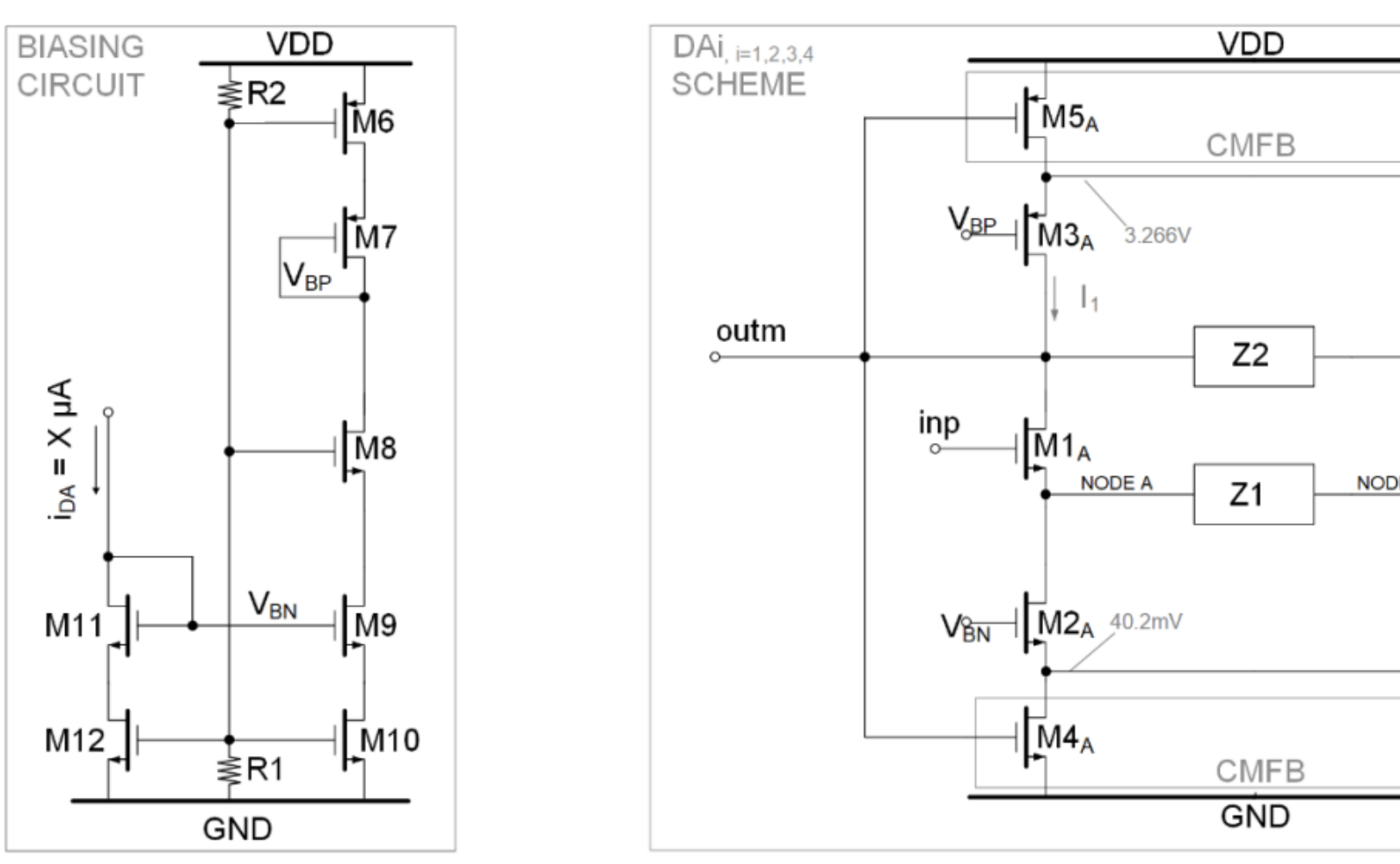

*Figure 7 - Differential Amplifier Scheme (right) with its biasing circuit (left)*

*Table 6 - Component values for the amplifiers DA1 to DA3 in the circuit topology of Figure 7*

| PARAMETER | DA1 | DA2 | DA3 |
|---|---|---|---|
| Z1 | 1.33 kΩ | $2.44\ k\Omega \parallel \left(2.4\ k\Omega + \frac{1}{s \cdot 42.8\ pF}\right)$ | $2.4\ k\Omega + \frac{1}{s \cdot 42.8\ pF}$ |
| | | The parallel of R=2.44 kΩ and the series R-C of 2.4 kΩ and 42.8 pF | Series of R=2.4Ω and C= 42.8 pF |
| Z2 | 6.26 kΩ | 6.26 kΩ | 9.47 kΩ |
| I1 = I2 | 283 µA | 295 µA | 293 µA |
| $I_{TOT}$ | 746 µA | 748 µA | 745 µA |
| $P_{TOT}$ @ $V_{DD}$ = 3.3 V | 2.46 mW | 2.47 mW | 2.45 mW |

### 3.4.3 Pre-discriminator Gain Stage

The DA3 shaper output is AC coupled to one additional differential amplifier, DA4, referred to as pre- discriminator gain stage, which provides additional gain to the discriminator. The circuit topology of this stage is identical to the one of the previously presented differential amplifiers (Figure 7), with the Z1 and Z2 impedances equal to 0 Ω and 2.7 kΩ. NODE A and NODE B of Figure 7 are shorted to maximize gain and bandwidth at the expense of higher sensitivity of the gain to process variation. Since, however, the threshold is applied at its input, the gain sensitivity to process variation is irrelevant. Instead, the smaller load resistance (Z2) provides lower driving impedance to the subsequent discriminator stage.

### 3.4.4 The Discriminator

The discriminator (Figure 8), is a high DC-gain differential amplifier with symmetrical current-mirror loads and a main differential pair, $M_{IN_A}/M_{IN_B}$, biased at about 500 µA. Two current-mirror "loops" provide a differential gain of about 500 with no hysteresis.

Hysteresis is provided by the M1a/M1b pair, which unbalances the static current through the main differential pair by a variable external current. The main bias current is provided by R1 (poly-resistor) as shown in the yellow box of Figure 8.

It should be noted that the setting of the hysteresis value leads to an increase of the effective trigger threshold (see section 4.1.6).

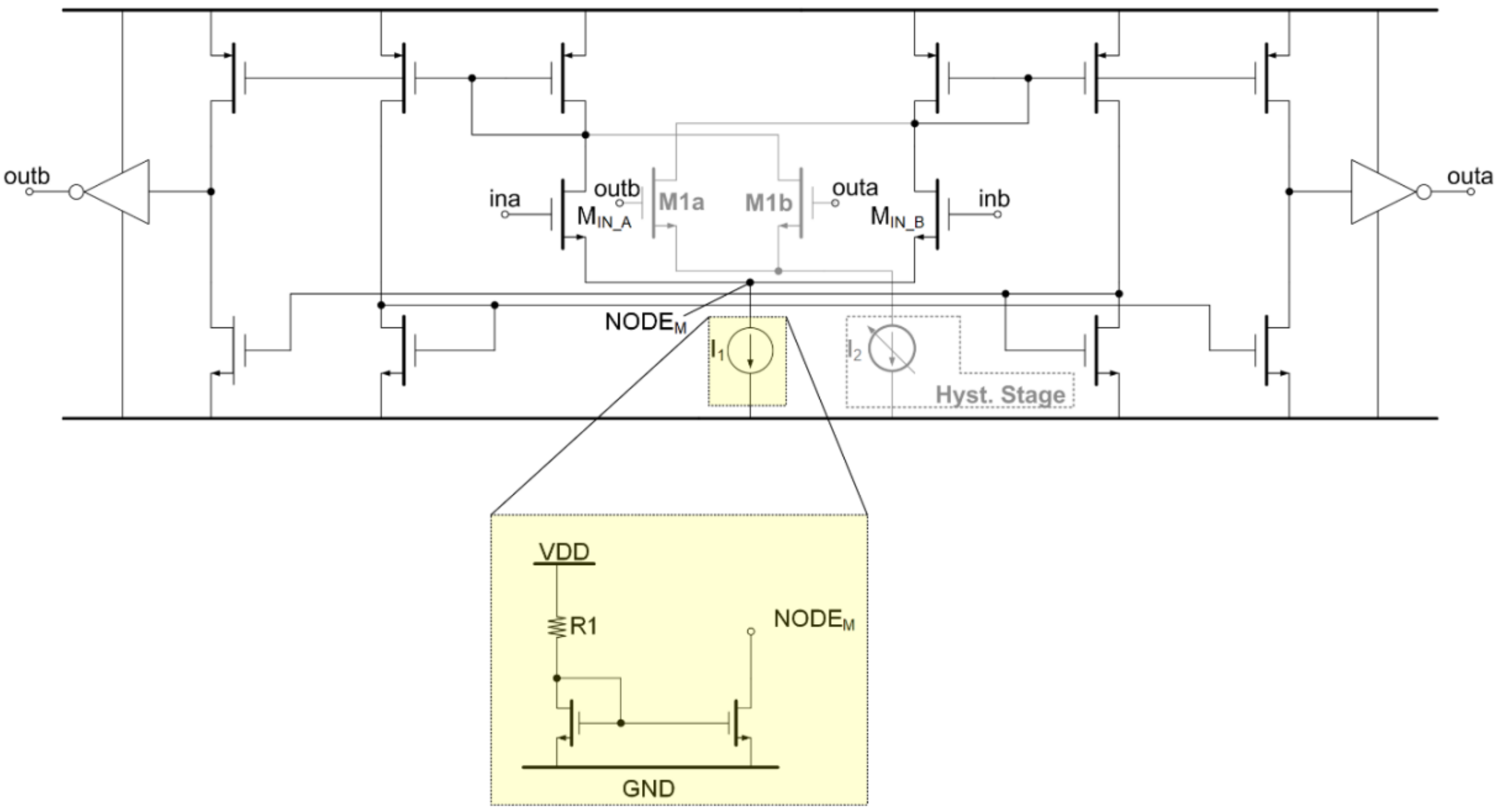

*Figure 8 - Schematic of the Discriminator Stage*

### 3.4.5 The Analog Pad Driver and the LVDS Output Cell

The last channel (CH7) in each chip has been implemented with an additional test function, making the output of DA3 available outside of the chip for diagnostic purposes. The driver for the DA3 output is called Test Point Buffers or Analog Pad Drivers (APD) as shown in Figure 4. The circuit, given in Figure 10 is based on a set of cascaded source followers. The ADP works with a gain of - 2.6 dB. This means that readings must be multiplied by 1.35 to correspond to the DA3 output. The two switches, marked S1, disable the stage when not in use to limit mismatch with other channels and to save power. Both switches are controlled by the same signal S1, shown in the figure.

The output of the Low Voltage Differential Signals (LVDS) (Figure 9) can be selected by a multiplexer to be either the direct Discriminator output, i.e. the Time over Threshold (ToT) or else the output of the Wilkinson ADC, where the time difference between leading and trailing edge corresponds to the charge deposited on the Integration Gate, as described in section 3.3. Either option can be selected by the Chip Mode bit in the shift register, see Table 8.

The LVDS output works with a nominal swing of 180 mV into 100 Ω centered at 1.23 V. This range is compatible with the 'reduced range link' described in IEEE 1596.3 [22]. The input nodes ($IN_A$ and $IN_B$) come from a previous MUX stage and go to a pair of moderate-sized inverters ($M5_A/M6_A$ and $M5_B/M6_B$). These inverters drive the output stage, which is essentially a pair of inverters ($M2_A/M3_A$ and $M2_B/M3_B$) with their output current limited by transistor pairs ($M1_A/M1_B$ and $M4_A/M4_B$) operating in their resistive region. $M1_A/M1_B$ and $M4_A/M4_B$ constitute the common mode feedback fixing the common mode output voltage. The DC characteristics are set entirely by transistor sizes.

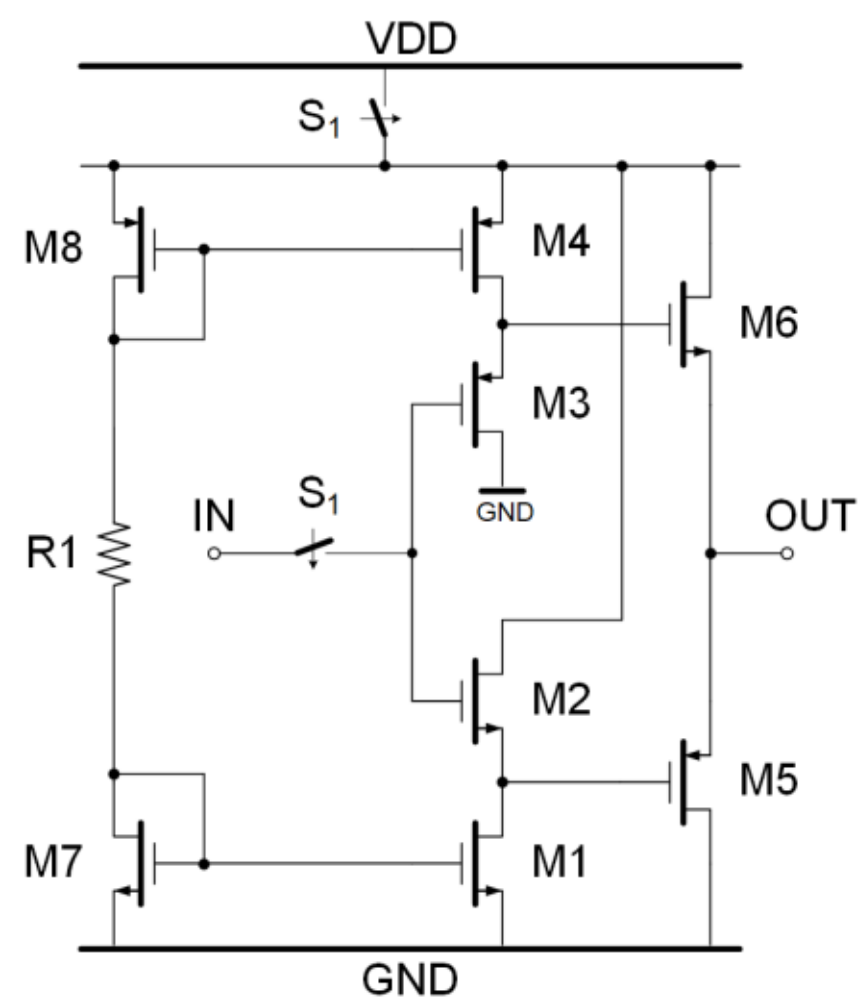

*Figure 10 - Scheme of the Analog Pad Driver in Ch 7*

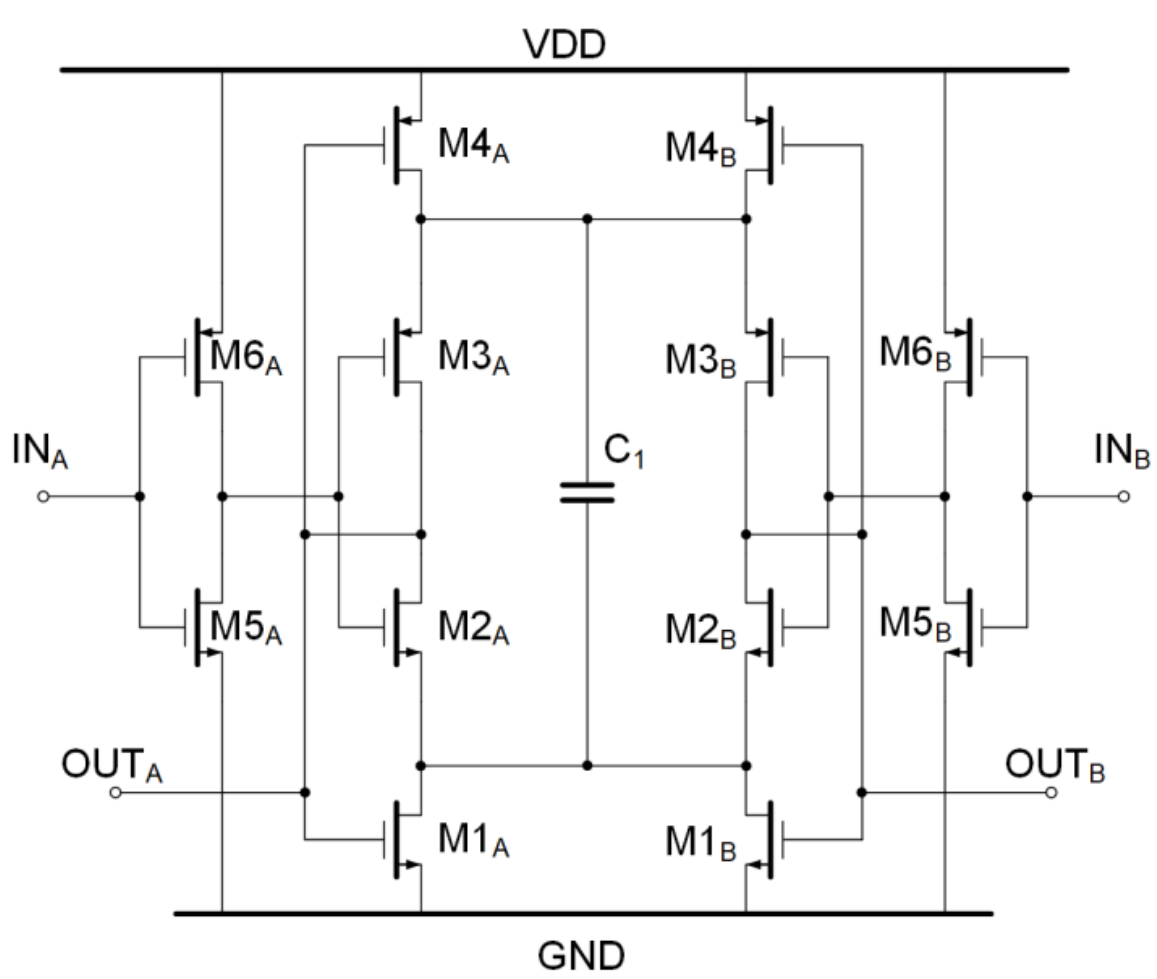

*Figure 9 - Schematic of the LVDS stage*

### 3.4.6 The Wilkinson ADC

The Wilkinson ADC (WADC) performs an approximate measurement of the amplitude of the incoming signal by generating a discriminator output pulse width proportional to the amplitude of the signal. Knowing this amplitude allows for a correction of the raw trigger time for time slewing (as already mentioned in section 3.3) and, in addition, provides diagnostics for monitoring the gas gain of the chamber.

The WADC performs a Voltage-to-Time conversion by charging and discharging the capacitor $C_H$ shown in Figure 11. The conversion starts when an input signal is detected and the DISC1 stage provides the Start-of-Conversion signal. The rising edge of the DISC1 signal fires the FF1 flip-flop and the D_GATE delay element controlled by the "Integration Gate Width" bits. The SG signal is output of FF1 and determines the charge on the sampling capacitor $C_H$ for the time TGW, as defined by the D_GATE delay elements. At the same time, the $V_A$ (t) – $V_B$ (t) signal reaches the $V_{AB_MAX}$ voltage, which is proportional to the input charge. Detection of the falling edge of SG fires the second flip flop (FF2), starting the discharge phase through the SW signal (FF2 output). The $C_H$ discharge occurs with a constant current, controlled by the Rundown Current bits and finishes when $V_A$ (t) – $V_B$ (t) signal crosses zero, corresponding to the falling edge of the DISC2 output. The time of the $C_H$ discharge is called $T_{DISCH}$ and is proportional to $V_{AB_MAX}$. It results in a WADC output width given by TGW and $T_{DISCH}$ and is controlled by the $\Phi_1$ and $\Phi_2$ signals, i.e. on ON–OFF phases.

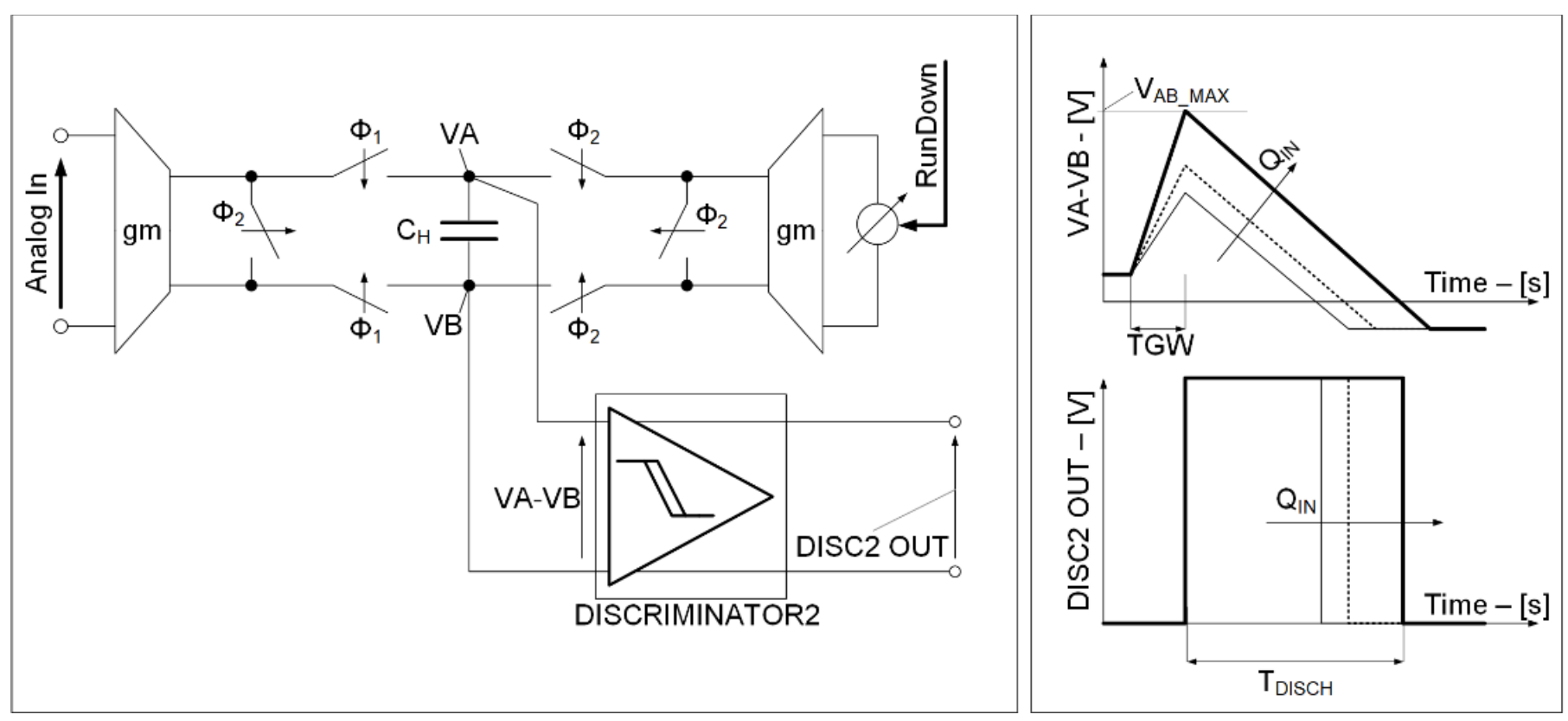

*Figure 11 - Wilkinson Analog to Digital Converter (W-ADC) Scheme*

The timing depends on a specific set of programmable control bits for Integration Gate Width, Rundown Current and DeadTime. The corresponding data format is defined in section 3.5. The DeadTime (DT) signal is generated from the third flip flop (FF3) by the rising edge OUT when the delay elements (DT_1, DT_2 and DT_3) are activated. All delay elements are based on complementary current sources, charging appropriately sized capacitors, driving logic gates with hysteresis.

The currents are programmable by a binary-weighted switched resistor string. The discriminator DISC1 is disabled for the time DT, such that incoming signals, crossing the threshold, will not create output towards the subsequent stage, the TDC. A proper differential logic (as described in [20]) controls the complementary phases $\Phi_1$ and $\Phi_2$ as well as the corresponding MOS switches.

The gm-transconductors are DA stages based on the scheme shown in Figure 7, where polysilicon resistors replace Z1 and Z2. They are used like in the shaper stage as a floating current source, both for the integration current source and for the rundown current sink.

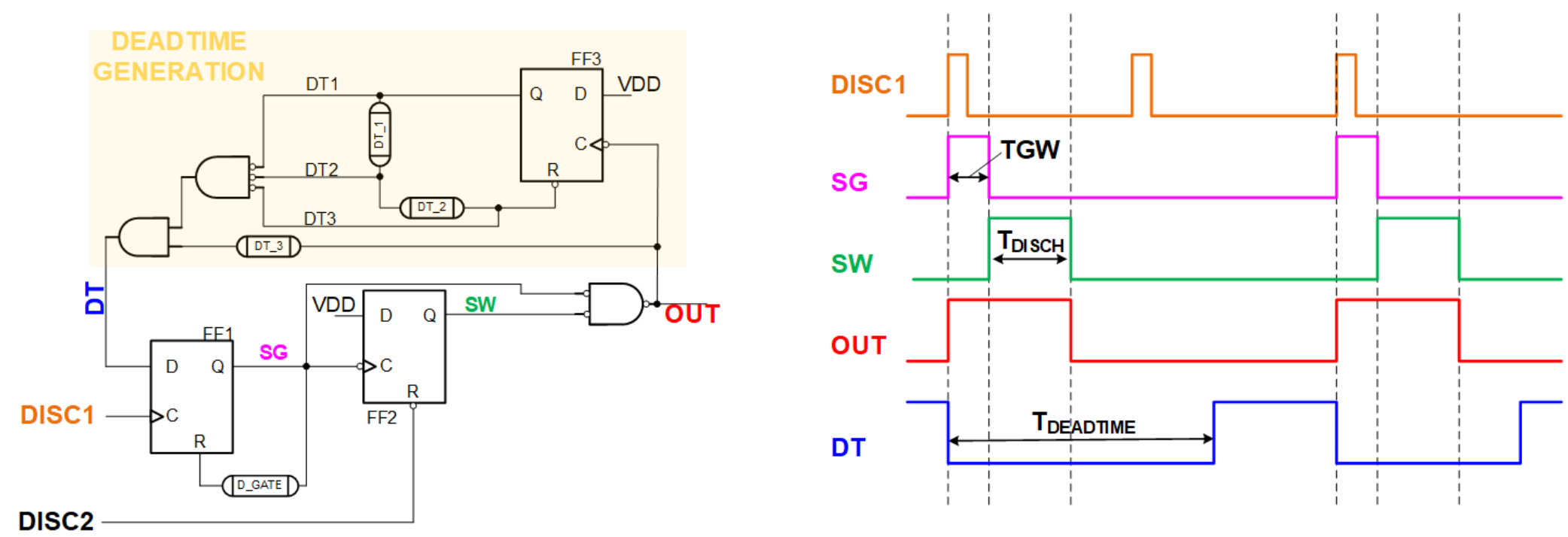

*Figure 12 - Control logic of ADC and Deadtime generator and corresponding signals*

### 3.4.7 Protection of the CSP Input against Overvoltage

The central wire in the standard MDT tubes is at 3090 V DC. The inputs of the preamplifiers must therefore be protected against spontaneous discharges of the wire, such that the amplitude at the ASD input does not exceed the allowed maximum of about 12 V.

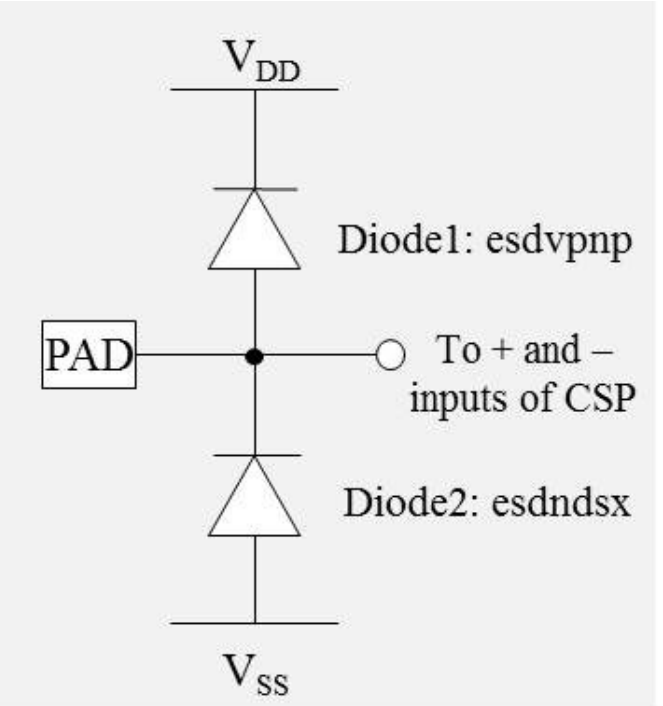

*Figure 13 - Simplified Scheme of the Integrated Protection Circuit*

As a standard design feature, all inputs in this technology are protected against accidental discharges, typically cause by Human Body Discharge (HBD). Crossed protective diodes are implemented as illustrated in Figure 13. Here the CSP inputs ($IN_{CSP-}$ and $IN_{CSP+}$ in Figure 4) are connected to each pad.

The two protection diodes esdvpnp (P+/NW ESD) and esdndsx (N+/PW ESD) are from the ESD library of IBM013. Their breakdown voltage is 11.5 –12 V with 1 µA of reverse current at 25° C. The diodes esdvpnp and esdndsx are robust, as the foundry guarantees sufficient hardness to high HBD (> 4 kV).

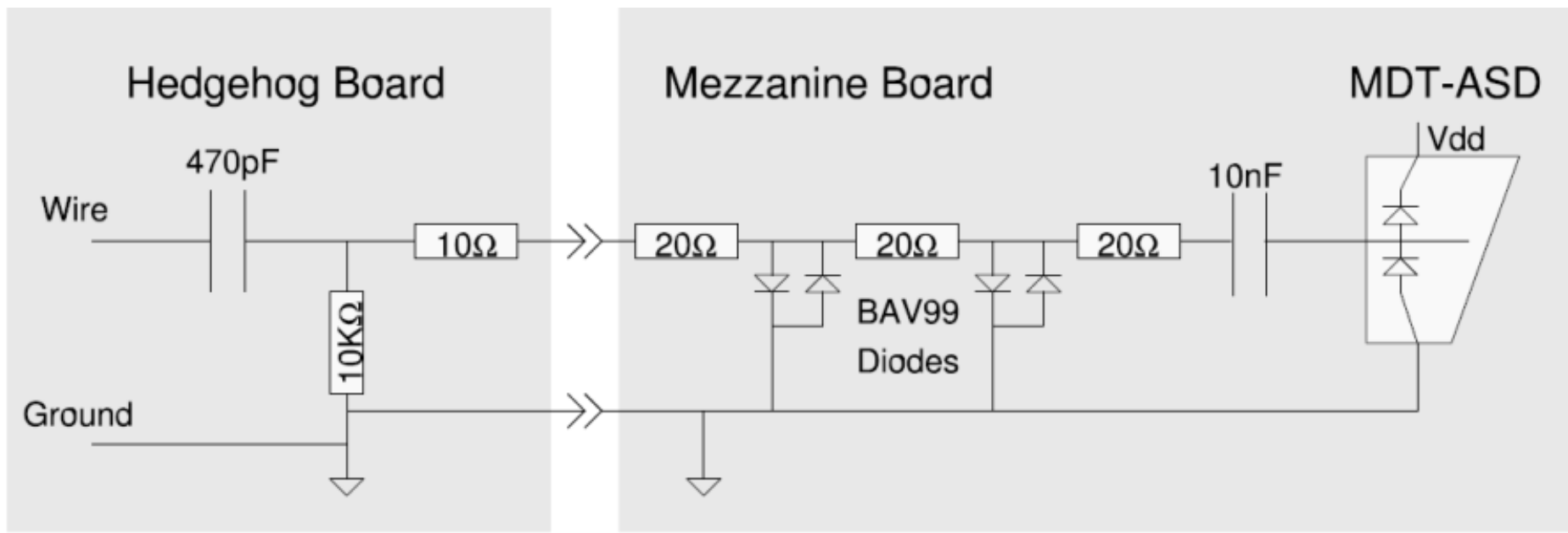

*Figure 14 - Protective circuit against HV discharges applied to the input of the legacy ASD*

In addition to internal protection, equivalent circuitry is implemented on the carrier PCB. Figure 14 shows the protective network as presently implemented on the frontend readout boards (“mezzanines”). The HV on the wire is decoupled from the readout by a 470 pF capacitor. AC-wise, the input of the ASD is protected by two sets of 20Ω resistors followed by crossed diodes. A series of tests was done to assure this protection circuit also to be sufficient for ASD2.

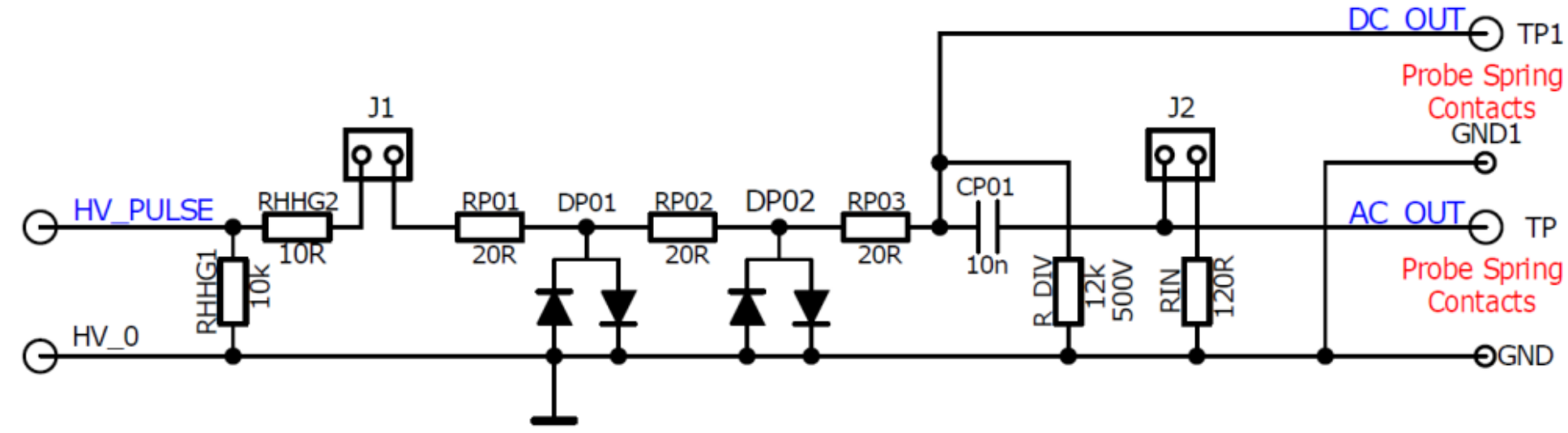

*Figure 15 - HV circuit for tests of the protective network*

Figure 15 shows the circuit for the corresponding test. HV pulses with up to ± 3 kV amplitude were applied to the network but did not lead to amplitudes at the AC output outside the specified voltage range.

In a first test, switch J2 was closed and resistor $R_{IN}$ simulated the resistive load of the amplifier. Subsequently, ASD2 prototypes were connected to the AC output (TP) and a large number of HV pulses was applied to the network without any damage found in the tested devices.

### 3.4.8 Simulation of the Signal Processing Chain

The ASD2 chip has been simulated at the schematic level using Cadence Tools and the GF 130 nm (CMRF8RF_LM) Process Design Kit (PDK). The configuration used for simulation is presented in Figure 16, while Figure 21 shows the main input/output voltage signals of the Analog Pad Driver (see section 3.4.5).

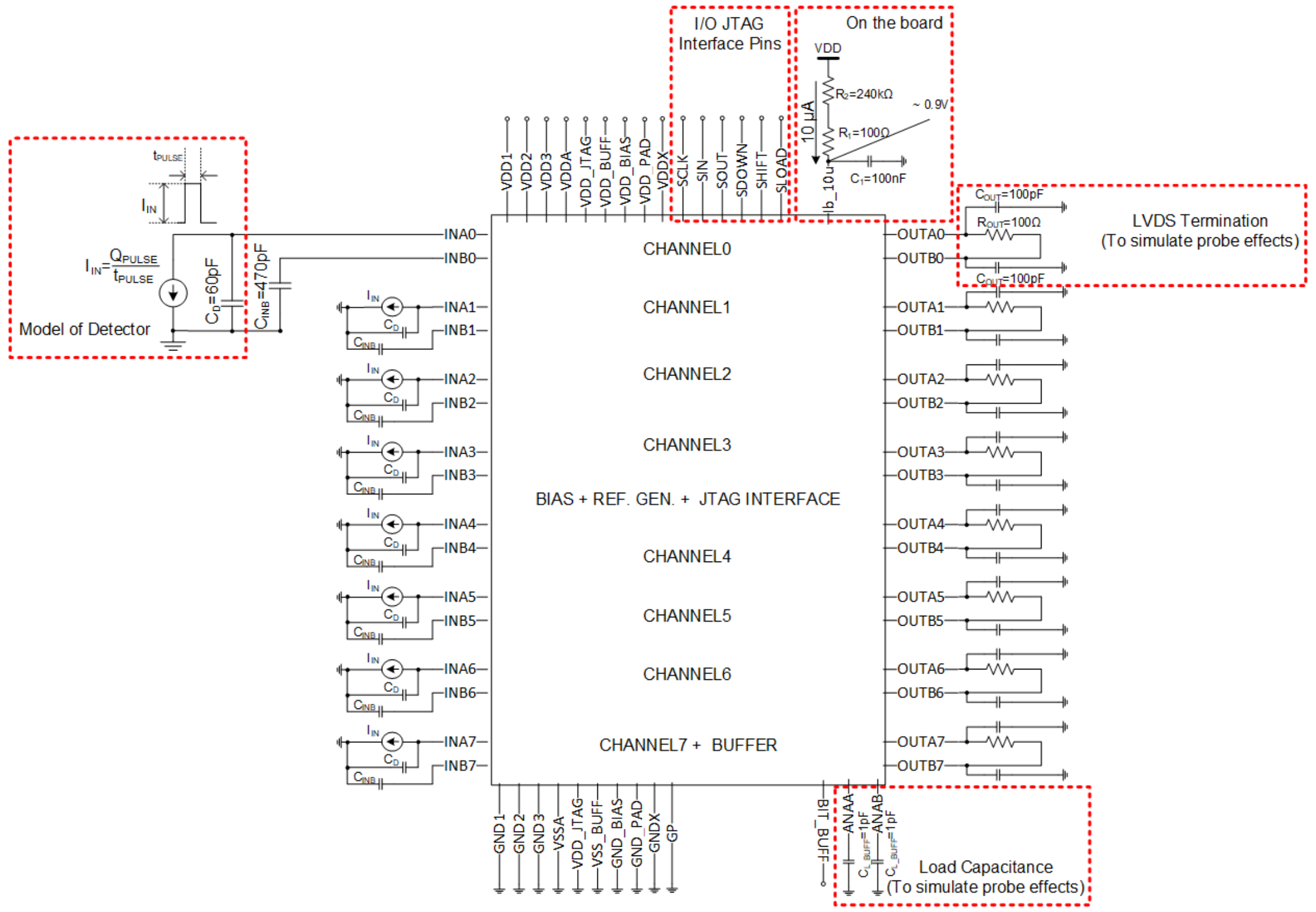

*Figure 16 - ASD2 simplified Simulation Scheme*

#### 3.4.8.1 Simulation in the Time Domain

The ASD2 signal processing chain was designed to detect input charges $Q_{IN}$ up to 100 fC, but the range of simulation has been extended to 175 fC to also cover the saturation region. The input charge has been simulated with a parasitic capacitance of 60 pF in parallel to an ideal current pulse ($I_{PULSE}$). $I_{PLUSE}$ has a fixed duration ($T_{PULSE}$) of 3 ns and an amplitude given by $Q_{IN}/T_{PULSE}$. For the sensitivity of the CSP, simulation predicts 1 mV/fC, for the peaking time about 7 ns, see Figure 17 (top).

The differential stages DA1 to DA3 amplify and shape the output of the CSP, as shown in Figure 17. At the end of the analog chain, i.e. at the output of DA3, the peak amplitudes for input charges of 25, 50, 75 and 100 fC gain values are 17.6, 16.8, 14.2 and 12.2 mV/fC. The small-signal gain of about 20 mV/fC is thus considerably higher than the 8.9 mV/fC, specified Table 3. Likewise, the peaking time of 12 ns is faster than the specified 15 ns.

A characterization in terms of peak input vs peak output has been done for each differential amplifier. DA1 has a purely resistive feedback and its output changes linearly with the input. The shaping stages, in contrast, show increasing non-linearity at high input charge (Figure 18). Measurements on the finished chip have verified the corresponding saturation effect of the DA3 output (Figure 30).

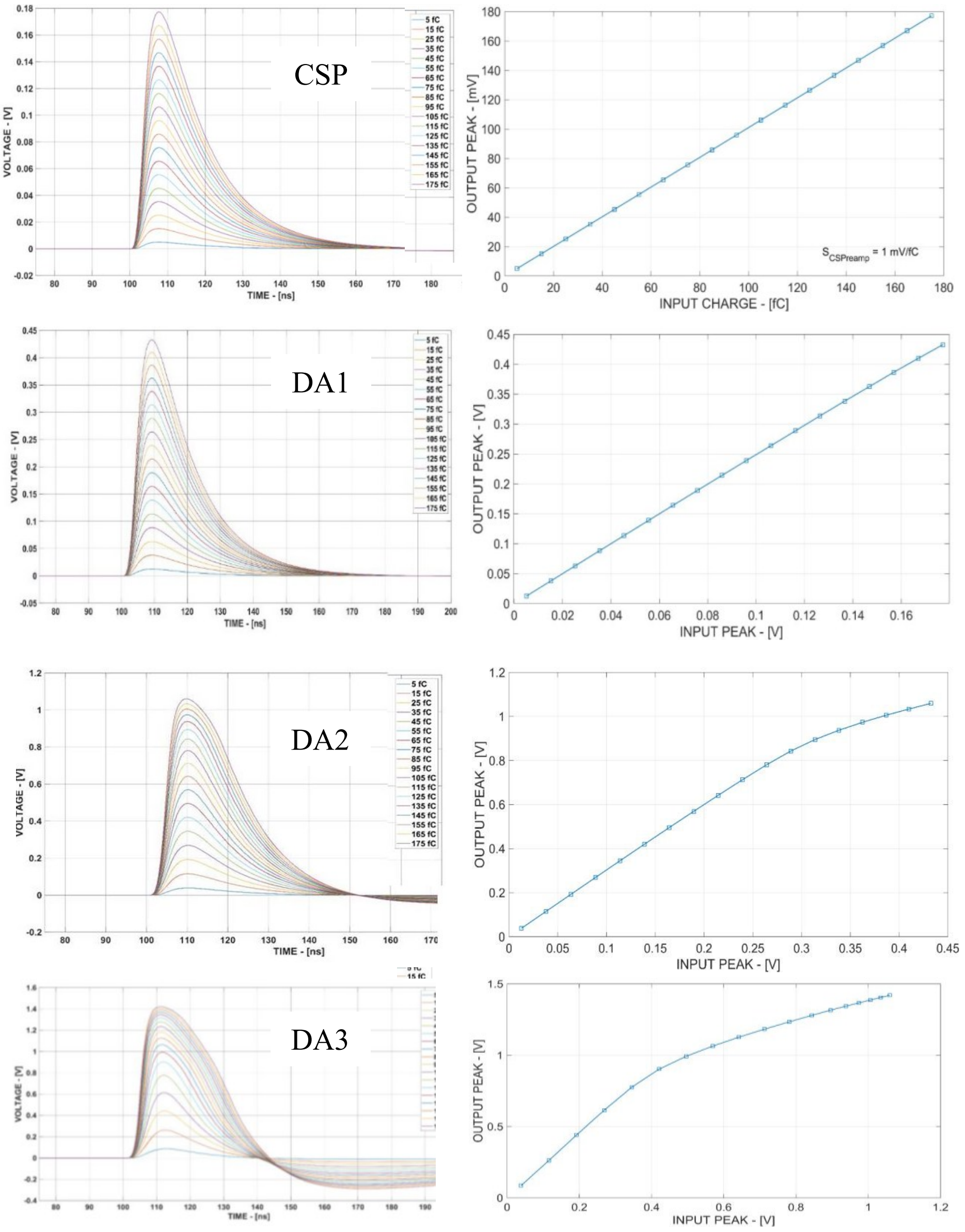

*Figure 17 - Signal shape and gain of CSP and DA shaping stages vs. input charge 5 to 175 fC*

#### 3.4.8.2 Simulation in the Frequency Domain

The AC characteristics of the CSP and the differential amplifiers DA1-DA3 are shown in Figure 18. The DA1-DA3 amplifiers serve both as gain and as shaping stages (section 3.4.2). The topology for all amplifiers is identical, while the feedback network differs according to the desired frequency characteristics and the corresponding values of the feedback elements Z1 and Z2 (Table 6).

- The CSP is designed as a voltage amplifier for the capacitive input and the feedback network. The ideal gain is given, approximately, by the ratio of the coupling capacitor at the MDT tube (470 pF, see Figure 1) to the capacitor $C_F$ in the feedback loop of the CSP (0,6 pF, see Figure 5), corresponding to about 58 dB. The ~ 8 dB difference observed in the simulation relative to this ideal value could be attributed to impedance variations, parasitic effects and to attenuation caused by the finite loop gain.
- DA1 is a simple gain stage with purely resistive feedback. The gain is 8.3 dB with a 3dB bandwidth of 129.5 MHz (Figure 18).

- the DA2 and DA3 stages implement the intended bipolar shaping. Their frequency response is achieved replacing the impedance Z2 with the R-C nets specified in Table 6.

The maximum gain of DA2 is 9.9 dB between 750 kHz and 184 MHz, while the DA3 provides 8 dB in the range 1.1 MHz – 65 MHz. Thus, the total gain of DA1-DA3 is 26.2 dB, corresponding to an amplification factor of 20,4 relative to the output of the CSP, which (given the gain of the CSP of 1 mV/fC) leads to a total charge gain of 20,4 mV/fC. This compares well with the gain measured in section 4.1.4.

Figure 18 shows, notably, the roll-off of the signal amplitude of the CSP, which is a measure of its "speed", determining the peak time of the signal chain. In this plot, the attenuation of the signal magnitude is about 16 dB in the range 10-100 MHz. This number can be compared to the value of 25 dB of ASD1 (see in Fig. 12 in the ASD1 Manual [15]). The much higher roll-off of ASD1 explains its reduced peak time compared to ASD2, as shown in Figure 29. The reason for the higher "speed" of ASD2 lies in the improvements, when chip production moved from HP 500 nm to IBM 130 nm technology, see section [1].

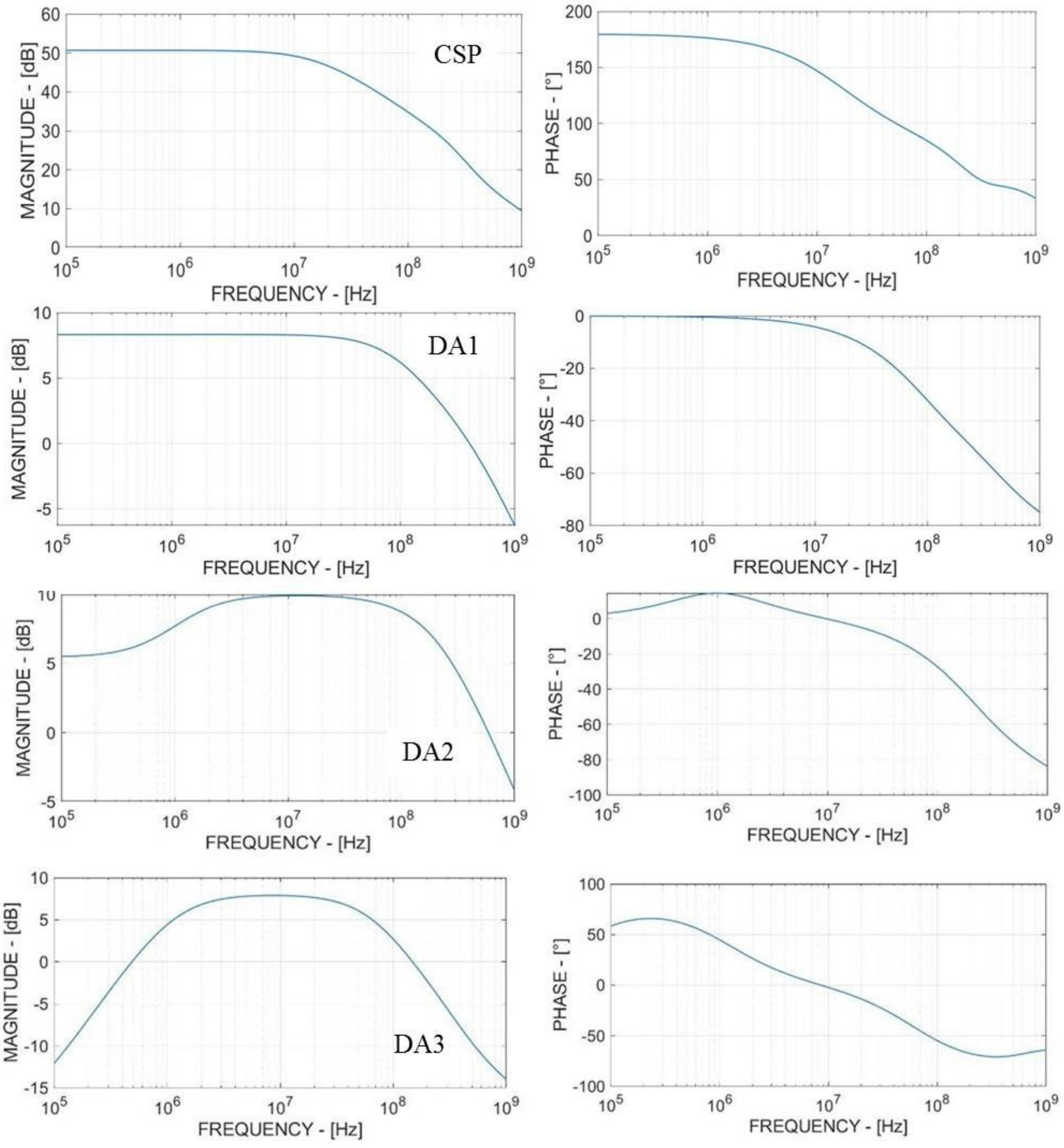

*Figure 18 - Gain and Phase of the CSP and DA shaping stages vs. frequency*

Figure 19 shows gain vs. frequency for all four analog blocks as well as their superposition. The pass-band frequency response is characterized by a bandwidth of about 16 MHz (from 1.4 MHz to 17 MHz) with a maximum gain of 76 dB. The high-pass cutoff frequency (1.4 MHz) depends only on the frequency response of DA3, while all stages contribute to the low-pass cutoff (17 MHz), as DA3 is the only stage with AC coupling in the chain (see Figure 4).

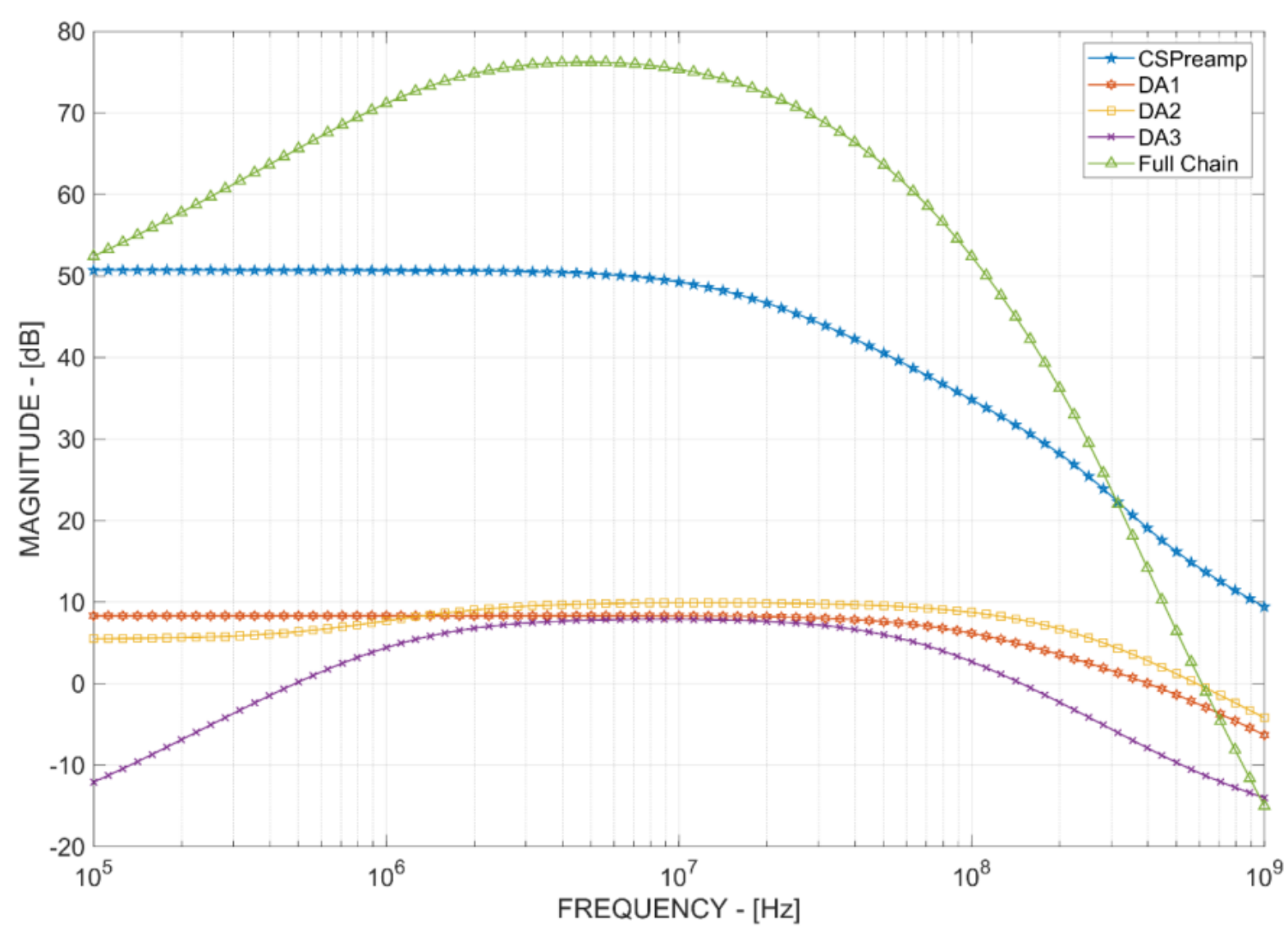

*Figure 19 - Frequency response of the full analog chain and response of each single stage*

The Analog Pad Driver (APD), being only implemented on channel 7, plays a particular role for tests of the ASD. It allows to test the output of DA3, controlling performance parameters like gain, peak time or pulse shape. It may also be used to detect digital-to-analog interference or crosstalk among adjacent channels.

In order not to distort the signal source, it must have a somewhat wider bandwidth than DA3. Figure 20 show its Bode diagram with a roll-off of about 3 dB between 10 and 100 MHz, which is considerably less than the one of the DA3 output, which is about 16 dB in Figure 18. Consequently, it does not modify the shape of the DA3 output at any significant level. The *absolute value* of the DA3 output, however, is modified by a slight attenuation, i.e. a gain of - 2.6 dB, corresponding to Vout/Vin = 0.741. The amplitudes observed at the APD output must therefore be multiplied by 1.35 to correspond to the input seen by the Discriminator DA4.

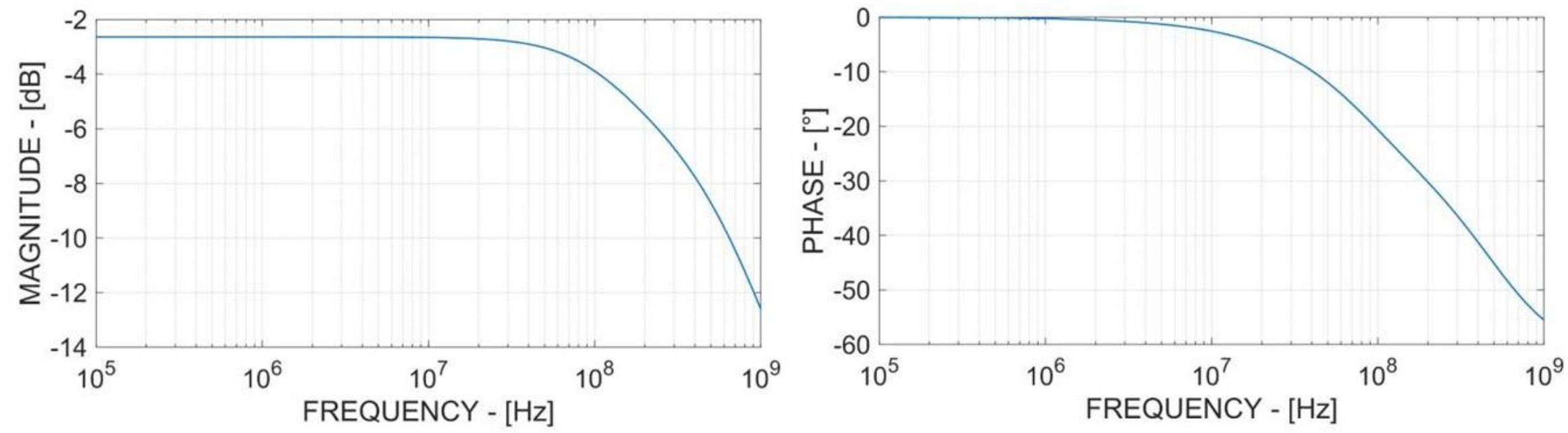

*Figure 20 - Gain and Phase of the Analog Pad Driver vs. frequency.*

As the input to the APD represents a non-negligible load to the output of DA3 and, in addition, consumes a significant amount of power, the signal input and the supply voltage can be disconnected when the pad driver is not in use. This way, all 8 channels have the same electrical behavior and operate under identical conditions, see section 3.4.5 and Figure 10.

Figure 21 shows the input signals to the CSP (left), where the input INB7 is tied to ground. The top-right diagram shows the simulated pulse shape at the outputs of the LVDS-Driver (top-right) and the corresponding ones of the APD. The CSP signal, with a peak amplitude of about 0,05 mV, is the residual voltage variation at the CSP virtual ground node (INA7 pin), which is the applied input voltage step reduced by the open-loop gain of the amplifier. Therefore, the actual voltage step, generated by the external pulse generator is about 2 orders of magnitude higher.

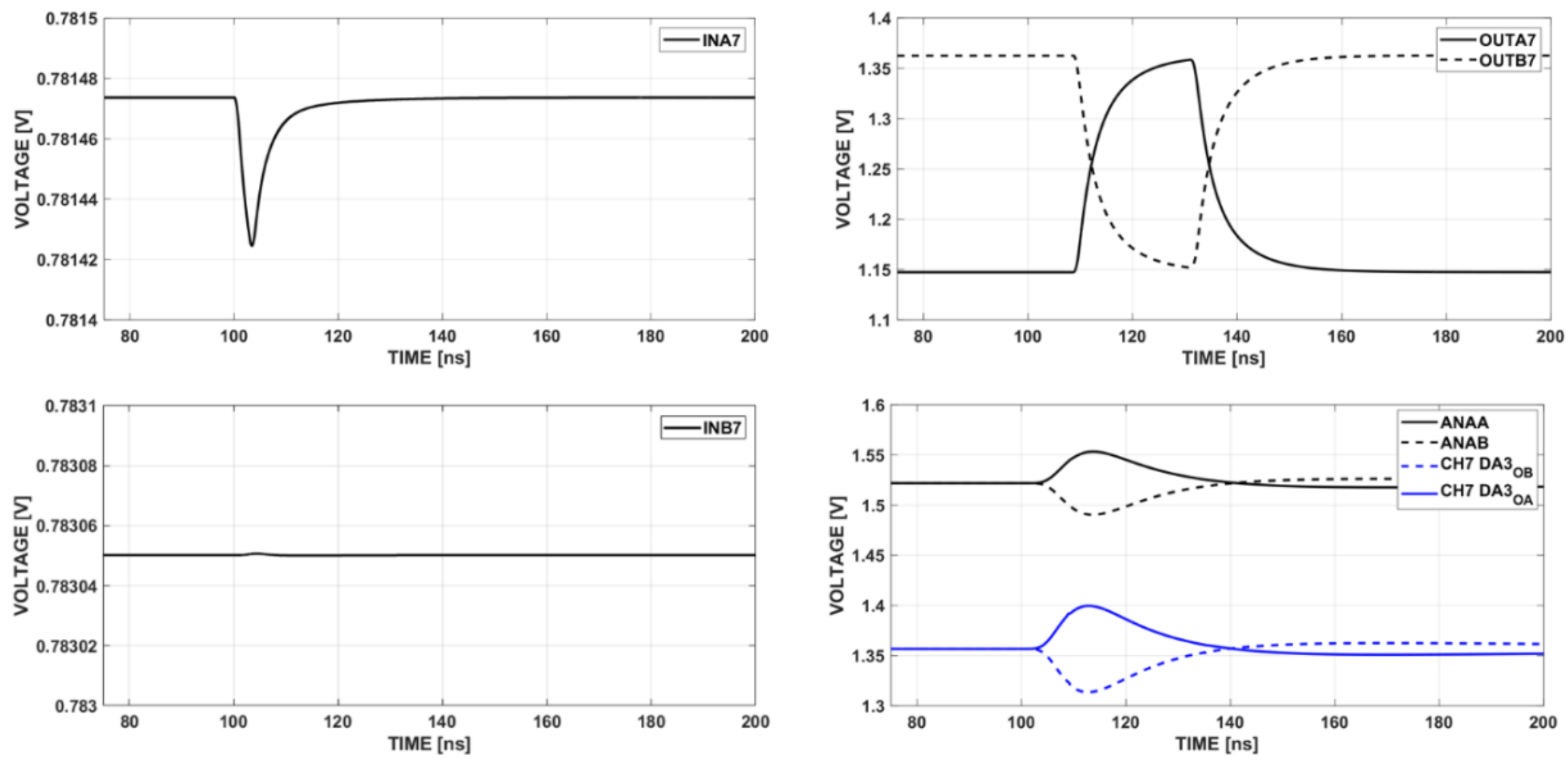

*Figure 21 - Input and Output Voltage signals of test channel (CH7)*

### 3.4.9 Post-Layout simulation (PEX) simulation

Cadence tools allow to investigate the chip performance with respect to Process/Voltage/-Temperature (PVT) variations with the aim to detect potential mismatch among devices of identical design. Our result w.r.t. process variations was that the variation of transistor parameters (e.g. threshold voltage, transconductance, output impedance) and resistor and capacitor values may cause performance variations within ± 20 %. for parameters like peak voltage, peak time, width of the ADC output and deadtime.

Simulation was based on the following sets of parameters:

- five different processes of fabrication: typical (TT), slow slow (SS), fast fast (FF), slow fast (SF) and fast slow (FS). In addition, a variation of the supply voltage of 10% is assumed.
- a temperature range between – 40° and 120°

An example for the variation of device parameters is shown in Table 7. For this purpose, four typical devices have been chosen: an input NMOS transistor (W/L = 220 µm / 400 nm), a PMOS load (W/L = 500 µm / 400 nm), a resistor (W = 600 nm, L = 16.42 µm) and a capacitor (W = 11.48 µm, L = 25 µm). Obviously, the variation of performance is given by a combination of all relevant parameters. This justifies the variation of the peak voltage and the peaking time delay at the DA3 output in the range 60 – 120 mV and 11 – 15 ns, respectively.

Table 7 - Range of parameter variation used in simulation for critical components

| PARAMETER | MIN | NOM | MAX | UNIT |
|---|---|---|---|---|
| NMOS Threshold Voltage | 396.5 | 514.5 | 611.6 | mV |
| NMOS Transconductance | 3.744 | 4.504 | 5.248 | mA/V |
| NMOS Output Impedance | 3.468 | 6.971 | 11.01 | kΩ |
| PMOS Threshold Voltage | 371.9 | 363.9 | 447.6 | mV |
| PMOS Transconductance | 7.029 | 8.203 | 9.628 | mA/V |
| PMOS Output Impedance | 5.886 | 7.893 | 10.62 | kΩ |
| Resistor | 9.823 | 10 | 10.2 | kΩ |
| Capacitor | 599.2 | 599.8 | 600.7 | fF |

### 3.4.10 The Wilkinson ADC

The WADC performs a voltage-to-time conversion, resulting in an approximate measurement of the input charge [15]. An example of the W-ADC output pulse generation is shown in Figure 21.

The width of the WADC output (Figure 21, bottom) is the sum of the charge and discharge time of the capacitor CH (Figure 11). The signal at CH during charge and discharge is shown in Figure 21, top. It is obtained with 32 mV of threshold voltage (VTH2), an integration gate time (i.e. charge phase) of about 26 ns and a rundown current of 3.4 µA. The integration gate and rundown signals are shown in the center of the same Figure.

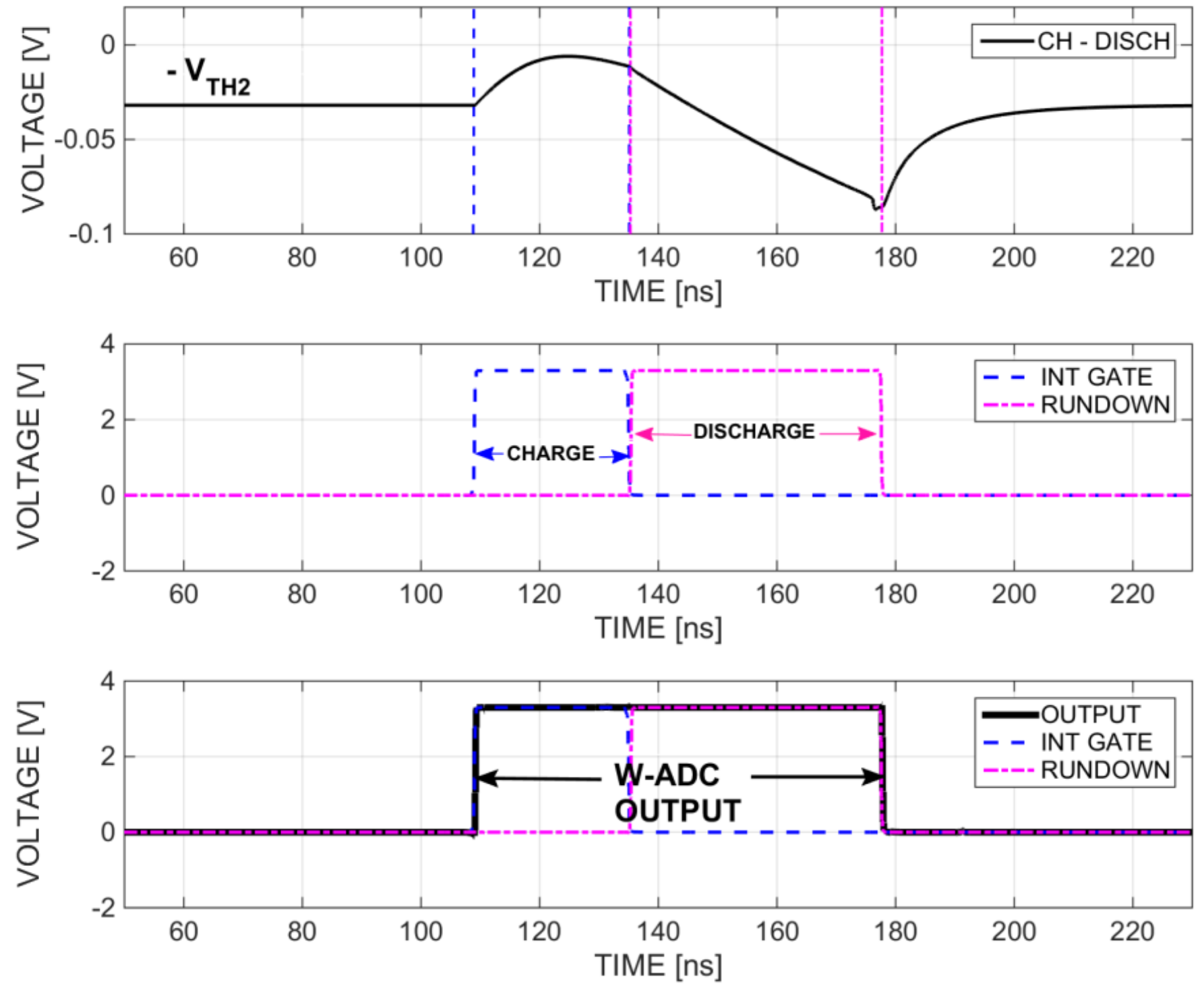

Figure 22 - WADC Output from its internal signals

## 3.5 Programmable Parameters

As indicated in the block diagram of Figure 4, the Programmable Parameters (PPs) section on the chip controls the proper functioning of the ASD by defining parameters like threshold, gate width, run-down current and dead time. A 55-bit digital word is transmitted to the ASD using a simplified serial protocol, while an interface, based on a chain of register, transmits the word code associated to each parameter to Digital-to-Analog Converters (DACs).

Table 8 gives a summary of the PPs together with the number of bits, corresponding range, and resolution/LSB. Analog PPs control Discriminator (DISC1) and Wilkinson ADC (WADC) stages. The other PPs are functional and manage channel/chip operation mode.

*Table 8 - Summary of the Programmable Parameters*

| STAGE | PARAMETER | RANGE | # of bits | LSB | UNIT |
|---|---|---|---|---|---|
| DISC1 | DISC1 Threshold – $V_{TH1}$ | -127 to 128 | 8 | ~ 3 | mV |
| | DISC1 Hysteresis | 0 - 320 | 4 | 20 | µA |
| WADC | Wilkins. Integration Gate | 8 - 45 | 4 | ~ 2.5 | ns |
| | DISC2 Threshold –$V_{TH2}$ | 32 - 256 | 3 | 32 | mV |
| | Run-down current | 2.4 – 7.3 | 3 | ~ 0.7 | µA |
| | Dead Time | 13.8 - 785 | 3 | ~ 100 | ns |
| LVDS OUTPUT | Channel Mode | ON, HI, LO | 2 | - | - |
| | Chip Mode | ADC, ToT | 1 | - | - |

### 3.5.1 Programmable analog Parameters

#### 3.5.1.1 Discriminator control

The Discriminator Stage compares the analog DA3 output with a **threshold voltage** and uses a **hysteresis** to avoid multiple threshold crossings due to noise.

The threshold voltage ($V_{TH1}$) is applied at the AC coupled input of the pre-discriminator differential amplifier stage (DA4) and is generated by two complementary 8-bit dual resistor divider voltage DACs. The final differential output threshold can vary from about – 380 mV to + 380 mV with an LSB of ~3 mV. This range of thresholds covers signal input charges up to about 20 fC (cf. Figure 30), more than enough for any practical application.

The range 0 to 255, the "absolute" code, is sometimes transformed into the range -127 to 128, the code "relative" to the center value of zero (at a threshold of 0 mV), while the center value of the absolute code is 127. The *absolute* code 114, currently used for the ASD2, thus equals a *relative* code of -13, corresponding to a threshold of -39 mV, i.e. about 2 fC.

Hysteresis is applied through a scaled-transistor current source DAC with a resolution of 4 bits. The range of the DAC is 320 µA with an LSB of 20 µA, which is 0 - 85 mV at the discriminator input. It should be noted that the "effective" threshold of the discriminator DA4 is influenced by the hysteresis setting (see section 4.1.6).

#### 3.5.1.2 Wilkinson ADC Control

The WADC output pulse width depends on the following programmable parameters:

- Integration Gate Width (IGW)
- Threshold ($V_{TH2}$) for the WADC comparator (DISC2)
- Rundown Current (RC)
- Dead Time (DT)

The Integration Gate Width (IGW) determines the 16 different charge timings of the capacitor $C_H$ (section 3.4.6) through 4-bit words. The settable value range is 8 ns – 45 ns, with steps of ~ 2.5 ns.

The DISC2 threshold ($V_{TH2}$) is applied to the differential threshold terminals of the WADC comparator by two coupled resistor divider voltage DACs with 3-bit resolution and a range of 32 mV to 256 mV.

The Rundown Current (RC) controls the discharge phase of the capacitor $C_H$. It is set by a binary-weighted switched resistor chain between 2.4 µA and 7.3 µA. The pulse width range of the ADC is jointly determined by the Integration Gate Width (IGW), the discriminator threshold $V_{TH2}$ and the Rundown Current of the switched resistor chain.

The Dead Time (DT) defines an independent time window after each hit, during which the logic does not accept new hits. It is controlled by a 3-bit field and can be selected in the range 50 ns to 800 ns (DT-code 0 to 7). For the range and spread of values reached in production, see section 0.

### 3.5.2 Programmable functional Parameters

Chip mode

The ASD2 can operate in ADC or Time-over-Threshold (ToT) mode, selectable by bit number 16. When bit 16 is zero (ADC mode) the LVDS output corresponds to the output of the WADC, otherwise (ToT mode) it corresponds to the output of the discriminator.

Channel mode

For diagnostic purpose it might be useful to force the LVDS output to a logic HIGH or LOW. One of the following channel operation modes can be selected:

- ON MODE: default working setting; the output depends on the Chip Mode bit.
- HI MODE: the output is forced to HIGH or 'Logic 1' (regardless of what happens in the analog part of this channel).
- LO MODE: the output is forced to LOW or 'Logic 0' (regardless of what happens in the analog part of this channel).

In HI or LO MODE the channel is disabled. This option may be used to disconnect noisy MDT tubes from the Readout, which otherwise might saturate the readout bandwidth with meaningless data.

## 3.6 The Serial data interface

### 3.6.1 Architecture

The ASD2 serial interface architecture employs separate shift and working registers. The shift register is connected directly to digital input (SIN) and output (SOUT) pads, respectively. The data can be uploaded any time (asynchronously) to the shadow registers, which control the DACs, multiplexers etc. The architecture allows downloading the whole set of active bits from the shadow to the shift register in order to send them back to the controller for diagnostic or monitoring purposes. This can be done any time and does not interfere with the normal operation of the ASD.

The interface for each data bit consists of the shift register-cell, implemented as a static master-slave D flip-flop, the shadow register cell realized as a static transparent latch and 2 two-in-one multiplexers (Figure 23). ASDs thus can be daisy-chained to form a closed JTAG-like control loop. Shift and shadow registers have a length of 56 bits, where 55 are actual data bits. The last shift register cell is clocked with an inverted clock, making the output change at the falling edge of the clock.

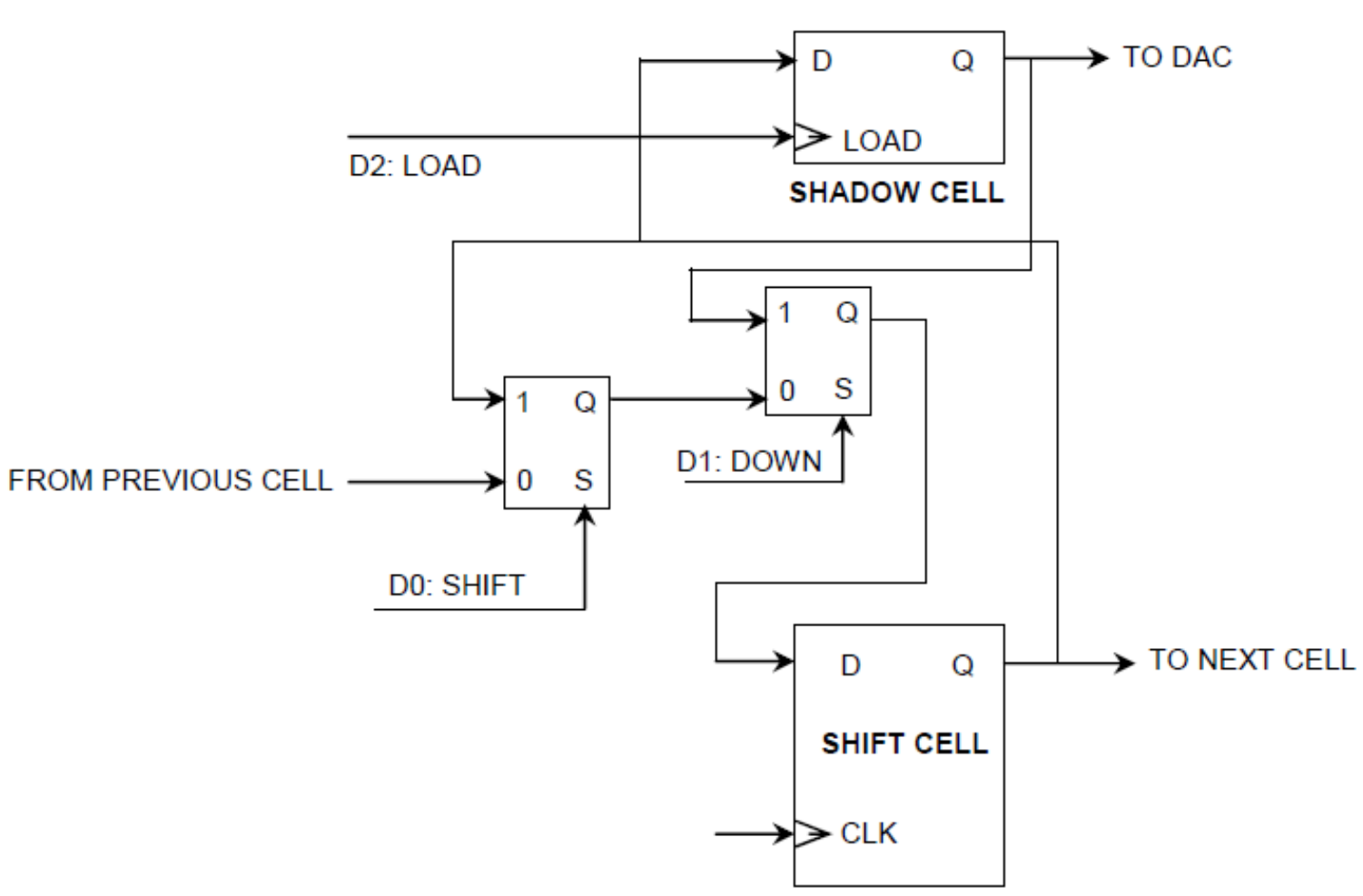

*Figure 23 - Serial data interface*

The protocol requires 2 data lines (SIN, SOUT), 3 control lines (SHIFT, DOWN, LOAD) and one clock line (CLK). The configuration allows extensive control over the data flow (Table 9). HOLD mode keeps the data in the shift register by feeding back each cell with its own content. SHIFT mode shifts data right at the rising edge of the clock. LOAD active asynchronously copies the bits in the shift register to the shadow register at any time. DOWN active copies the contents of the shadow registers to the shift register at the next rising clock edge (overrides SHIFT & HOLD). The ASD serial input expects the data to be stable at the rising edge of the clock. The controller will change data bits at the falling clock-edge. Thus, data are stable at the input for the entire clock period with the sensitive rising edge in the middle. Data bits at the ASD serial data output also change state at the falling edge of the clock.

*Table 9 - Serial interface instruction encoding*

| Mode | SHIFT | DOWN | LOAD | Description |
|---|---|---|---|---|
| Shift | 1 | 0 | X | Shift right at rising edge of CLK |
| Hold | 0 | 0 | X | Keep shift register contents (self-feedback) |
| Down | X | 1 | X | Copy contents of shadow register to shift register @ rising edge of CLK |
| Load | X | X | 1 | Load shadow registers with contents of shift register (asynchronous) |

## 3.6.2 Shift register bit Assignment

The bit assignment of the shift register is given in Table 10. Bit #0 is the last bit to enter the shift register as illustrated in Figure 24. Data words for the parameters are loaded LSB first. Channel 0 is the topmost channel when looking at the chip with the analog inputs on the left-hand side. It should be noted that in the field "Hyst. DAC (DISC1)" the lsb is on the right, in contrast to the other fields.

*Table 10 - Bit assignment in the shift register*

| JTAG BIT # | DESCRIPTION | LSB/code |
|---|---|---|
| [0:1] | Channel Mode – Channel 0 (TOP) – [0:1] | 00 → ON (ACTIVE)<br>10 → ON (ACTIVE)<br>01 → LO (FORCED)<br>11 → HI (FORCED) |
| [2:3] | Channel Mode – Channel 1 – [0:1] | |
| [4:5] | Channel Mode – Channel 2 – [0:1] | |
| [6:7] | Channel Mode – Channel 3 – [0:1] | |
| [8:9] | Channel Mode – Channel 4 – [0:1] | |
| [10:11] | Channel Mode – Channel 5 – [0:1] | |
| [12:13] | Channel Mode – Channel 6 – [0:1] | |
| [14:15] | Channel Mode – Channel 7 (BOTTOM – with Test Points) – [0:1] | |
| [16] | Chip Mode | 0 → ADC Mode<br>1 → ToT Mode |
| [17:19] | DeadTime – [2:0] | Bit 19 LSB |
| [20:23] | WADC Integration Gate – [3:0] | Bit 23 LSB |
| [24:26] | WADC Rundown Current – [2:0] | Bit 26 LSB |
| [27:30] | Hysteresis DAC (DISC1) – [0:3] | Bit 27 LSB |
| [31:33] | WADC threshold 2 (DISC2) – [2:0] | Bit 33 LSB |
| [34:41] | Threshold 1 (DISC1) – [7:0] | Bit 41 LSB |
| [42:54] | NOT USED | – |

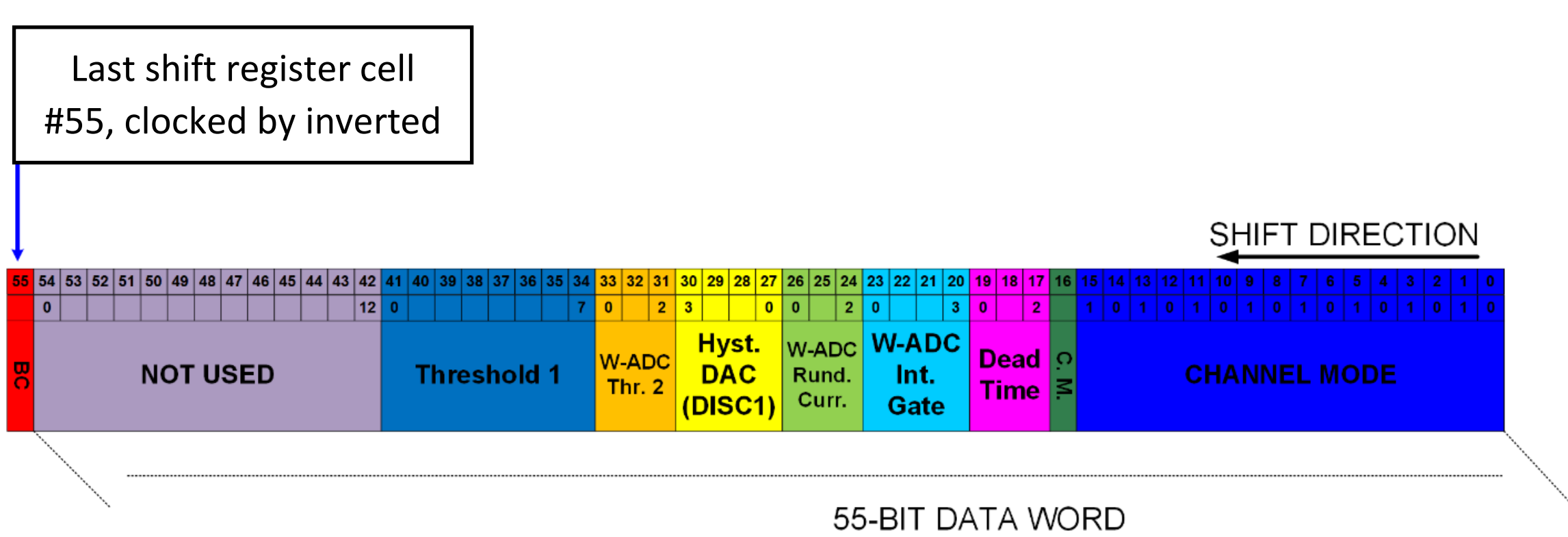

*Figure 24 - Shift Register Image*

## 3.7 IC layout and I/O pad assignment

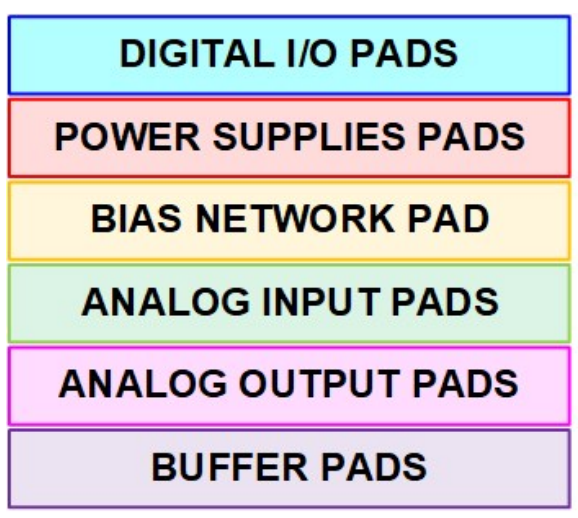

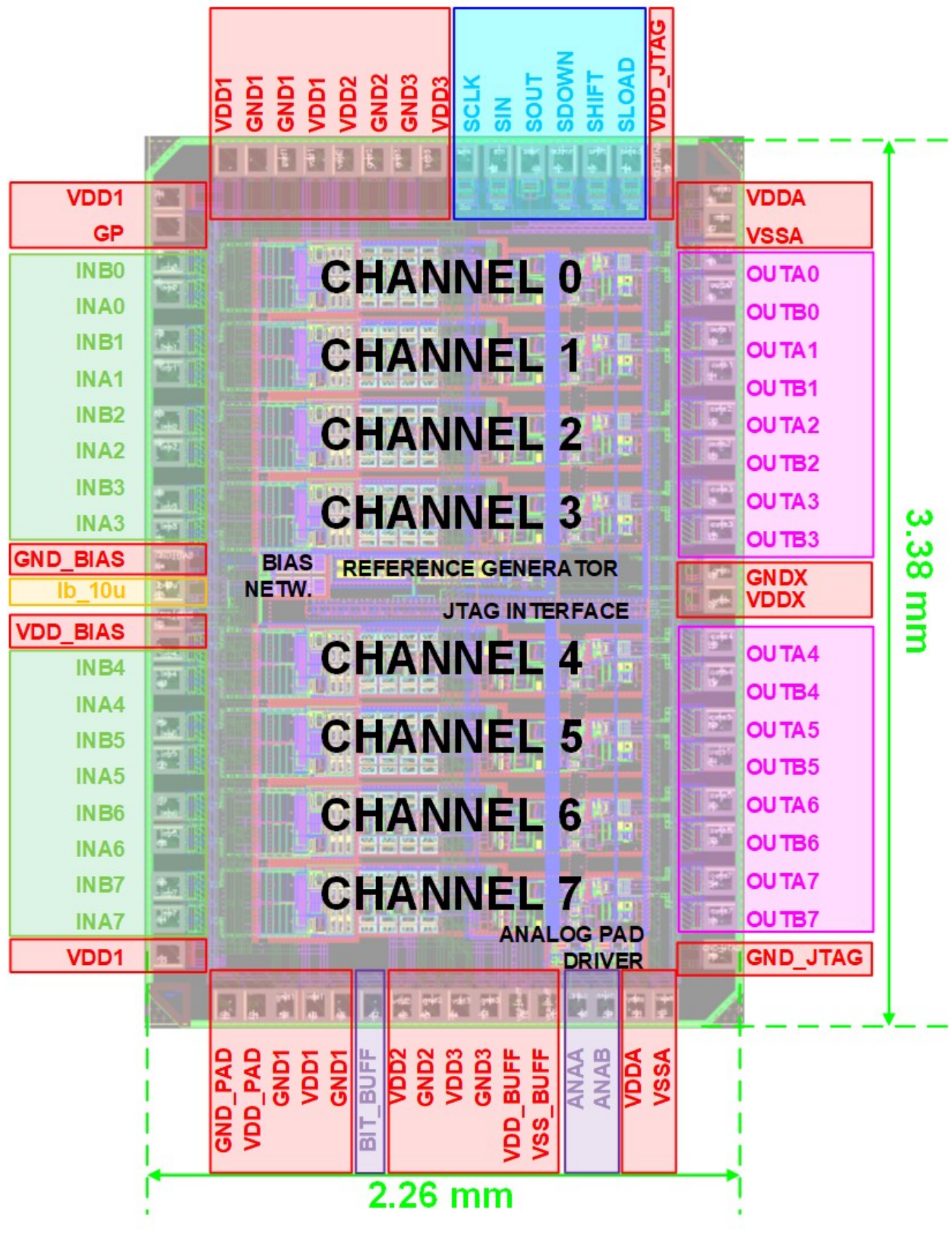

*Figure 25 - Layout and floor plan of the ASD2 chip fabricated in IBM 130 nm technology*

ASD2 occupies an area of 3380 µm x 2260 µm. The overall structure is shown in Figure 25. The padring is composed by 74 pads, according to the following segmentation:

- Digital I/O pads
- Power supply pads
- Bias network pads
- Analog input pads
- Analog output pads
- Buffer (i.e. analog pad driver) pads

The individual function of the 74 signal and power pads is given in Table 11. The various VDD power domains, kept separate inside the chip, will remain separate on the carrier PCB in order to avoid supply voltage coupling between digital and analog circuitry. In contrast, the VDD_JTAG/GND_JTAG and VDD_PAD/GND_PAD domains will be connected to the power and ground planes of the PCB. One should note in this respect, that there will be no digital activity on JTAG related pads during normal ASD operation, discarding the possibility of signal interference during data taking.

The ASD2 core includes the eight channels arranged to preserve the matching. For the same reason, all shared reference (bias network and reference generators) are in the center, see Figure 25. Power supply pads are at top and bottom of the chip to guarantee an accurate supply for each channel, minimizing resistive voltage drop with distance from the supplied channel.

*Table 11 - List and description of the I/O pads*

| | PAD NAME | DESCRIPTION | TYPE | # | I/O INTERFACE |
|---|---|---|---|---|---|
| JTAG (8 PADs) | SCLK | Clock Line | Digital Input | 1 | Vpulse Generator |
| | SIN | Load Control Line | Digital Input | 1 | Vbit Generator |
| | SDOWN | Down Control Line | Digital Input | 1 | Vpulse Generator |
| | SHIFT | Shift Control Line | Digital Input | 1 | Vpulse Generator |
| | SLOAD | Data Line | Digital Input | 1 | Vpulse Generator |
| | SOUT | Data Line | Digital Output | 1 | noConn Istance |
| | VDD_JTAG | JTAG Supply Voltage | Analog Input | 1 | DC-Volt. Generator of 3.0V |
| | GND_JTAG | JTAG Ground Voltage | Analog Input | 1 | DC-Volt. Generator of 0V |
| ASD-MDT-Channel (55 PADs) | <ina0:ina7> | Positive Input of Channel | Analog Input | 8 | - 1nF Cap. if the channel is not used<br>- Current generator in parallel with 60pF capacitance, if the channel is used for charge detection |
| | <inb0:inb7> | Negative Input of Channel | Analog Input | 8 | 470pF Cap. in simulation<br>Floating on the PCB |
| | <outa0:outa7> | Positive Output of Channel | Analog Output | 8 | Input 'A' of LVDS Termin. |
| | <outb0:outb7> | Negative Outp. of Channel | Analog Output | 8 | Input 'B' of LVDS Termin. |
| | VDD1 | CSP Supply Voltage | Analog Input | 5 | DC-Volt. Generator of 3.0V |
| | GND1 | CSP Ground Voltage | Analog Input | 4 | DC-Volt. Generator of 0V |
| | VDD2 | Shaper Section Supply Volt. | Analog Input | 2 | DC-Volt. Generator of 3.0V |
| | GND2 | Shaper Section Grnd. Volt. | Analog Input | 2 | DC-Volt. Generator of 0V |
| | VDD3 | Digital Supply Voltage | Analog Input | 2 | DC-Volt. Generator of 3.0V |
| | GND3 | Digital Ground Voltage | Analog Input | 2 | DC-Volt. Generator of 0V |
| | VDDA | MUX-LVDS Supply Voltage | Analog Input | 2 | DC-Volt. Generator of 3.0V |
| | VSSA | MUX-LVDS Ground Voltage | Analog Input | 2 | DC-Volt. Generator of 0V |
| | VDDX | Common Block Supply Volt. | Analog Input | 1 | DC-Volt. Generator of 3.0V |
| | GNDX | Common Block Grnd. Volt. | Analog Input | 1 | DC-Volt. Generator of 0V |
| CH7 – Buffer (5 PADs) | VDD_BUFF | Buffer Supply Voltage | Analog Input | 1 | DC-Volt. Generator of 3.0V |
| | VSS_BUFF | Buffer Ground Voltage | Analog Input | 1 | DC-Volt. Generator of 0V |
| | ANAA | da3_oa of CH7 | Analog Output | 1 | 1pF Capacitance |
| | ANAB | da3_ob of CH7 | Analog Output | 1 | 1pF Capacitance |
| | BIT_BUFF | Active High Enable for ana-log test output | Digital Input | 1 | Vpulse Generator |
| External Bias (3 PADs) | VDD_BIAS | Ext.l Bias Supply Voltage | Analog Input | 1 | DC-Volt. Generator of 3.0V |
| | GND_BIAS | Ext. Bias Ground Voltage | Analog Input | 1 | DC-Volt. Generator of 0V |
| | Ib_10u | Input Reference Current | Analog Input | 1 | Current generator of 10uA |
| PadRing (3 PADs) | VDD_PAD | PadRing Supply Voltage | Analog Input | 1 | DC-Volt. Generator of 3.0V |
| | GND_PAD | PadRing Ground Voltage | Analog Input | 1 | DC-Volt. Generator of 0V |
| | GP | PadRing Ground Plane | Analog Input | 1 | DC-Voltage Generator of 0V |

# 4 Measured Performance of the ASD2

## 4.1 Measurements on MPW prototypes

The ASD2 was designed to replace the ASD1, the "legacy" ASD1 in the upgrade program of the LHC, the High-Luminosity-LHC. The much higher luminosity and the corresponding increase in hit rates and readout bandwidth required the replacement of the entire readout chain of the MDT. For the frontend electronics, this allowed to profit from new developments in ASIC technology. The ASD1 design, conceived in the late 1990's, was based on the 500 nm HP technology, while the ASD2 design, started about a decade later, could use the 130 nm IBM technology, which, besides smaller feature size, provided numerous technical innovations, like Cu- instead of Al- interconnects and thinner gate-oxides. This was expected to improve, among others, high frequency behavior, noise and process control.

The practical development of the ASD2 went along a series of prototype chips from MPW runs. Aim was to optimize critical parameters like gain, peak time and noise. Other important design aims were the equality of thresholds among the 8 channels of a given chip, the relation between pulse length at a given input charge and the programmed run-down-current code (RDC) as well as the length of dead time versus programmed code. As the ASD2 is a "mixed-mode" device, containing analog and digital circuitry, the absence of digital noise in the analog signal had carefully to be controlled (4.1.2).

To test the chips from the MPW and production steps, a setup has been developed, which allowed to measure the response to a given test input at varying values of the programmable parameters with automatic recording and storage of the resulting measurements.

In the following sections we report on:

- the structure and operation of the ASD test environment
- the measurement of peak time, linearity and saturation behavior, as observed at the Analog Pad Driver APD), which corresponds to the signal seen by the DA4 discriminator (Figure 4). These measurements also allow to verify the absence of digital-to-analog interference
- the comparison between the peak times of ASD2 and ASD1 and the resulting difference in time slewing as a function of input charge (see section 4.1.3, below).
- the comparison between the gains of 2 versions of the ASD2, simulation and the gain of ASD1
- the measurement of the intrinsic noise of the ASD, where the discriminator DA4 response to a fixed test pulse is measured at varying settings of the DA4 threshold
- the hysteresis setting of the DA4 and its influence on the "effective" threshold of DA4
- the determination of the voltage differential at the DA4 input corresponding to one step of the threshold setting
- the performance of the digital part, i.e. the signal length produced by the WADC as a function of the programmed rundown-current code as well as the dead time generated with respect to the programmed dead time code
- equality of the DA4 response to threshold settings among the 8 ASD channels: This is important because all channels are controlled by one 8-bit word in the serial register, which does not allow for corrections of individual channels. A detailed inquiry of this problem with the method of the S-curve scan is presented in section 4.1.5. The same is true for the equality of run-down current (which controls the length of the WADC output) where all channels are controlled by one 3-bit word.

### 4.1.1 The Test Setup

The automatic evaluation system for ASD chips testing is shown in Figure 26. The signal generator injects test pulses with variable repetition rate into any combination of the 8 channels of the ASD. The amplitude of the signal is controlled by a programmable attenuator. Test signals are voltage steps applied to a capacitor to generate charge injection into the ASD frontend (step function leading to a delta-charge). Optical insulation between the measuring setup and the PC is used to avoid injection of digital and/or switching noise into the ASD frontend.

The operating parameters of the ASD are controlled by the "ASD serial interface board", which defines, controlled by software, the shift register bits (cf. Table 10). This way, a programmed sequence of measurements can be obtained with minimal personal intervention. A threshold scan for all 8 channels, e.g., generates an important volume of data, so automatic operation of the test is a necessity.

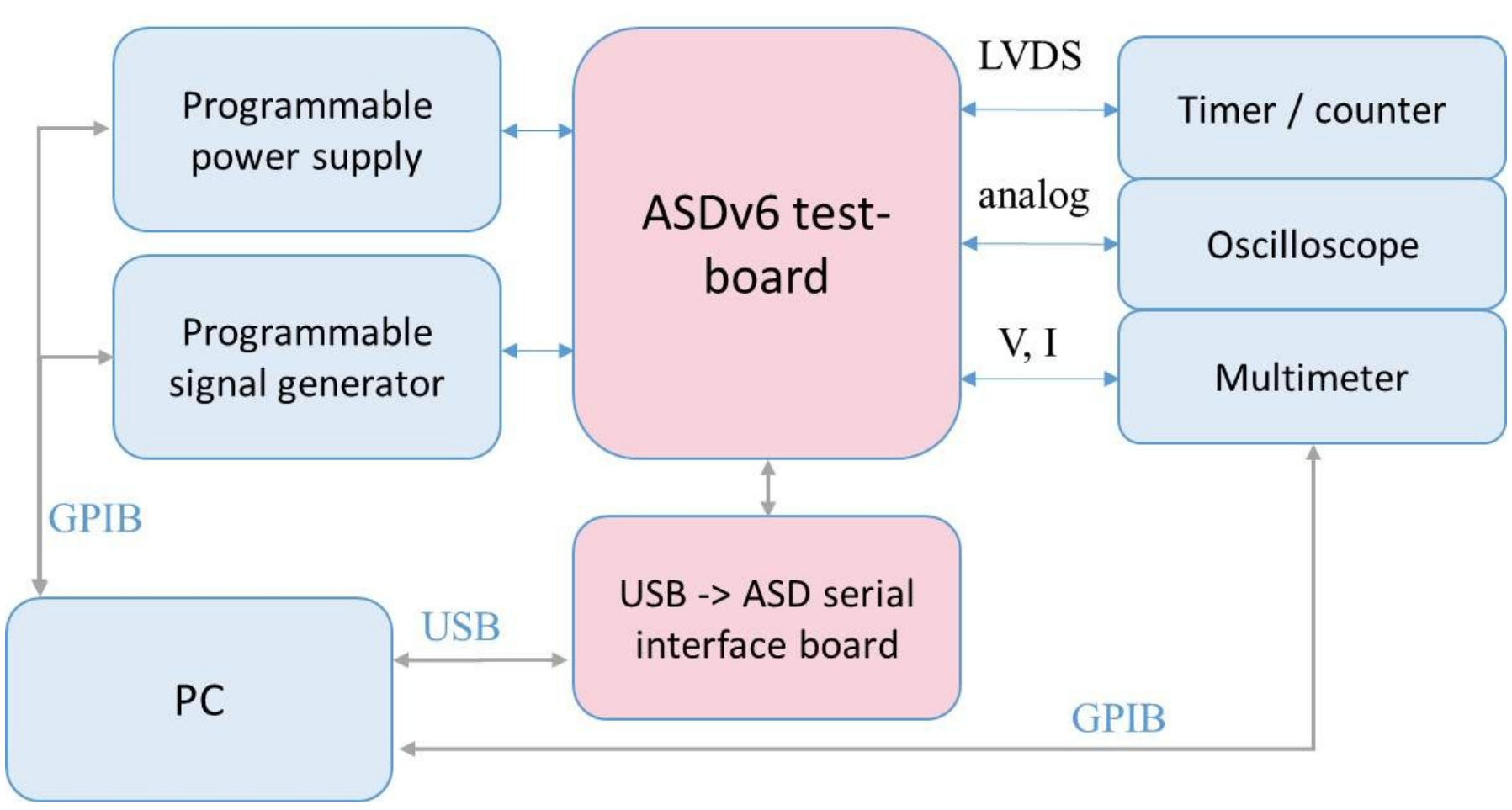

*Figure 26 - The ASD2 test environment*

While data from the timer unit, connected to the LVDS outputs of the ASD, can be stored and processed in the controlling PC, screenshots on the oscilloscope, recorded from the Analog Pad Driver on channel 7 (section 3.4.5) can equally be stored on the PC for later analysis.

### 4.1.2 Peak time, Linearity, Saturation behavior of the DA3 output

Figure 27 (left) shows the response of the shaper stage DA3 to a delta charge injection into the CSP input. The signal shown is the positive output[4] of the Analog Pad Driver of channel 7 (section 3.4.5). The peak time of about 12 ns and the zero-crossing time of 40 ns correspond to the results from simulation, see Figure 17. The screenshot on the right shows the corresponding results for the injection of much higher charges, which may occur due to heavily ionizing particles or converted neutrons. As the signal returns below the trigger threshold after about 65 ns and reaches zero after 400 ns, there is no significant loss of efficiency, as the amplifier is immediately ready to record the next signal.

The ASD chip is operating in "mixed mode", i.e. sensitive analog amplifiers are operating close to digital circuitry. Coupling from the digital part into the analog part may happen via supply voltage lines and/or grounds, but also via the substrate of the chip. Separate external supply lines for voltages

[4] ) The positive output for this measurement was selected for practical reasons. The *differential* one would have produced twice the amplitude in the screen shots, shown in Fig. 27.

and local grounds have been foreseen in the design for the CSP stage, the DA1-DA3 amplification stages, the digital section and the LVDS output part, see Table 11. Substrate coupling has been suppressed by an implantation of highly insulating boron fluoride (“BF2 moat”), which separates the domains of analog and digital activity from each other.

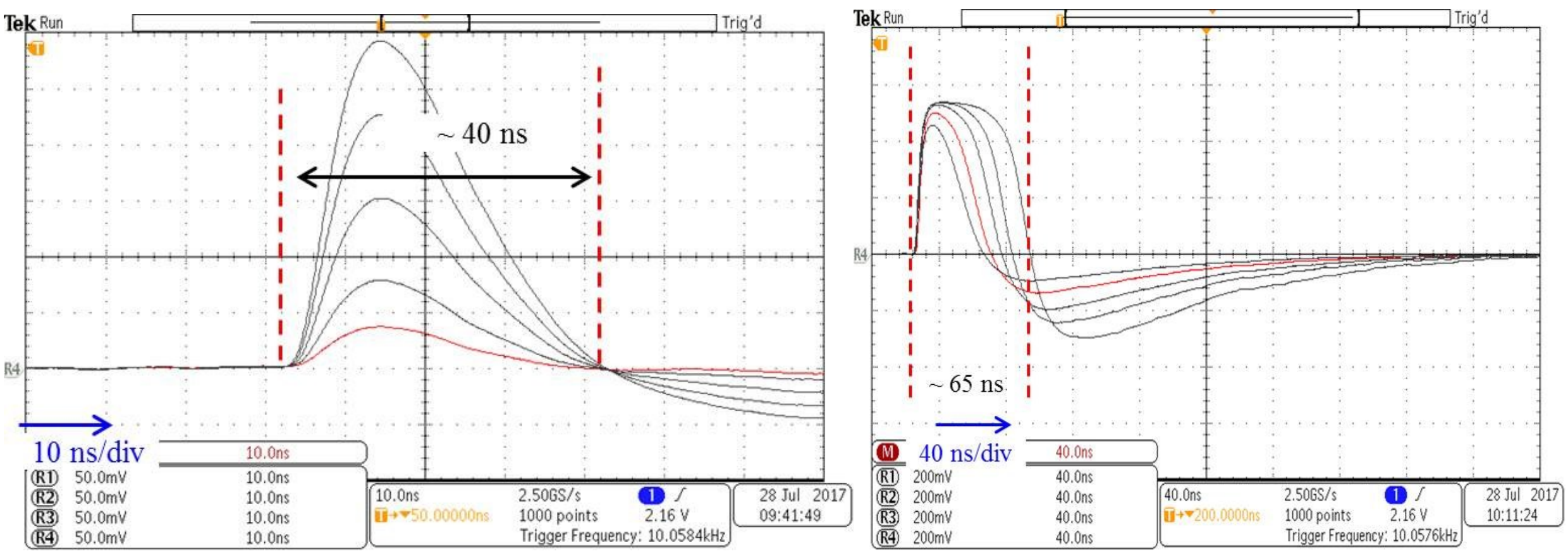

*Figure 27 - The step response to injected charges of 5,10,20,30, 40 fC behind DA3, showing a peak time of 12 ns (left). In overload, like at charges of 100, 200, 500, 1000 and 2000 fC, the DA3 output quickly returns to the base line.*

The absence of digital-to-analog coupling can be verified in the screenshot of Figure 28. It displays two analog signals at the DA3 and the corresponding digital output at the LVDS pads. As can be seen, the output of DA3 is not affected by the rising or falling edges of the digital output, demonstrating good separation of the analog and digital circuitry. The figure also illustrates the operation of the WADC: the ratio of the lengths of the 2 LVDS output signals of 1,48 corresponds to the one of the amplitudes of 1,67 at the level of about 13 %.

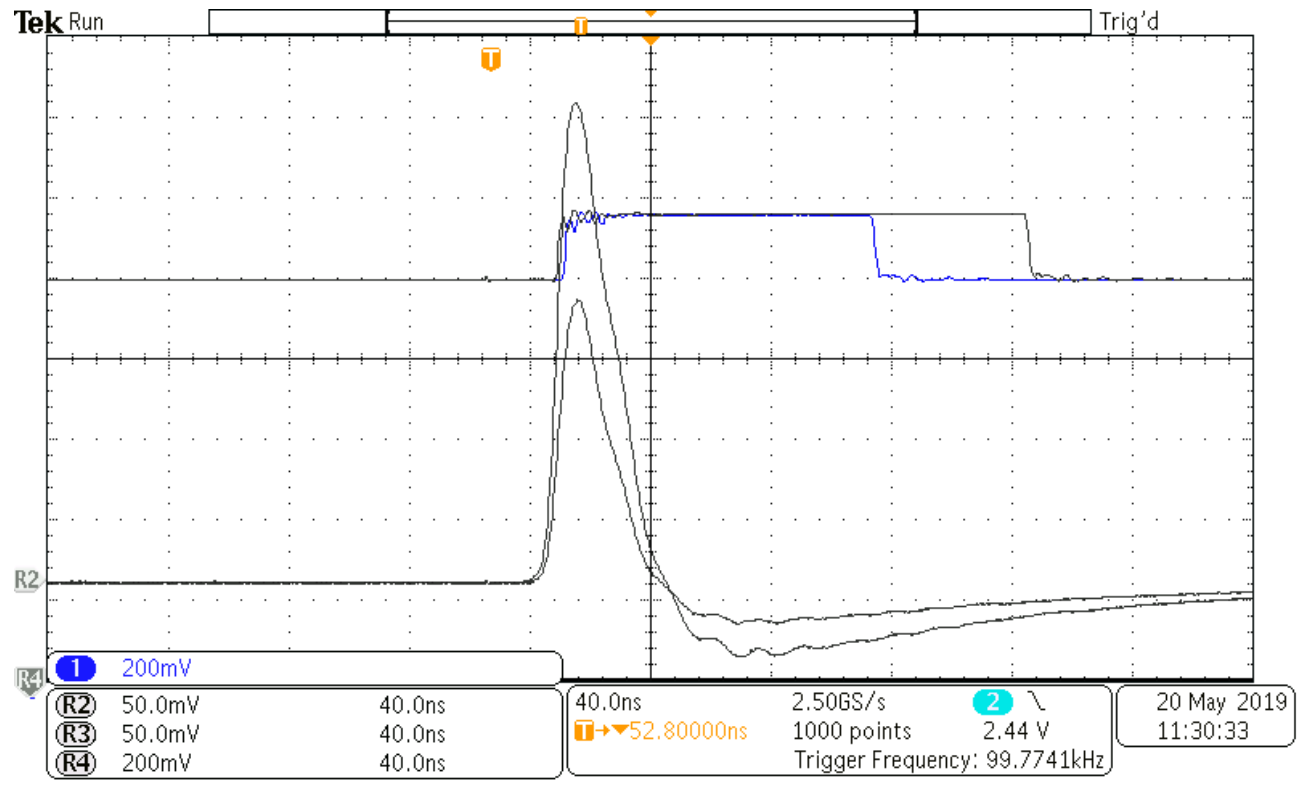

*Figure 28 - Two signals at the DA3 output (bottom) and the corresponding LVDS outputs (top)*

### 4.1.3 Time Slewing versus Input Charge

The most relevant performance parameter for the determination of the MDT coordinates is the accuracy of the arrival time measurement of the electrons closest to the MDT wire, as discussed in section 0 and 3.4.8.1. The main error source does not come – by far – from the accuracy of the TDC, but from the fact that the amplitude of the signals, produced by the MDT tube, is varying over a wide range, depending on the distribution of primary electrons along the track. Small signals take longer to reach the trigger threshold than larger ones. A short peak time and/or high gain of the DA3 output - at a given threshold - tends to mitigate this problem, the “time slewing effect”.

As discussed in the previous section, the small-signal gain of ASD2 of 19.6 mV/fC, derived from Figure 30 is in fair agreement with the simulated value of 18 mV/fC, presented in sect. 3.4.8.1 and exceeds the gain, specified in Table 3 of 8.9 mV/fC as a minimum requirement by a factor of two.

Figure 29 shows the measured time slewing for ASD2 (red) and for the “legacy” ASD1 (blue). Due to higher gain and shorter peak time, ASD2 provides less time slewing at a given input charge. At 20 fC, e.g., the slewing delay is about 2 ns compared to about 4 ns in ASD1. Given the average electron drift velocity of about 20 μm/ns, this corresponds to measuring errors of 40 μm and 80 μm for ASD2 and ASD1, respectively. Part of this effect can be corrected off-line, using the charge measurement of the WADC.

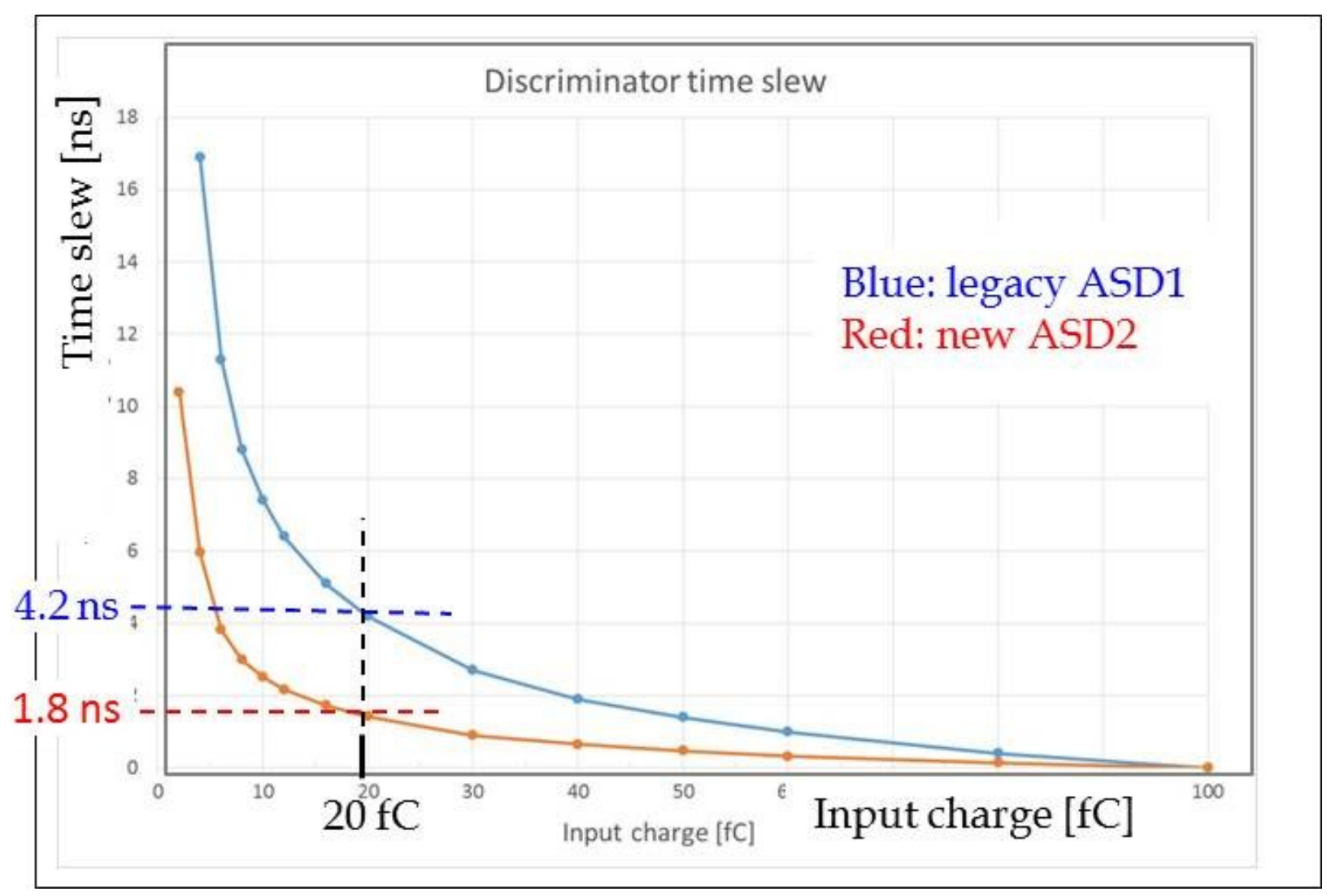

*Figure 29 - Time slewing vs. Input Charge for ASD2 (red) and ASD1 (blue)*

Figure 29 indicates that offline corrections for “time-slewing” will be less important once the new readout electronics with the ASD2 will be implemented in the Muon Spectrometer.

### 4.1.4 Gain of the ASD

Figure 30 shows the peak amplitudes at the Analog Pad Driver (APD) versus input charge for two ASD2 chips and a legacy ASD1. The curves for the two ASD2 only show a small mismatch of up to 3 % at input charges > 50 fC, illustrating good process control of the 130 nm GF technology.

The hatched line comes from simulation (at the schematic level), showing fair agreement with the measurements. As explained in section 3.4.5 the peak amplitudes presented by the APD have to be multiplied by a factor 1,35 in order to correspond to the output of DA3 (see section 3.4.5). Thus, at an Input Charge of 20 fC the reading of 300 mV in Figure 30 is equivalent to a Peak Amplitude of 400 mV at the input of DA4 and the gain at DA4 is 20 mV/fC.

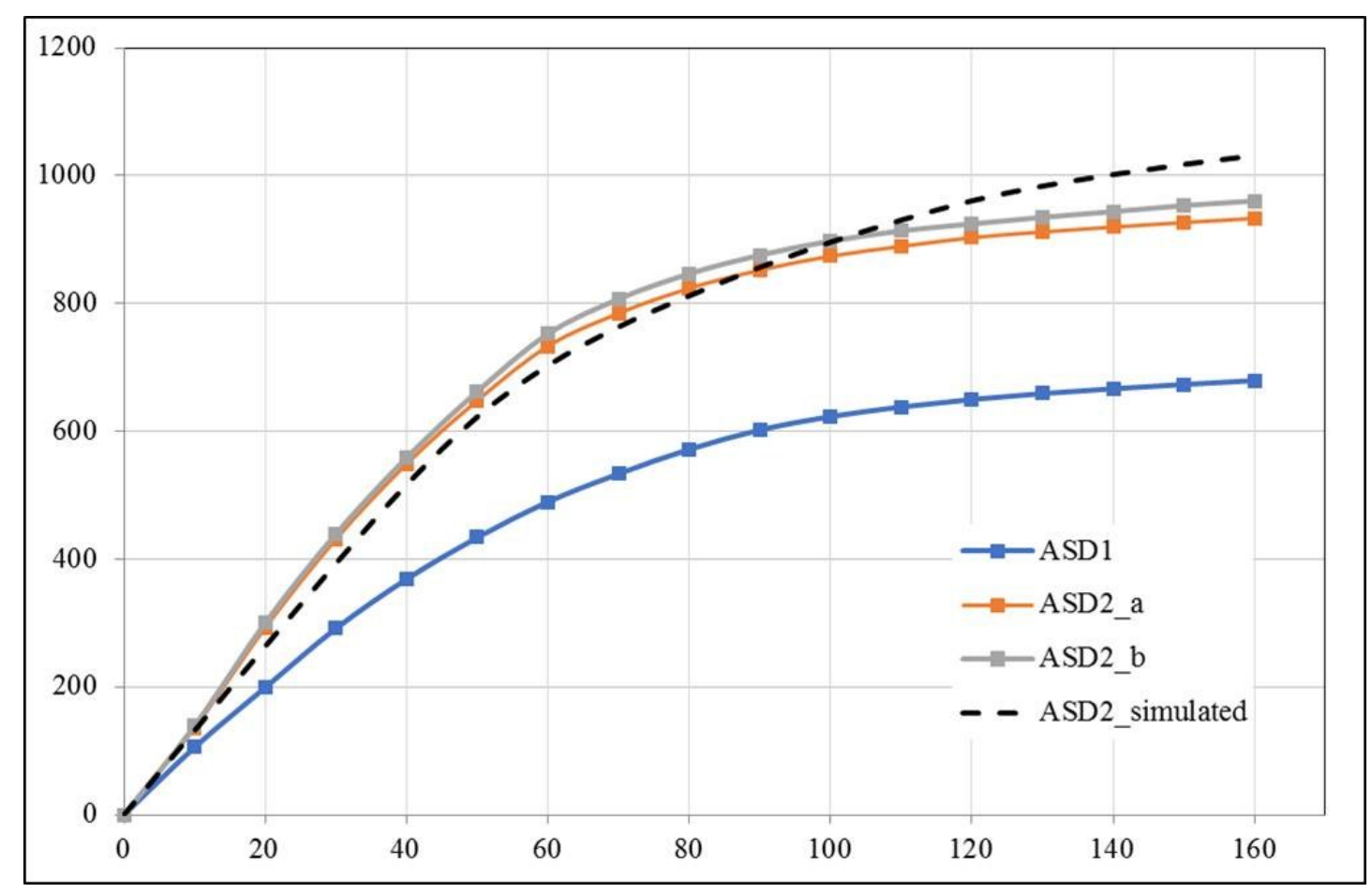

*Figure 30 - Peak amplitude at the APD (mV) vs. Input Charge (fC) for two prototypes of the ASD2 (red, grey) and for an ASD1 chip*

### 4.1.5 Threshold Scans

The following procedure was used to measure the noise of an ASD channel. A pulse with constant amplitude was injected into the input of the ASD with a constant frequency, like 10 kHz, while the frequency of the discriminator *output* was recorded as a function of the programmed threshold. Figure 31 shows results for one channel of the ASD2. Scanning the pulses by threshold steps from high to low, the output frequency is changing from zero to 10 kHz. At 5 kHz recorded rate, half of the hits is below and half above the applied threshold. The width of the S-curve corresponds to the noise on top of the test pulses, which is mainly due to the noise of the CSP and the rest of the analog chain[5].

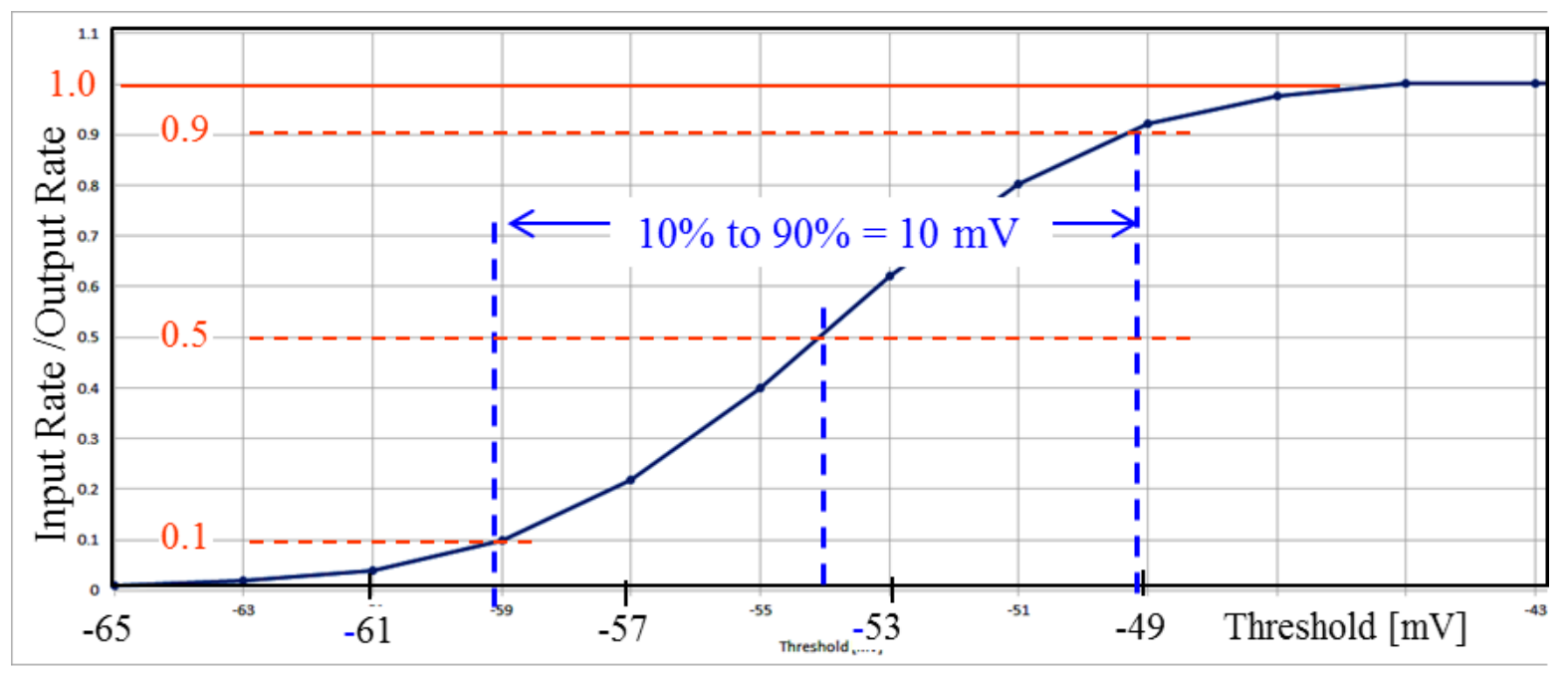

*Figure 31 - Threshold scan for one channel of an ASD2 chip*

If the pulse height distribution at the discriminator corresponds to a gaussian, the S-curve in Figure 31 represents the integral over its area. The FWHM of the gaussian is about 0.94 of the 10-90% range of the integral, and $\sigma_{RMS}$ about 0.4 of this range. Looking at Figure 31, the 10-90% voltage range is (at face value) 10 mV, leading to a FWHM of 9.4 mV and a $\sigma_{RMS}$ of 4 mV. Based on the measured gain of 15 mV/fC, the $\sigma_{noise}$ is about 0.27 fC, corresponding to 1700 electrons.

Ideally, all 8 channels of an ASD chip should produce an output signal, when more than 5 primary electrons (equivalent to a charge of about 1 fC) arrive at the wire of an MDT drift tube. As the discriminator thresholds, however, cannot be adjusted individually for each channel, a sufficiently small threshold spread at a given threshold setting is essential for uniform detector operation. Threshold deviations to a higher value may reduce efficiency and increase slewing delays, while deviations in the opposite sense may increase the rate of noise hits.

[5] If there was “zero noise”, the distribution would be a step, going from zero to one at the discriminator threshold.

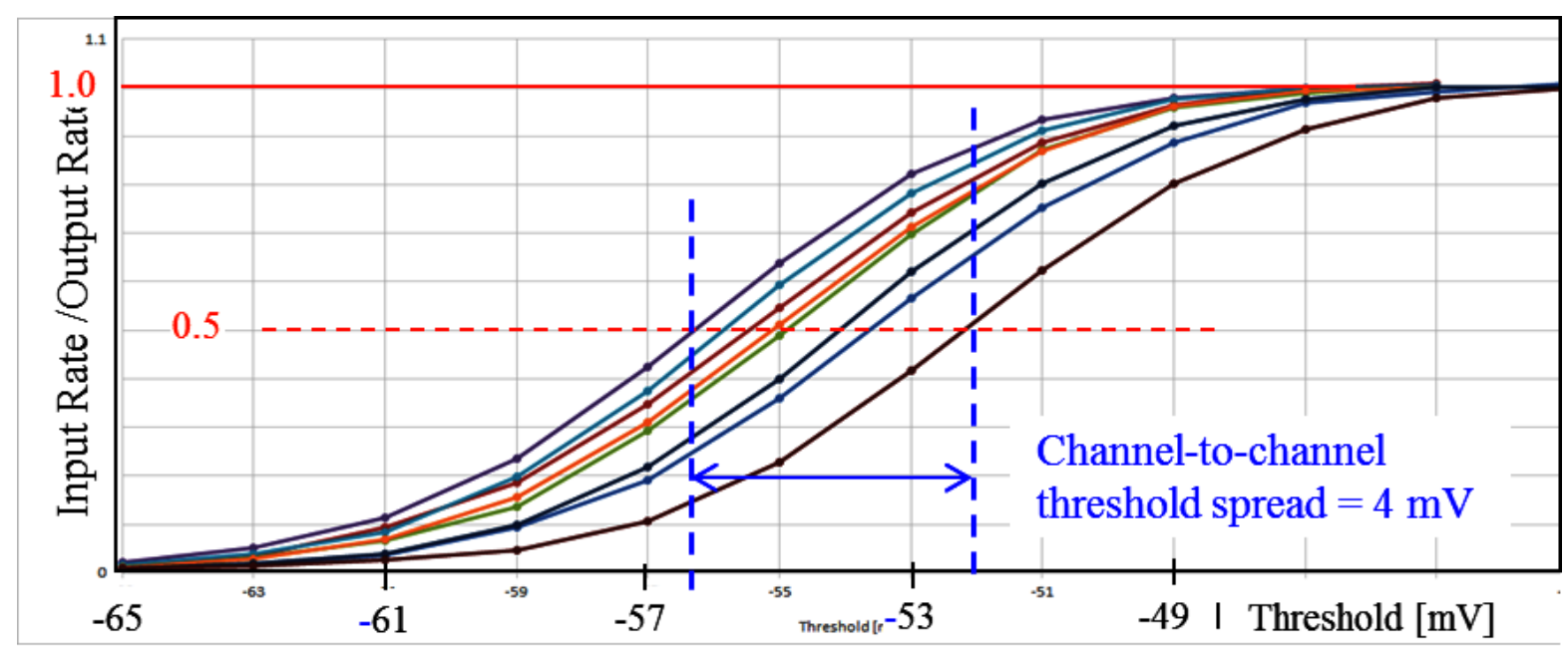

*Figure 32 - Threshold scan for all 8 channels of an ASD2 chip*

The spread of the discriminator thresholds in the ASD2 is about a factor of 3-4 lower than in the case of the "legacy" ASD1. This improvement of the threshold matching is thought to be due to improved technology and process control in the 130 nm IBM/GF technology compared to the 500 nm HP technology, used for ASD1.

### 4.1.6 Calibration of Discriminator Codes w.r.t. Input Charge

The voltage differential corresponding to "*one step*" of the DISCR threshold codes can be derived from the gain of the amplifier in the small signal region, as shown in Figure 30, combined with the threshold codes for two input charges, as displayed in Figure 33. The number of about 15 mV/fC derived from the slope of the ASD2 curves in the low signal region has to be corrected for the signal attenuation of 2.6 dB in the APD (see section 3.4.5), giving a gain of 20 mV/fC at the input of DA4. Figure 33 shows that 10 fC correspond to 62 threshold counts, which leads to about 2,9 mV per step of the discriminator. It should be noted that this correspondence comes from internal design choices. The exact step size of the discriminator in terms of mV is not relevant for setting of the programmable parameters.

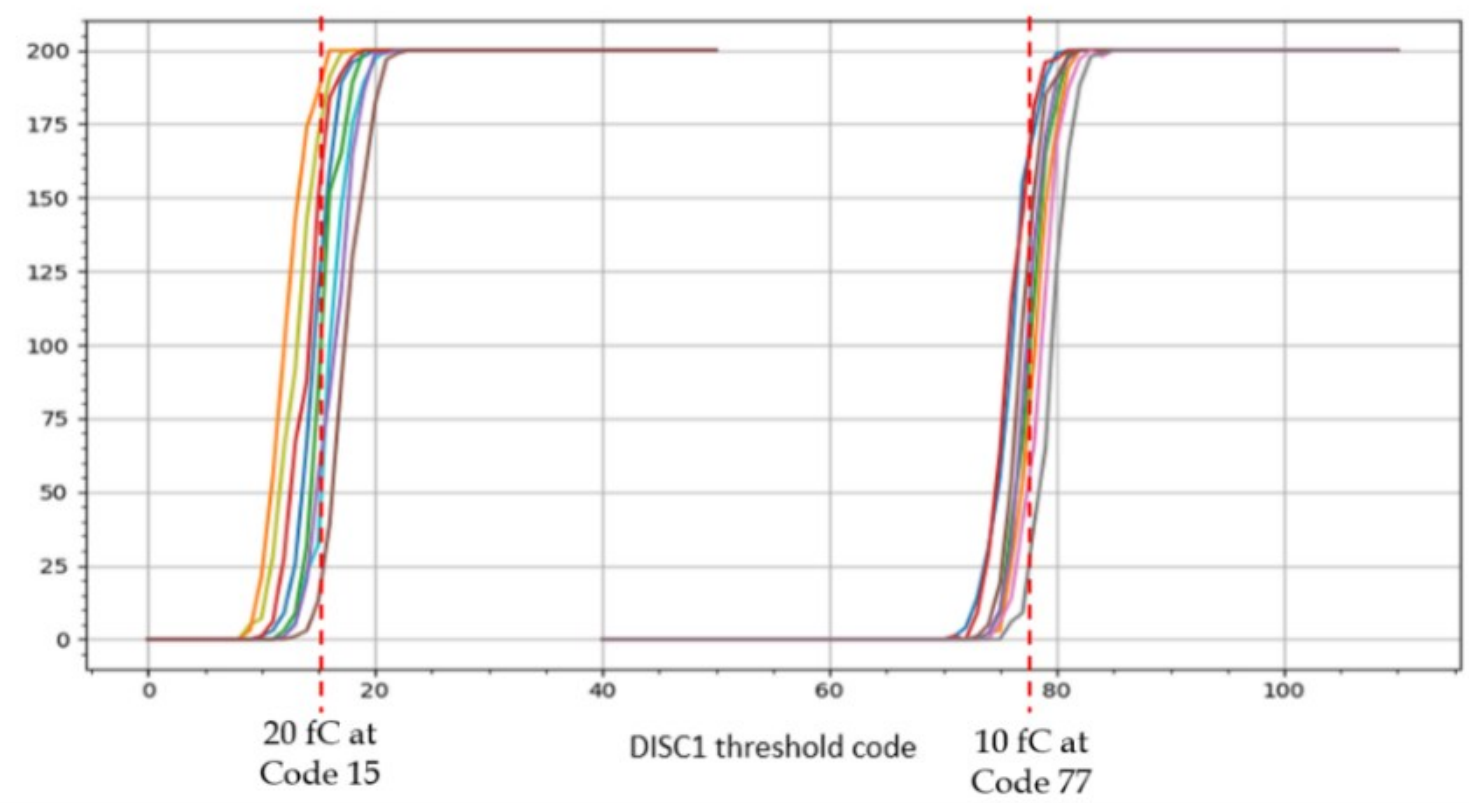

*Figure 33 - Threshold scans for 10 and 20 pC input charge, corresponding to DISC reading 77 and 15, resp. All 8 channels of one chip have been tested in this scan.*

### 4.1.7 The hysteresis setting of the Discriminator

The discriminator of the ASD compares the signal amplitude with a fixed voltage, which is controlled by the 8-bit "threshold code", defined in the shift register (Fig.24). When the input signal is exceeding the threshold, the discriminator is triggered, generating a digital output. The discriminator, however, is not immediately ready for a second trigger, but first requires that the input amplitude falls below a second threshold, lower than the main trigger threshold. The difference between the two thresholds, the hysteresis, is controlled by the 4-bit code in the shift register.

The purpose of the comparator design with hysteresis (Schmitt Trigger) is to reduce the rate of digital outputs in a situation where the amplitude is fluctuating in a narrow band around the trigger threshold due to a high noise environment. In our application, the function of the hysteresis is redundant to the one of the programmable Dead Time, which, on its own, provides sufficient protection against high-frequency toggling of the discriminator output.

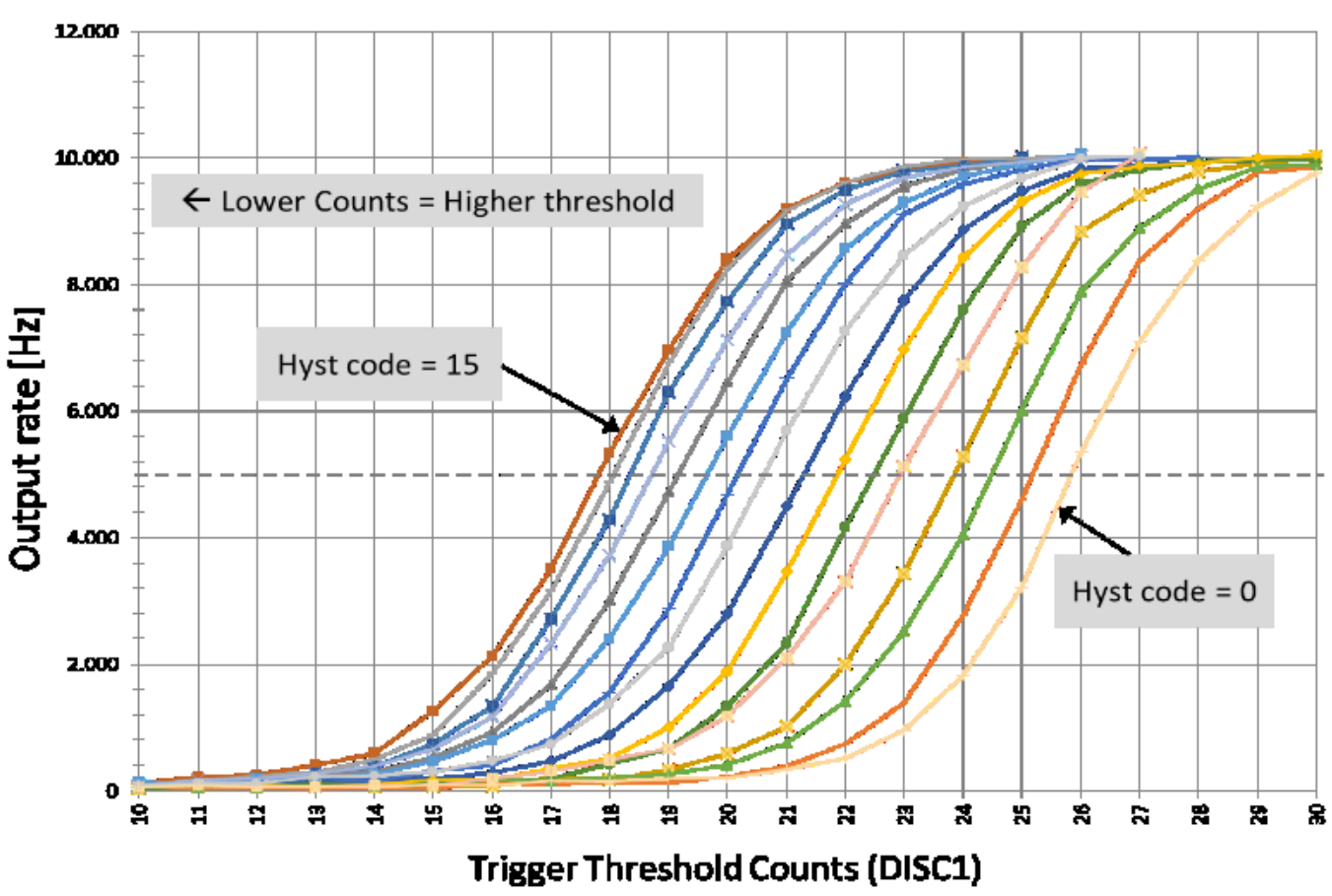

*Figure 35 - Threshold scans at 20 fC input charge vs.Hyst code. The effective threshold varies between 18 to 26 counts.*

*Figure 34 - Variation of the effective threshold vs. Hyst code.*

This “redundancy” between hysteresis and Dead Time, however, is not fully implemented in the ASD1 due to a design imperfection. In the ASD1, small pulses could trigger the discriminator DISC1 *but not the discriminator DISC2*, which starts the loading of the WADC integration gate and Dead Time generation (cf. Table 8), resulting in DA4 outputs, with very short signal length, this length being determined by the Time over Threshold. These “events”, due to noise and showing up in ADC spectra as spikes at “zero”, could be suppressed by a hysteresis setting > 0. In the experimental runs 2009-2025

(where only ASD1 chips were used) a hysteresis value of 14 was selected. In the ASD2, however, this problem in small signals processing does not exist, and a particular hysteresis setting is not required to suppress frequent re-triggering of the DA4 output.

Using a non-zero hysteresis setting has a direct influence on the *effective* trigger threshold, as shown in Figure 34 and Figure 35. In this Threshold Scan a charge of 20 fC was injected, while the hysteresis code varied from 0 to 15. The resulting variation of the *effective* threshold ranged from about 18 to 26 counts (Figure 34).

In practical chamber operation, the threshold value should be low for optimum efficiency but high enough to keep noise hits below an acceptable level. Using the ASD2, it may be good practice to keep the hysteresis code at a fixed value (once and for all) and “play” with the Trigger Threshold to find the best compromise between hit efficiency and hit rate from noise.

### 4.1.8 The digital part: the WADC and Dead Time generation

#### 4.1.8.1 The performance of the WADC

The charge detected by the WADC is translated into the time delay between leading and trailing edge of the discriminator output of the corresponding channel. The time delay is approximately proportional to the run-down current which, in turn, is inversely proportional to the resistance of the discharge resistor. This resistor can be selected to have 8 values, being controlled by a 3-bit code (bits [24:26] in Table 10.

Figure 36 shows the width of the LVDS output versus the 3-bit run-down current code for the 8 channels of an ASD2 chip and for the 24 channels of three chips (right). Ideally, all 8 trace would be on top of each other. Due to process variations there is a spread, increasing with increasing code (i.e.

decreasing rundown current). At a typical operating code = 2 the spread among the 8 channels is 120 to 150 ns, which is acceptable for our application, as the measured charge is only used to correct for the slewing effect. The screen shot on the right of Figure 36 shows the same dependence for 3 different ASD2 chips. The spread among these 3 chips at operating code = 2 is again 120 to 150 ns.

#### 4.1.8.2 The Linearity of the Dead Time versus DT code

As mentioned in 0, the discriminator can be disabled after a threshold crossing for a programmable time, which is controlled by a 3-bit code. This feature of the chip allows to suppress multiple threshold crossings from the same track, leading to a reduction of the data load to the DAQ. Figure 37 presents the response of dead time vs. 3-bit code for the 8 channels of an ASD2 chip, showing a satisfactory match among the 8 channels. Depending on the code, the dead time can be selected between 40 and 800 ns in steps of 100 ns. As the maximum drift time of electrons in the MDT tubes (30 mm diameter) is about 750 ns, the programmable dead time covers the whole range of the drift time. In the case of the small-diameter MDT tubes (sMDT) with a maximum drift time of 180 ns, a programmed dead time of 200 ns would be an appropriate choice (Dead Time code = 1).

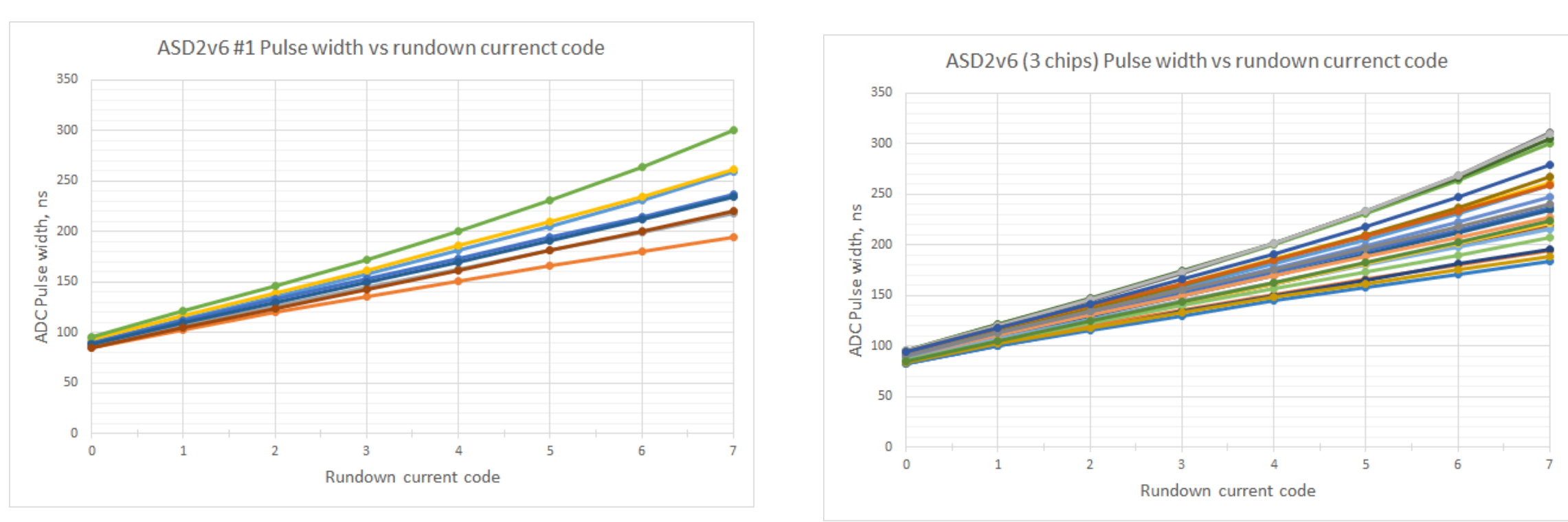

*Figure 36 - Pulse width vs. Run-down code for the 8 channels of one (left) and of 3 chips (right)*

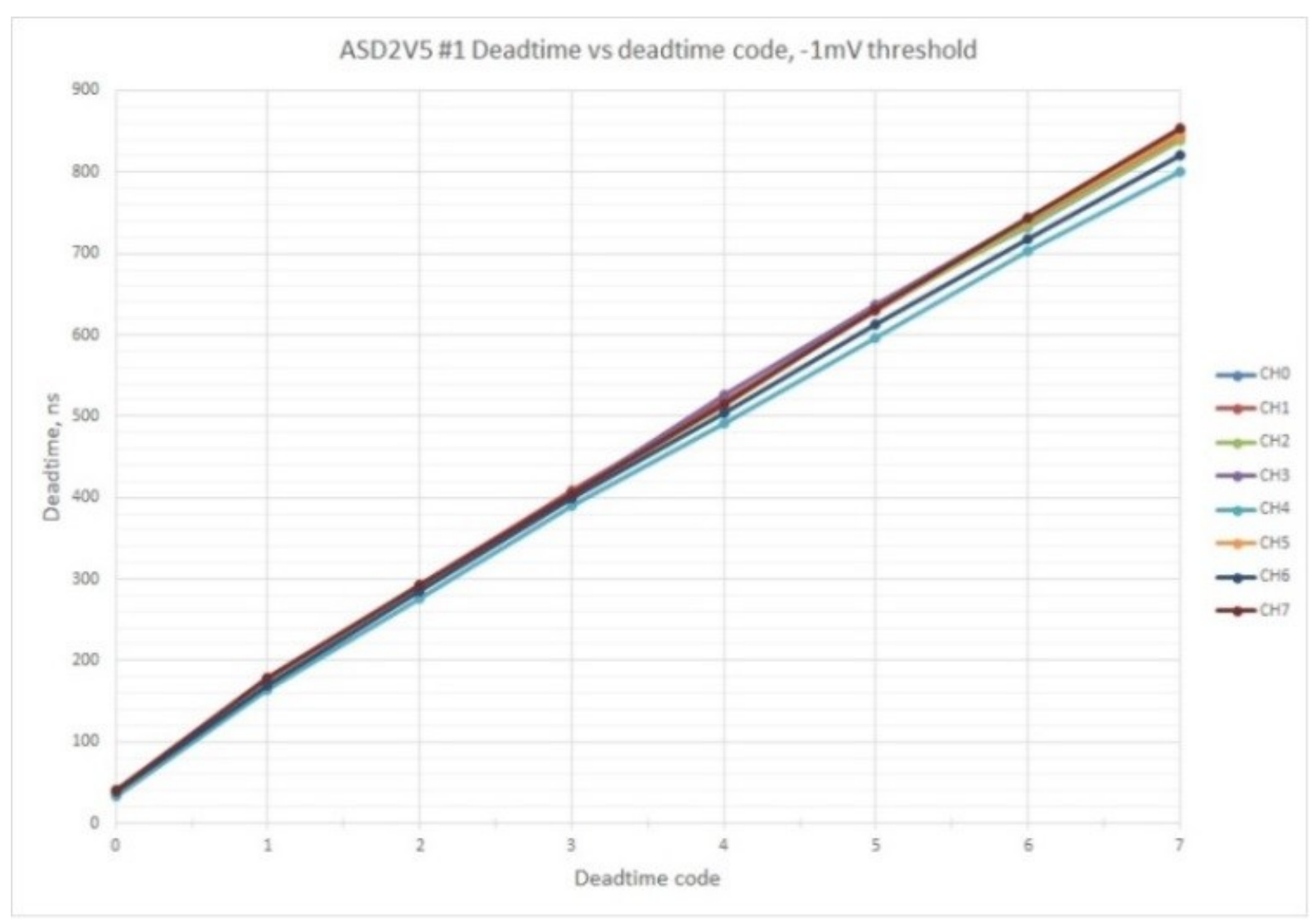

*Figure 37 - Dead time (ns) vs. 3-bit Dead Time code for the 8 channels of an ASD2 chip*

## 4.2 Performance of chips from the Engineering Run

A production order of 27 wafers was placed with Global Foundry. Each of the 8'' wafers were expected to yield about 3500 ASD2 chips. After delivery of 3 preproduction wafers ("engineering run"), packaging was ordered with the IMEC company. Wafers were thinned to 100 μm thickness, diced into individual chips and bonded into a QFN88 package. Bonding wires were Cu with 20 μm diameter and an average length of 2,7 mm. The detailed bonding plan is shown in Figure 45. Five pins of the package are bonded to the package ground.

### 4.2.1 The Automatic Test Facility

While Multi Project Wafer runs (MPW) only delivered small quantities of chips, the *engineering run* gave about 10000 chips, which allowed for higher statistical significance in the evaluation of chip performance. Test results of a subset of 1400 chips are presented in [25], while tests of a similar quantity of chips from the *production run* are described in [26]. For the test of the whole quantity of about 80.000 chips from the volume production, a specific test facility was developed, where all test functions could be operated automatically, see Figure 46. In this facility, measurements are done by digitally controlled commercial components, all on-board functions being steered by a micro-processor. Communication for parameter setting and test sequencing is achieved with a high-level control language. Test results are automatically transferred to an external data base for further analysis [6].

### 4.2.2 Parameters Tested

The design of the ASD for production was identical to the one for the MPW prototypes, however, for volume production, new masks were needed to cover a full wafer. A careful comparison between chips from production and MPW is needed, as process variations might lead to different performance of the same design in different production cycles.

For use in the experiment, it had to be tested, whether all relevant parameters are inside an acceptable margin around the following parameters:

- The total current drawn by the ASD: This is a simple but quite revealing parameter, as malfunction or outside-limit behavior is often reflected in a deviation of at least one of the 4 supply currents from the nominal value.
- Gain and noise: Threshold scans at a fixed value of the injected charge yield the gain of each of the 8 channels of the ASD, while the variation among the 8 values is a measure of the channel-to-channel threshold spread as well as of the noise (see section 3. 1.6). These two parameters are the most relevant ones for the accurate determination of the drift time and thus for the momentum resolution of the Muon Spectrometer.
- The programmable Dead Time allows to disable the input for a predefined time span in the range 180 – 900 ns. The dead time prevents the generation of multiple hits from the same track. Typical settings are 750 ns for Large MDT tubes and 180 ns for Small sMDT tubes.
- The pulse height measurement: As discussed in 3.1.7, the Wilkinson-ADC provides an approximate measurement of the signal slope when crossing the threshold. This allows for a correction of the delay, when small pulses are crossing the threshold, compared to large ones, see section 4.1.3. While this correction of the order of 1 to 2 ns is only relevant for a small fraction of pulses, it can improve the overall spatial resolution of MDT tubes by up to 10 %.
- The signal swing of the lvds outputs: the amplitude of these differential digital signals has to fulfill the requirement of the TDC inputs (cf. section 3.4.5).

[6] This work was done in cooperation between LMU Munich, MPI Munich and a commercial company.

While the ASD2 chip performance, presented in sections 4.1.2 to 4.1.8 was only based on a small number of devices from MPW runs, the tests of chips from engineering and production runs provided much higher statistics and allowed to compare the performance of chips from volume production with the one from MPW runs. In this campaign, about 1400 chips from the engineering run and 1175 chips out of the volume production were tested and the results compared to the values obtained from the MPW runs [26][26].

The first step in a volume test is a measurement of the currents drawn by the 4 power domains of the chip. Significant deviations from nominal currents often point to defaults in other parameters, so faulty chips could be discarded before further tests. Figure 38 shows the currents drawn by the 4 power domains in a test of 1175 chips, see [26]. VDD1 is the supply current of the CSP, VDD2 the one for the shaping Stages DA1-DA3. VDD3 supplies the discriminator DA4 and the WADC, while VDDA supplies the lvds drivers. On this basis, 48 chips have been discarded from further consideration on the basis of the loose and additional 29 chips on the basis of the tight cuts.

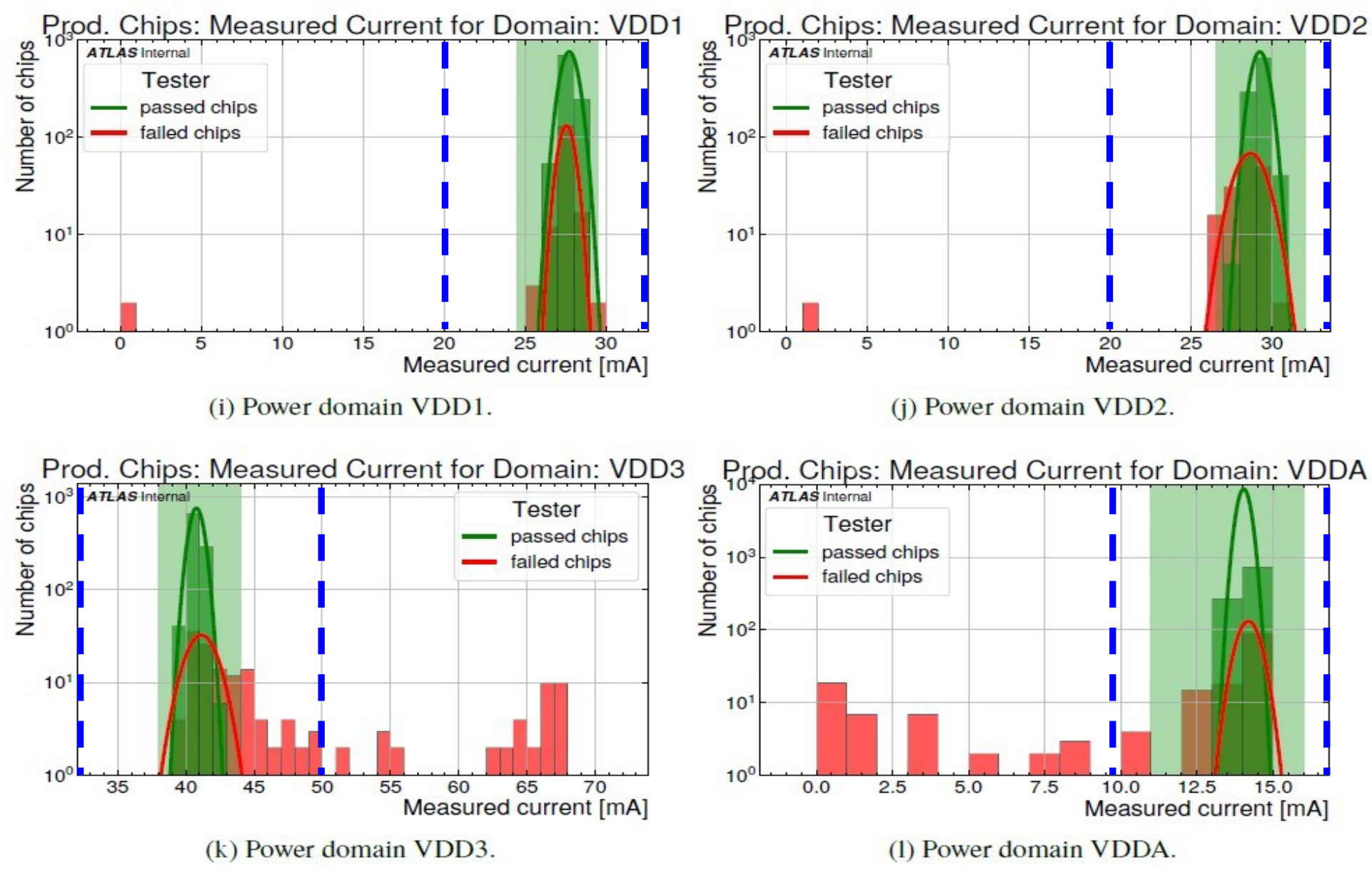

*Figure 38 - Cuts applied to the 4 power domains for acceptance for further, more detailed tests as presented in [26]. Green area: tight cuts, blue lines: loose cuts. (Note the logarithmic scale.)*

In a 2nd step, the setting of the serial string, which controls the function of the chip, was tested by a write-read cycle, which did not show any failures. The same is true for the lvds signal levels, which require a voltage swing in the range 150-230 mV for both polarities (cf. section 3.4.5).

The next steps were:

- test of the threshold spread inside each chip
- equality and absolute value of the dead time in response to the corresponding DT code
- the pulse width of the Wilkinson ADC in response to a test pulse with fixed amplitude at 2 values of the programmable run-down current code.

The results, derived from Table 6.2 in [26], are presented in Table 12 in a slightly simplified form, while Table 13 gives the cuts applied to the different parameters.

In summary, 833 chips out of the 1175 had all measured parameters inside tight cuts and an additional 175 chips inside loose cuts, corresponding to about 71% and 86% of all 1175 chips, respectively. 167 chips, i.e. 14% were rejected.

The volume production tested by the external company gave the following results: Among 73400 chips in the volume test, 57000 (77,6%) were inside tight and 11400 (15,6%) inside loose cuts. A quantity of 5000 chips (6,8%) was rejected (see [27]).

It should be noted that there appears to be a inconsistency between the rejection rates given in table 6.2 in [26] and the distibution of the measured parameters given in the same paper (cf. Figure 39). Thus, 68 chips were discarded in the test of threshold uniformity (6,2%), while Fig. 6.1 (c) shows only very few chips outside the limit of < 3 counts. Fig. 6.1 (d), giving the threshold distribution, is probably meant to be the basis of the rejection criterium of (10.0,30.0) at 18 fC injection charge. However, these cuts seem not to correspond to the numbers 32 to 46 given on the x-axis of this figure.

*Table 12 - Yields of accepted chips w.r.t. various parameters from the ASD2 production run. Categories A and B correspond to chips fulfilling the tight and loose quality cuts, respectively.*

| **1175 production chips** | A | B | C | A + B |
|---|---|---|---|---|
| Currents drawn OK | 1098 | 29 | 48 | 1127 |
| Threshold uniformity | 1030 | 3 | 142 | 1033 |
| Dead Time | 979 | 29 | 167 | 1008 |
| Wilkinson-ADC | 833 | 175 | 167 | 1008 |
| *yield* | *70,9%* | *14,9%* | *14,2%* | *85,8%* |
| **73.416 production chips** | 57.001 | 11.432 | 4.983 | 68.433 |
| *yield in volume test* | *77,6%* | *15,6%* | *6,8%* | *93,2%* |

This observation has to be kept in mind, in the case, chips from class B or C might be required for a later detector upgrade or for the repair of critical components like mezzanine cards, as a substantial fraction of chips in category B and C might still be acceptable for these applications. An additional test of the remaining chips might be advisable in this case.

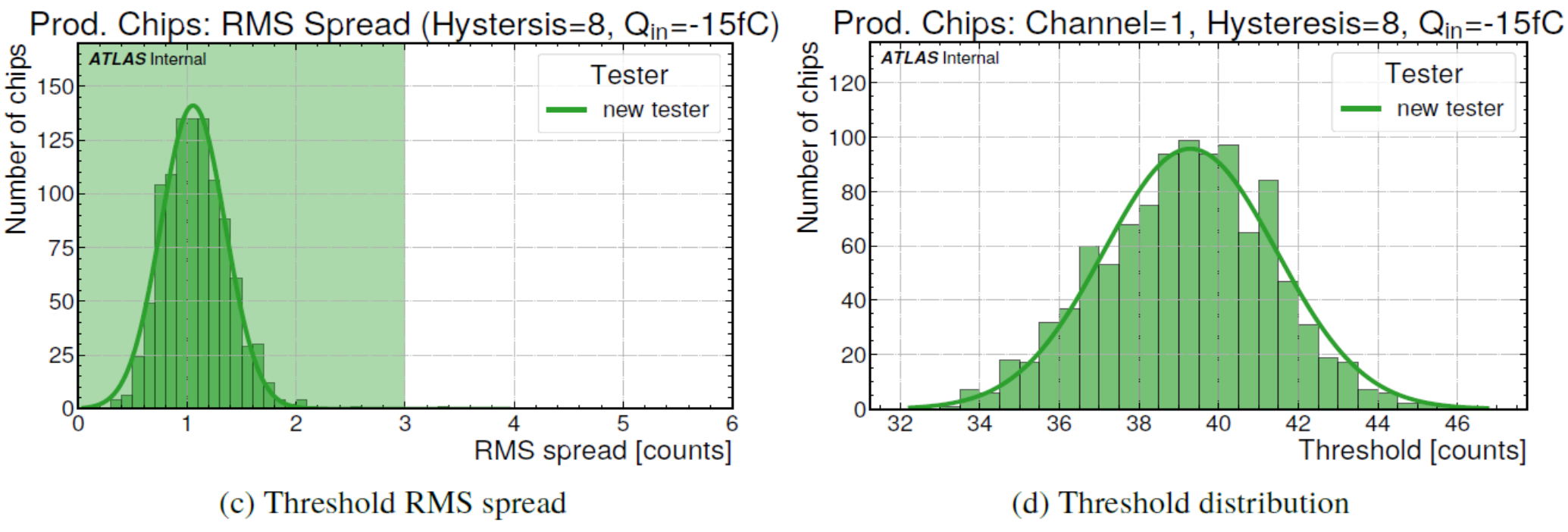

*Figure 39 - Threshold plots from Fig. 6.1 in [26] to be compared to the classification table 6.2 in the same paper, reproduced in Table 13, above*

*Table 13 - Limits of acceptance for the ASD2 chips in the volume test*

| | | Cat. A | Cat. B |
|---|---|---|---|
| **Currents drawn [mA]** | VDD1 (CSP) | 24 - 30 | 20 - 38 |
| | VDD2 (shaping stages) | 26 - 32 | 20 - 40 |
| | VDD3 (DISCR, WADC) | 38 - 44 | 30 - 50 |
| | VDDA (lvds driver) | 12 - 16 | 10 - 20 |
| **Threshold** | RMS spread | < 3 counts | |
| | Distribution at 18 fC | 10 - 30 counts | |
| **Pulse width from WADC** | RMS spread [ns] | RDC=2: < 35 | RDC=4: < 70 |
| | Distribution at 31 fC [ns] | RDC=2: 50-250 | RDC=4: 80-300 |
| **Dead Time** | RMS spread [ns] | DTC=2: < 20 | DTC=7: < 60 |
| | Distribution [ns] | DTC=2: 250-550 | DTC=7: 750-1050 |

## 4.3 Immunity against Environmental Effects

### 4.3.1 Radiation Levels in the Experimental Hall: Tolerance against Total Ionizing Dose

The frontend electronics, containing the ASD2, is exposed to ionizing radiation caused by conversion of neutrons and $\gamma$-rays. The Total Ionizing Dose (TID), accumulated in the Muon Spectrometer during 10 years of HL-LHC operation (i.e. for an event rate of about 4000/fb) is expected to be in the range 2-20 krad, depending on the location of the device in the detector. These numbers already contain "Safety factors" of about 15, representing uncertainties due to variations in dose rate ("low dose rate effect"), chip-to-chip variation and others (cf. appendix A in [4]).

To certify the ASD2 for sufficient tolerance against ionizing radiation, four chips were exposed in the CERN X-ray irradiation facility. Two chips were irradiated up to 150 krad, while two others were irradiated up to 1 Mrad, the maximum dose available in this campaign. Subsequent measurements did not reveal any changes w.r.t. pre-irradiation in any performance parameter, like peak time, gain, noise, ADC functionality or LVDS output amplitudes.

As the tested dose of 1 Mrad exceeds the required TID level of 20 krad by a large factor, TID tolerance of the ASD2 for use in the Muon Spectrometer has been successfully demonstrated. This result is in accordance with predictions in a publication by F. Faccio, CERN, about radiation damage in deep submicron CMOS [23].

### 4.3.2 Packaging, Power Dissipation and Junction Temperature

The ASD2 chip is packaged in a QFN-88 package with 10 * 10 mm$^2$ outer diameter and 0.4 mm pad pitch. A unique device number is attached to each chip, using a label with a 2D-bar. The rear side of the chip contains, inside the square of solder pads, a blank surface for good heat transmission through tight contact to the PCB. Figure 41 shows a readout card for the sMDT with three ASDs mounted.

Figure 40 presents the top and bottom view of the ASD2 package and a picture of the package taken with an X-ray tomograph, where the wiring between the 74 bond pads of the dye and the 88 soldering pins of the QFN can be seen. Blue color corresponds to material with high X-ray absorption, like copper, yellow to low-Z materials like silicon and aluminum.

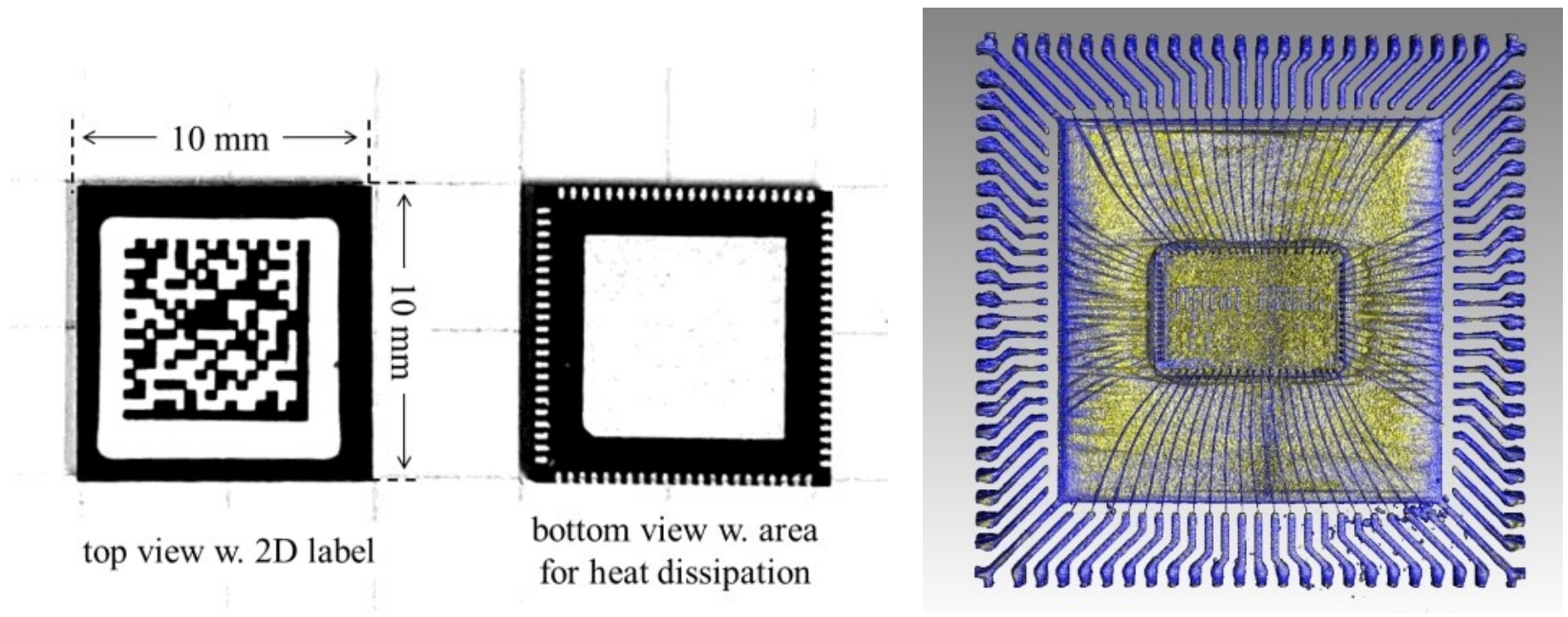

*Figure 40 – Photo of the package(left) and X-ray picture of an ASD2 dye (right).*

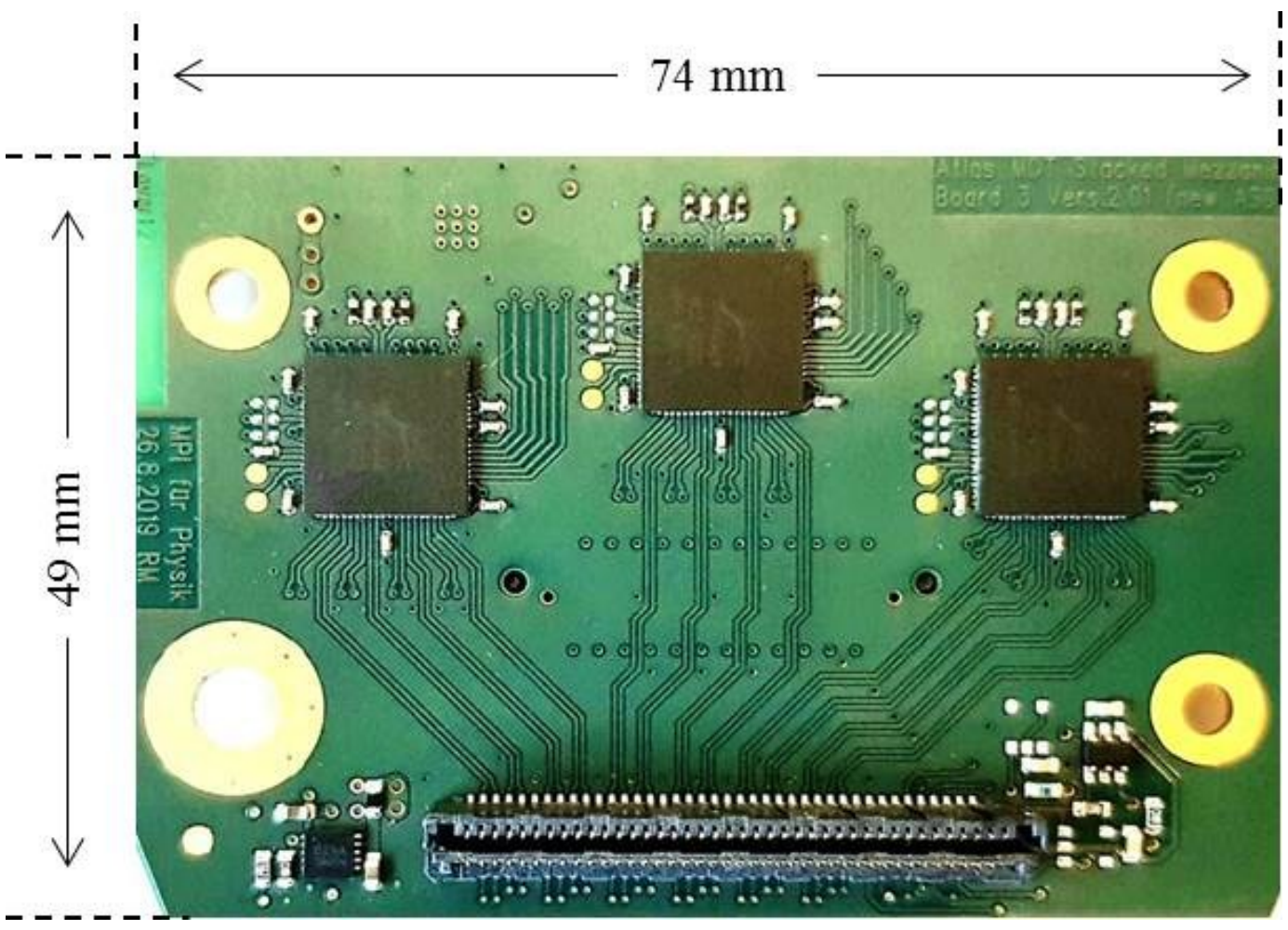

*Figure 41 - 3 ASD2 mounted on a PCB for the sMDT readout*

Power consumption and expected junction temperature of the ASD2 were evaluated for supply voltages 3.3, 3.0 and 2.7 V. Figure 42 shows the corresponding temperatures of the QFN package and the dissipation of the heat in the PCB, measured with an IFR camera. Good heat dissipation in the PCB is essential for keeping a low junction temperature of the chip. It requires tight contact between package and PCB as well as sufficient copper inside the board. As seen in the Figure 42, temperatures of ASD case are around 45 °C. Conservatively, one can expect the real junction temperature to be below 50°C as the thermal resistance junction-to-case for this type of package is about 1 °C/W, see Ref.[24]. With the dissipated power of the ASD of 0.3 W this leads to a negligible difference of 0.3 °C between the measured case temperature and the junction temperature. With the maximum operating temperature of the chip being specified as 120°C, the measured values around 50°C provide sufficient safety margin and concerns about chip lifetime can be safely discarded.

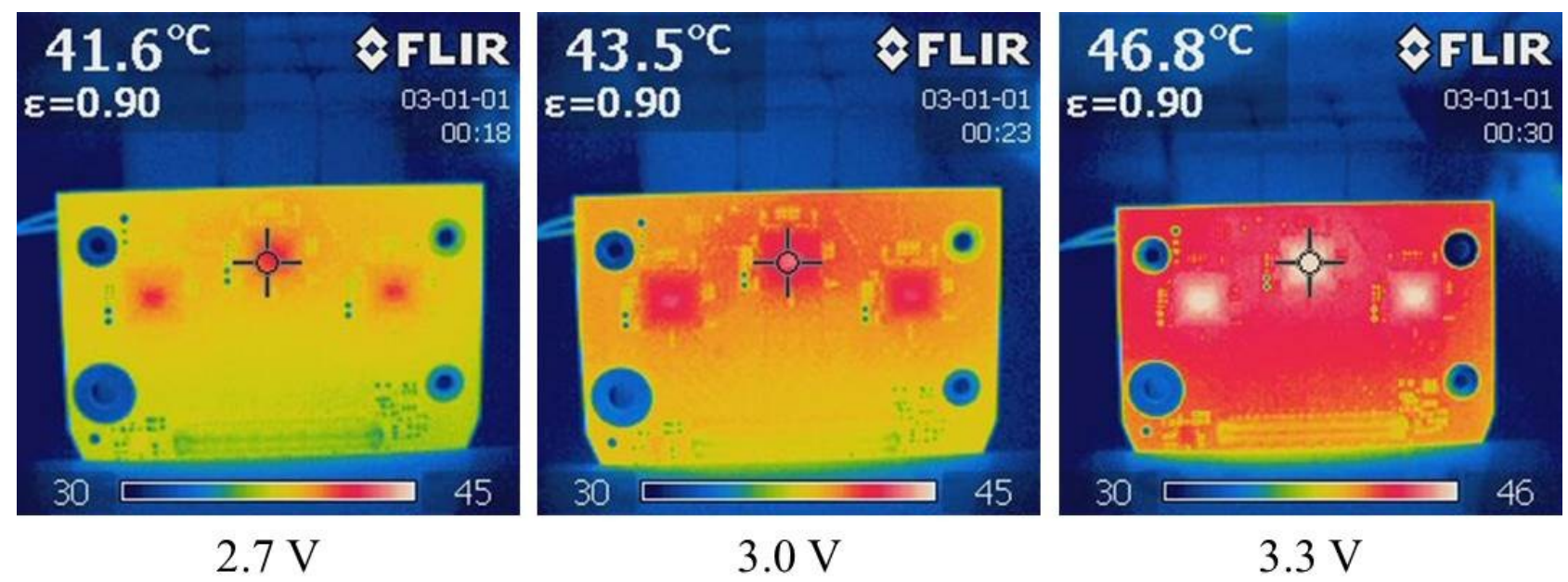

*Figure 42 - Temperature distribution on the PCB of Figure 41, operated at 3 supply voltages*

Table 14 gives power consumption and temperatures measured on the QFN88 package as shown in Figure 42. The operating voltage of 3.0V has been selected for $V_{cc}$. Tests have shown that this reduction of the supply voltage does not result in any significant performance degradation compared to operation at 3.3 V.

*Table 14 - Power consumption of the ASD2 vs. supply voltage Vcc*

| Vcc | Current | Power/ASD | | $T_{package}$ |
|---|---|---|---|---|
| *V* | *mA* | *mW* | *%* | *°C* |
| 3.3 | 116 | 382 | 49% | 47 |
| 3 | 85 | 256 | 0% | 43.5 |
| 2.7 | 70 | 189 | -26% | 41.5 |

The simulated power consumption of the ASD2 subsystems is listed in Table 15.

*Table 15 - Power consumption of the ASD components from simulation*

| Function | Name | **Operation @ 3.3 V** | | **Operation @ 3.0 V** | |
|---|---|---|---|---|---|
| | | I / chip | N / chip | I / chip | N / chip |
| | | *mA* | *mW* | *mA* | *mW* |
| Charge Sensing Pre- | $I_{VDD1}$ | 31.5 | 104.0 | 28.6 | 85.8 |
| Shaping Stages DA1- | $I_{VDD2}$ | 28.7 | 94.7 | 26.1 | 78.3 |
| Discriminators & | $I_{VDD3}$ | 43.1 | 142.2 | 39.2 | 117.6 |
| MUX & LVDS Drivers | $I_{VDDA}$ | 17.1 | 56.4 | 15.6 | 46.8 |
| Common Block | $I_{VDDX}$ | 1.1 | 3.6 | 1 | 3.0 |
| External Bias | $I_{BIAS}$ | 0.09 | 0.3 | 0.08 | 0.2 |
| Total Current & Power simulated | | 122 | 401 | 111 | 332 |
| Total Current & Power measured | | 121 | 399 | 102 | 306 |

# 5 Summary

Design, implementation and measured performance of the new ASD2 have been presented in this manual. Compared to the previous version of this chip (ASD1), presently used for the readout of the MDT chambers, the ASD2 shows improved performance in critical analog parameters like gain and peak time. Noise and threshold spread, both limiting for efficiency and time resolution, could be reduced by factors of about 2 and 3, respectively. In combination with the shorter peak time, this leads to a reduction of the time slewing effect for detector signals with small amplitude, resulting in substantially improved spatial resolution compared to ASD1. Figure 43 shows the spatial resolution in MDT chambers as measured with muon tracks of 150 GeV as a function of the γ-conversion rate.

The immunity of the chip against ionizing radiation has been verified at the CERN X-ray facility. While a Total Ionizing Dose (TID) of up to 20 krad is anticipated for the "hottest" regions of the muon spectrometer, tests up to 1 Mrad did not show performance degradation of the chip in any measured parameter.

The effectiveness of HV protection was tested with a generator, injecting negative and positive going pulses with up to 3 kV into the protective network, implemented inside and outside the chip, but no damage to the tested devices could be detected.

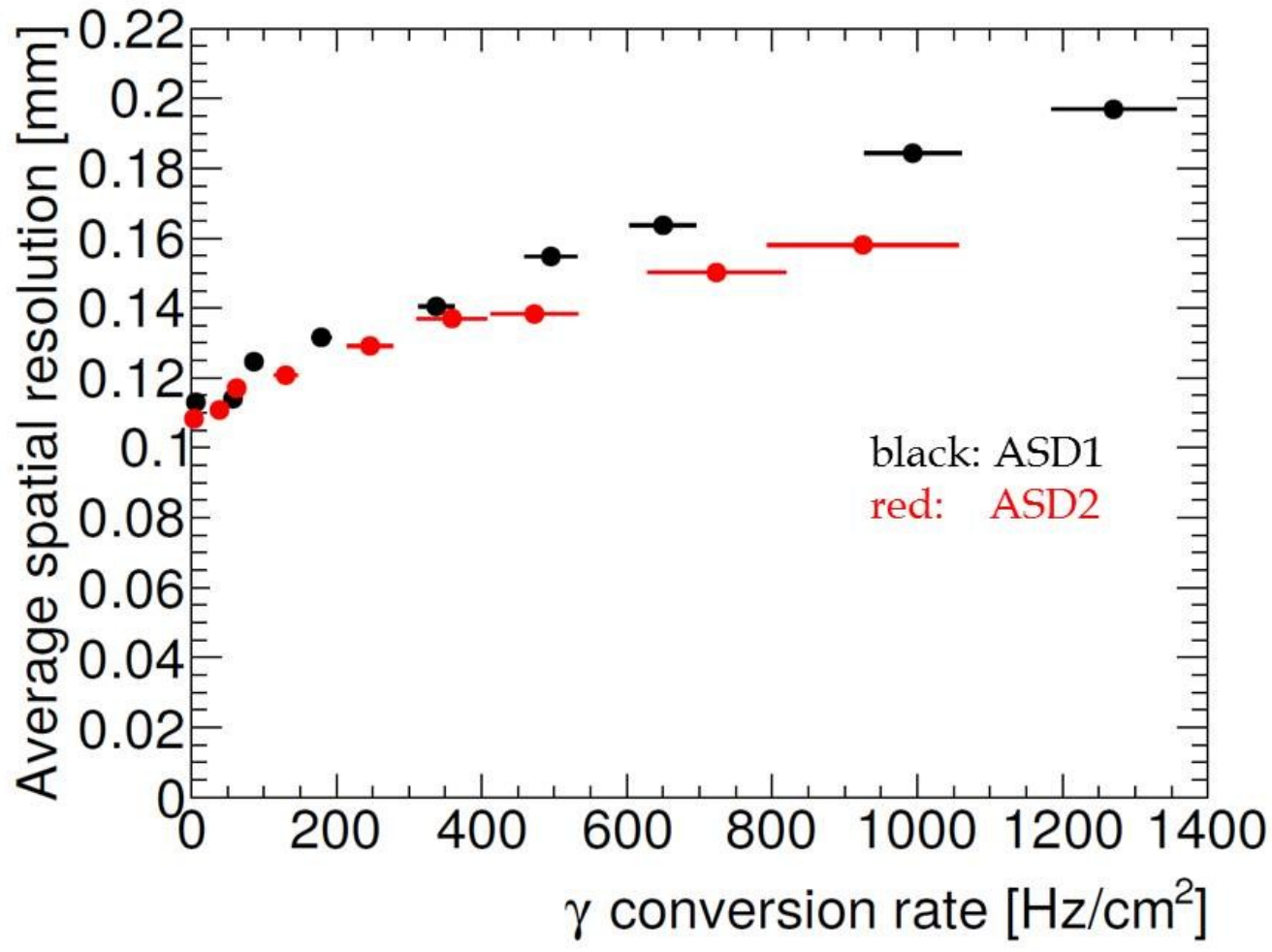

*Figure 43 - Spatial resolution of ASD1 and ASD2 vs. background rate in a test beam*

Known functional problems of ASD1, like amplitude instabilities of the LVDS outputs in response to noise hits (causing the "pair mode problem" in the TDC) and the failure to generate a Dead Time after a Discriminator trigger for small pulses (cf. section 4.1.7) have been corrected. In practical operation of ASD1, a high value of the hysteresis code was used to reduce the rate of faulty hits.

Full production was ordered with Global Foundry in 2019. After tests of 3500 chips from the engineering run, the full production of 24 wafers, corresponding to about 80000 chips was initiated. An automatic test facility was developed for the volume test. The resulting yield of fully functional chips was found to be > 90%.

# 6 Appendix – Packaging Information

# 7 Pin out Diagram of ASD2

The different groups of supply voltages and corresponding grounds have been chosen to minimize coupling among the functional blocks of the chip (cf. Figure 4), in particular between the analog and the digital part. Thus, VDD1 and GRND1 serve the CSP, while VDD2 supplies power to the amplification and shaping stages DA1, DA2 and DA3. VDD3, VDD_BUFF and VDD_JTAG serve the discriminator, the readings and the slow digital control system ("JTAG"). The pins ANAA and ANAB are the analog outputs of the APD, which allow to monitor the output of shaping stage DA3 (section 3.4.5). The Bonding Diagram of the ASD2 inside the QFN88 package is shown in Figure 44 - The Bonding Diagram of the ASD2 chi The corresponding bonding protocol (supplied by the GREATEK company) is shown in Figure 45.

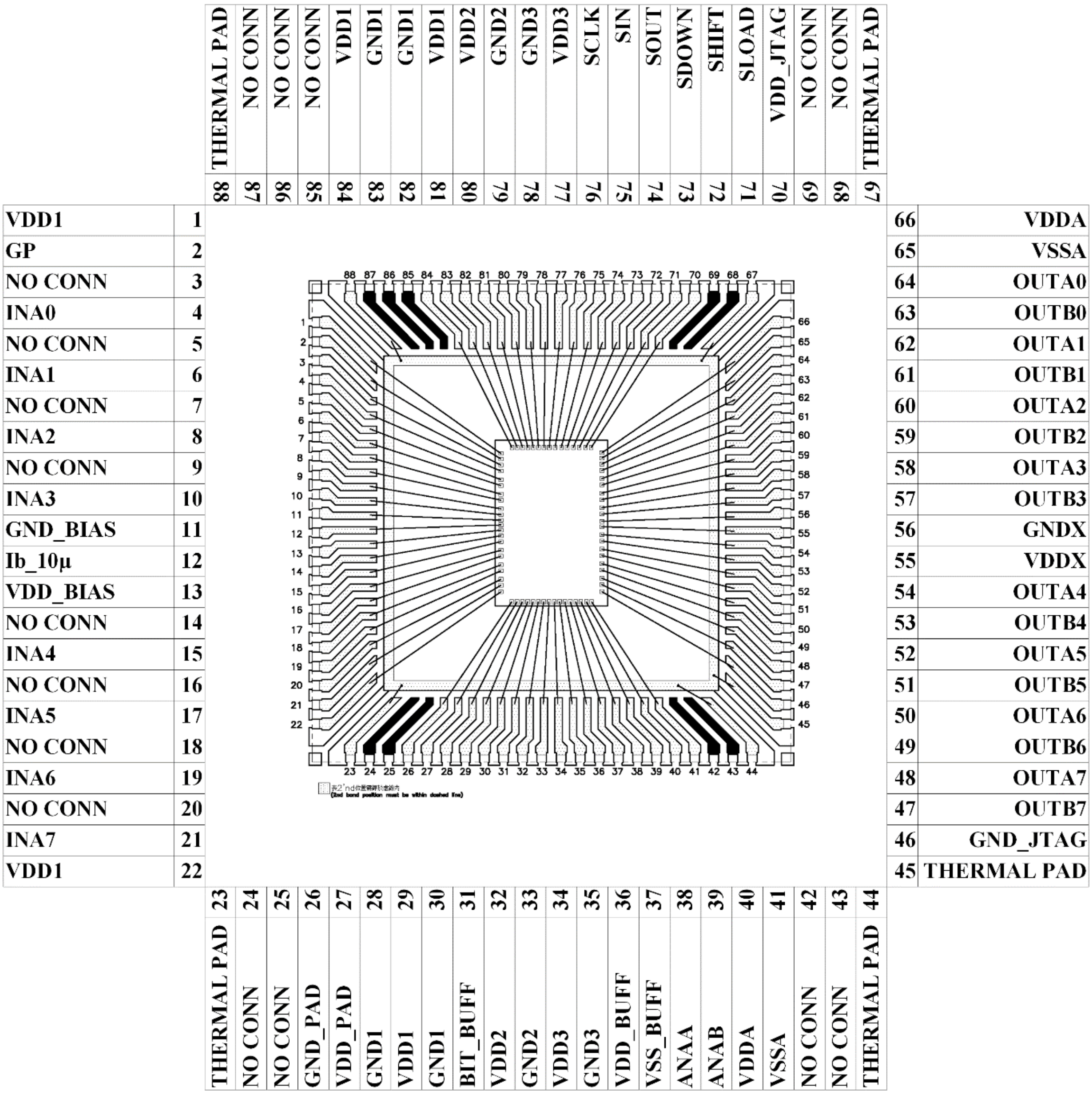

*Figure 44 - The Bonding Diagram of the ASD2 chip*

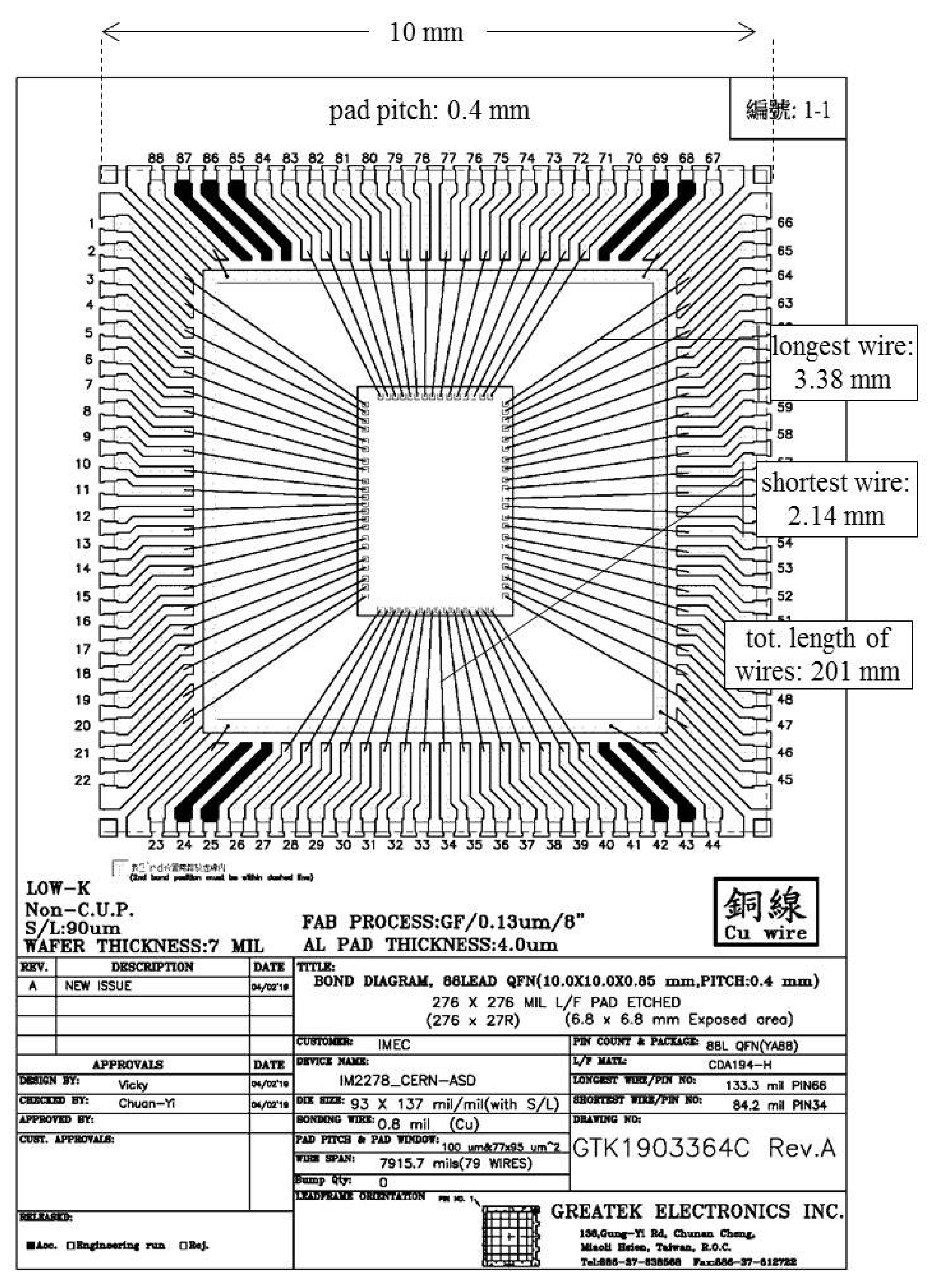

Figure 45- Bonding Protocol for ASD2

# 8 The Facility for Automatic ASD Testing

All functions for chip testing are integrated on one PC board and are controlled by a central micro-processor. Special care has been taken to avoid influence from external noise sources. As an example: test pulses are generated on-board by local pulse generators, the output of which is transferred to the input of the ASDs via capacitative coupling between adjacent layers of the PCB, while interfacing external devices goes via optical insulators to prevent ground loops.

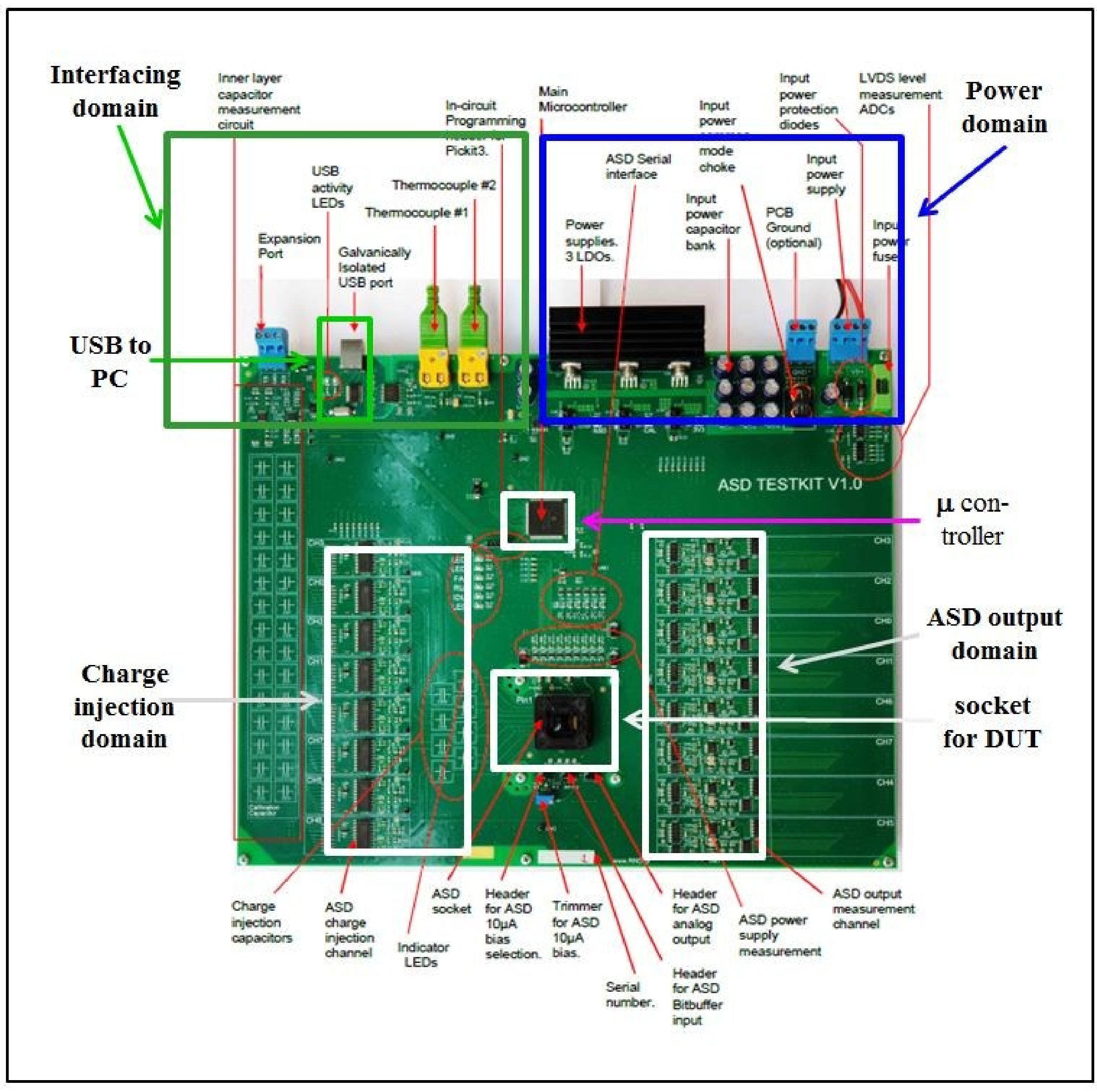

*Figure 46- The Automatic test Facility*